\documentclass[10pt,journal,compsoc]{IEEEtran}
\usepackage{url}
\usepackage[utf8]{inputenc}
\usepackage[table]{xcolor}
\usepackage{amsmath}
\usepackage{amssymb}

\usepackage[acronyms,nonumberlist,nopostdot,nomain,nogroupskip]{glossaries}
\usepackage{tablefootnote}
\usepackage{booktabs}

\usepackage{tabularx}

\usepackage{tikz}
\usepackage{pgfplots}
\pgfplotsset{compat=newest} 
\pgfplotsset{plot coordinates/math parser=false} 
\newlength\fheight
\newlength\fwidth
\usetikzlibrary{plotmarks,patterns,decorations.pathreplacing,backgrounds,calc,arrows,arrows.meta,spy,matrix}
\usepgfplotslibrary{patchplots,groupplots}
\usepackage{tikzscale}
\newif\ifexttikz
\exttikzfalse
\ifexttikz
	\usetikzlibrary{external}
\fi

\usepackage{multirow}
\usepackage{tkz-kiviat}

\usepackage[font=scriptsize]{subcaption}
\usepackage[font=footnotesize]{caption}

\usepackage{mathtools}

\usepackage{algorithm}
\usepackage{algorithmic}

\newacronym{arma}{ARMA}{Auto-Regressive Moving Average}
\newacronym{3gpp}{3GPP}{3rd Generation Partnership Project}
\newacronym{adc}{ADC}{Analog to Digital Converter}
\newacronym{5g}{5G}{5th generation}
\newacronym{aimd}{AIMD}{Additive Increase Multiplicative Decrease}
\newacronym{am}{AM}{Acknowledged Mode}
\newacronym{amc}{AMC}{Adaptive Modulation and Coding}
\newacronym{aqm}{AQM}{Active Queue Management}
\newacronym{awgn}{AGWN}{Additive White Gaussian Noise}
\newacronym{balia}{BALIA}{Balanced Link Adaptation}
\newacronym{bdp}{BDP}{Bandwidth-Delay Product}
\newacronym{bf}{BF}{Beamforming}
\newacronym{cc}{CC}{Congestion Control}
\newacronym{cdf}{CDF}{Cumulative Distribution Function}
\newacronym{cn}{CN}{Core Network}
\newacronym{cqi}{CQI}{Channel Quality Information}
\newacronym{cp}{CP}{Control Plane}
\newacronym{csirs}{CSI-RS}{Channel State Information - Reference Signal}
\newacronym{dc}{DC}{Dual Connectivity}
\newacronym{dce}{DCE}{Direct Code Execution}
\newacronym{dci}{DCI}{Downlink Control Information}
\newacronym{dl}{DL}{Downlink}
\newacronym{dmr}{DMR}{Deadline Miss Ratio}
\newacronym{dmrs}{DMRS}{DeModulation Reference Signal}
\newacronym{e2e}{E2E}{End-to-End}
\newacronym{ecn}{ECN}{Explicit Congestion Notification}
\newacronym{edf}{EDF}{Earliest Deadline First}
\newacronym{enb}{eNB}{evolved Node Base}
\newacronym{epc}{EPC}{Evolved Packet Core}
\newacronym{es}{ES}{Edge Server}
\newacronym{fdma}{FDMA}{Frequency Division Multiple Access}
\newacronym{fdd}{FDD}{Frequency Division Duplexing}
\newacronym[firstplural=Radio Access Technologies (RATs)]{rat}{RAT}{Radio Access Technology}
\newacronym{fs}{FS}{Fast Switching}
\newacronym{ftp}{FTP}{File Transfer Protocol}
\newacronym{gnb}{gNB}{Next Generation Node Base}
\newacronym{harq}{HARQ}{Hybrid Automatic Repeat reQuest}
\newacronym{hetnet}{HetNet}{Heterogeneous Network}
\newacronym{hh}{HH}{Hard Handover}
\newacronym{hol}{HOL}{Head-of-Line}
\newacronym{ia}{IA}{Initial Access}
\newacronym{imt}{IMT}{International Mobile Telecommunication}
\newacronym{iot}{IoT}{Internet of Things}
\newacronym{los}{LOS}{Line-of-Sight}
\newacronym{lte}{LTE}{Long Term Evolution}
\newacronym{m2m}{M2M}{Machine to Machine}
\newacronym{mac}{MAC}{Medium Access Control}
\newacronym{mc}{MC}{Multi-Connectivity}
\newacronym{mcs}{MCS}{Modulation and Coding Scheme}
\newacronym{mec}{MEC}{Mobile Edge Cloud}
\newacronym{mi}{MI}{Mutual Information}
\newacronym{mimo}{MIMO}{Multiple Input, Multiple Output}
\newacronym{mmwave}{mmWave}{millimeter wave}
\newacronym{mptcp}{MPTCP}{Multipath TCP}
\newacronym{mr}{MR}{Maximum Rate}
\newacronym{mss}{MSS}{Maximum Segment Size}
\newacronym{mtd}{MTD}{Machine-Type Device}
\newacronym{mtu}{MTU}{Maximum Transmission Unit}
\newacronym{nfv}{NFV}{Network Function Virtualization}
\newacronym{nlos}{NLOS}{Non-Line-of-Sight}
\newacronym{nr}{NR}{New Radio}
\newacronym{ofdm}{OFDM}{Orthogonal Frequency Division Multiplexing}
\newacronym{pdcch}{PDCCH}{Physical Downlink Control Channel}
\newacronym{pdcp}{PDCP}{Packet Data Convergence Protocol}
\newacronym{pdsch}{PDSCH}{Physical Downlink Shared Channel}
\newacronym{pdu}{PDU}{Packet Data Unit}
\newacronym{pf}{PF}{Proportional Fair}
\newacronym{pgw}{PGW}{Packet Gateway}
\newacronym{phy}{PHY}{Physical}
\newacronym{pbch}{PBCH}{Physical Broadcast Channel}
\newacronym[plural=\gls{mme}s,firstplural=Mobility Management Entities (MMEs)]{mme}{MME}{Mobility Management Entity}
\newacronym{prb}{PRB}{Physical Resource Block}
\newacronym{pss}{PSS}{Primary Synchronization Signal}
\newacronym{pucch}{PUCCH}{Physical Uplink Control Channel}
\newacronym{pusch}{PUSCH}{Physical Uplink Shared Channel}
\newacronym{rach}{RACH}{Random Access Channel}
\newacronym{ran}{RAN}{Radio Access Network}
\newacronym{red}{RED}{Random Early Detection}
\newacronym{rf}{RF}{Radio Frequency}
\newacronym{rlc}{RLC}{Radio Link Control}
\newacronym{rlf}{RLF}{Radio Link Failure}
\newacronym{rrc}{RRC}{Radio Resource Control}
\newacronym{rrm}{RRM}{Radio Resource Management}
\newacronym{rr}{RR}{Round Robin}
\newacronym{rs}{RS}{Remote Server}
\newacronym{rsrp}{RSRP}{Reference Signal Received Power}
\newacronym{rss}{RSS}{Received Signal Strength}
\newacronym{rtt}{RTT}{Round Trip Time}
\newacronym{rw}{RW}{Receive Window}
\newacronym{rx}{RX}{Receiver}
\newacronym{sa}{SA}{standalone}
\newacronym{sack}{SACK}{Selective Acknowledgment}
\newacronym{sap}{SAP}{Service Access Point}
\newacronym{sch}{SCH}{Secondary Cell Handover}
\newacronym{scoot}{SCOOT}{Split Cycle Offset Optimization Technique}
\newacronym{sdma}{SDMA}{Spatial Division Multiple Access}
\newacronym{sinr}{SINR}{Signal to Interference plus Noise Ratio}
\newacronym{sm}{SM}{Saturation Mode}
\newacronym{snr}{SNR}{Signal-to-Noise-Ratio}
\newacronym{son}{SON}{Self-Organizing Network}
\newacronym{ss}{SS}{Synchronization Signal}
\newacronym{srs}{SRS}{Sounding Reference Signal}
\newacronym{sss}{SSS}{Secondary Synchronization Signal}
\newacronym{tb}{TB}{Transport Block}
\newacronym{tcp}{TCP}{Transmission Control Protocol}
\newacronym{tdd}{TDD}{Time Division Duplexing}
\newacronym{tdma}{TDMA}{Time Division Multiple Access}
\newacronym{tfl}{TfL}{Transport for London}
\newacronym{tm}{TM}{Transparent Mode}
\newacronym{trp}{TRP}{Transmitter Receiver Pair}
\newacronym{tti}{TTI}{Transmission Time Interval}
\newacronym{ttt}{TTT}{Time-to-Trigger}
\newacronym{tx}{TX}{Transmitter}
\newacronym{ue}{UE}{User Equipment}
\newacronym{ul}{UL}{Uplink}
\newacronym{uml}{UML}{Unified Modeling Language}
\newacronym{um}{UM}{Unacknowledged Mode}
\newacronym{utc}{UTC}{Urban Traffic Control}
\newacronym{vm}{VM}{Virtual Machine}
\newacronym{rsrq}{RSRQ}{Reference Signal Received Quality}
\newacronym{rssi}{RSSI}{Received Signal Strength Indicator}
\newacronym{crs}{CRS}{Cell Reference Signal}
\newacronym{nsa}{NSA}{Non Stand Alone}
\newacronym{mrdc}{MR-DC}{Multi \gls{rat} \gls{dc}}
\newacronym{endc}{EN-DC}{E-UTRAN-\gls{nr} \gls{dc}}
\newacronym{5gc}{5GC}{5G Core}
\newacronym{si}{SI}{Study Item}
\newacronym{iab}{IAB}{Integrated Access and Backhaul}
\newacronym{wf}{WF}{Wired-first}
\newacronym{hqf}{HQF}{Highest-quality-first}
\newacronym{pa}{PA}{Position-aware}
\newacronym{mlr}{MLR}{Maximum-local-rate}
\newacronym{wbf}{WBF}{Wired Bias Function}
\newacronym{mib}{MIB}{Master Information Block}
\newacronym{sib}{SIB}{Secondary Information Block}
\newacronym{kpi}{KPI}{Key Performance Indicator}
\newacronym{ppp}{PPP}{Poisson Point Process}
\newacronym{du}{DU}{Distributed Unit}
\newacronym{cu}{CU}{Centralized Unit}
\newacronym{cran}{CRAN}{Cloud \gls{ran}}
\newacronym{sdn}{SDN}{Software Defined Networking}
\newacronym{ml}{ML}{Machine Learning}
\newacronym{qoe}{QoE}{Quality of Experience}
\newacronym{qos}{QoS}{Quality of Service}
\newacronym{imsi}{IMSI}{International Mobile Subscriber Identity}
\newacronym{bbu}{BBU}{Base Band Unit}
\newacronym{onap}{ONAP}{Open Network Automation Platform}
\newacronym[firstplural=Estimated Times of Arrival (ETAs)]{eta}{ETA}{Estimated Time of Arrival}
\newacronym{rmse}{RMSE}{Root Mean Squared Error}
\newacronym{brr}{BRR}{Bayesian Ridge Regressor}
\newacronym{gpr}{GPR}{Gaussian Process Regressor}
\newacronym{rfr}{RFR}{Random Forest Regressor}
\newacronym{ai}{AI}{Artificial Intelligence}
\newacronym{lstm}{LSTM}{Long Short Term Memory}
\newacronym{ru}{RU}{Radio Unit}
\tikzstyle{startstop} = [rectangle, rounded corners, minimum width=2cm, minimum height=0.5cm,text centered, draw=black]
\tikzstyle{io} = [trapezium, trapezium left angle=70, trapezium right angle=110, minimum width=3cm, minimum height=1cm, text centered, draw=black]
\tikzstyle{process} = [rectangle, minimum width=2cm, minimum height=0.5cm, text centered, draw=black, alignb=center]
\tikzstyle{decision} = [ellipse, minimum width=2cm, minimum height=1cm, text centered, draw=black]
\tikzstyle{arrow} = [thick,<->,>=stealth]
\tikzstyle{line} = [thick,>=stealth]
\tikzstyle{darrow} = [thick,<->,>=stealth,dashed]
\tikzstyle{sarrow} = [thick,->,>=stealth]
\tikzstyle{larrow} = [line width=0.1mm,dashdotted,->,>=stealth]

\makeatletter
\def\grd@save@target#1{%
  \def\grd@target{#1}}
\def\grd@save@start#1{%
  \def\grd@start{#1}}
\tikzset{
  grid with coordinates/.style={
    to path={%
      \pgfextra{%
        \edef\grd@@target{(\tikztotarget)}%
        \tikz@scan@one@point\grd@save@target\grd@@target\relax
        \edef\grd@@start{(\tikztostart)}%
        \tikz@scan@one@point\grd@save@start\grd@@start\relax
        \draw[minor help lines] (\tikztostart) grid (\tikztotarget);
        \draw[major help lines] (\tikztostart) grid (\tikztotarget);
        \grd@start
        \pgfmathsetmacro{\grd@xa}{\the\pgf@x/1cm}
        \pgfmathsetmacro{\grd@ya}{\the\pgf@y/1cm}
        \grd@target
        \pgfmathsetmacro{\grd@xb}{\the\pgf@x/1cm}
        \pgfmathsetmacro{\grd@yb}{\the\pgf@y/1cm}
        \pgfmathsetmacro{\grd@xc}{\grd@xa + \pgfkeysvalueof{/tikz/grid with coordinates/major step x}}
        \pgfmathsetmacro{\grd@yc}{\grd@ya + \pgfkeysvalueof{/tikz/grid with coordinates/major step y}}
        \foreach \x in {\grd@xa,\grd@xc,...,\grd@xb}
        \node[anchor=north] at (\x,\grd@ya) {\pgfmathprintnumber{\x}};
        \foreach \y in {\grd@ya,\grd@yc,...,\grd@yb}
        \node[anchor=east] at (\grd@xa,\y) {\pgfmathprintnumber{\y}};
      }
    }
  },
  minor help lines/.style={
    help lines,
    gray,
    line cap =round,
    xstep=\pgfkeysvalueof{/tikz/grid with coordinates/minor step x},
    ystep=\pgfkeysvalueof{/tikz/grid with coordinates/minor step y}
  },
  major help lines/.style={
    help lines,
    line cap =round,
    line width=\pgfkeysvalueof{/tikz/grid with coordinates/major line width},
    xstep=\pgfkeysvalueof{/tikz/grid with coordinates/major step x},
    ystep=\pgfkeysvalueof{/tikz/grid with coordinates/major step y}
  },
  grid with coordinates/.cd,
  minor step x/.initial=.5,
  minor step y/.initial=.2,
  major step x/.initial=1,
  major step y/.initial=1,
  major line width/.initial=1pt,
}
\makeatother

\definecolor{desireRed}{RGB}{230,57,60}%
\definecolor{darkPurple}{RGB}{59,31,43}%
\definecolor{springGreen}{RGB}{37,223,145}%
\definecolor{queenBlue}{RGB}{69,123,157}%
\definecolor{spaceCadet}{RGB}{29,53,87}%

\usepackage{dblfloatfix}    % To enable figures at the bottom of page
\usepackage{epstopdf}

\makeglossaries

\begin{document}

\title{Machine Learning at the Edge:\\A Data-Driven Architecture with Applications\\to 5G Cellular Networks}

\author{\IEEEauthorblockN{Michele Polese, \IEEEmembership{Member, IEEE}, Rittwik Jana, \IEEEmembership{Member, IEEE},\\Velin Kounev, Ke Zhang, Supratim Deb, \IEEEmembership{Senior Member, IEEE},\\Michele Zorzi, \IEEEmembership{Fellow, IEEE}}\vspace{-.5cm}
\thanks{Michele Polese was with the Department of Information Engineering (DEI), University of Padova, Padova, 35131 Italy, and is now with the Institute for the Wireless Internet of Things, Northeastern University, Boston, MA 02120 USA. Email: m.polese@northeastern.edu.}
\thanks{Rittwik Jana and Velin Kounev are with AT\&T Labs, Bedminster, NJ 07921 USA. Email: rjana@research.att.com, vk0366@att.com. Ke Zhang and Supratim Deb were with AT\&T Labs, and are currently at Dataminr and Facebook, respectively. Email: \{zhangke290, supratim.deb\}@gmail.com.}
\thanks{Michele Zorzi is with the Department of Information Engineering (DEI), University of Padova, 35131 Italy.
Email:zorzi@dei.unipd.it.}
\thanks{This work was supported in part by Supporting Talent in Research@University of Padua: STARS Grants, through the project ``Cognition-Based Networks: Building the Next Generation of Wireless Communications Systems Using Learning and Distributed Intelligence'', and has been presented in part in~\cite{polese2019exploiting}.}
}

% \makeatletter
% \patchcmd{\@maketitle}
%   {\addvspace{0.5\baselineskip}\egroup}
%   {\addvspace{-1.7\baselineskip}\egroup}
%   {}
%   {}
% \makeatother

\flushbottom
\setlength{\parskip}{0ex plus0.1ex}

\maketitle

\glsunset{nr}

\begin{abstract}

The fifth generation of cellular networks (5G) will rely on edge cloud deployments to satisfy the ultra-low latency demand of future applications. In this paper, we argue that such deployments can also be used to enable advanced data-driven and \gls{ml} applications in mobile networks. We propose an edge-controller-based architecture for cellular networks and evaluate its performance with real data from hundreds of base stations of a major U.S. operator. In this regard, we will provide insights on how to dynamically cluster and associate base stations and controllers, according to the global mobility patterns of the users. Then, we will describe how the controllers can be used to run \gls{ml} algorithms to predict the number of users in each base station, and a use case in which these predictions are exploited by a higher-layer application to route vehicular traffic according to network \glspl{kpi}. We show that the prediction accuracy improves when based on machine learning algorithms that rely on the controllers' view and, consequently, on the spatial correlation introduced by the user mobility, with respect to when the prediction is based only on the local data of each single base station.

\end{abstract}

\begin{tikzpicture}[remember picture,overlay]
\node[anchor=north,yshift=-10pt] at (current page.north) {\parbox{\dimexpr\textwidth-\fboxsep-\fboxrule\relax}{\centering \footnotesize This paper has been published on IEEE Transactions on Mobile Computing. \textcopyright 2020 IEEE.\\
  Please cite it as: M. Polese, R. Jana, V. Kounev, K. Zhang, S. Deb, M. Zorzi, ``Machine Learning at the Edge: A Data-Driven Architecture with Applications to 5G Cellular Networks,'' in IEEE Transactions on Mobile Computing, doi: 10.1109/TMC.2020.2999852.}};
\end{tikzpicture}%

% \vspace{-.5cm}

\begin{IEEEkeywords}
% \vspace{-.3cm}
5G, machine learning, edge, controller, prediction, mobility, big data.
\end{IEEEkeywords}
% \vspace{-.5cm}

\section{Introduction}
\label{sec:intro}

The \gls{5g} of cellular networks is being designed to satisfy the massive growth in capacity demand, number of connections and the evolving use cases of a connected society for 2020 and beyond~\cite{cisco2017}. In particular, \gls{5g} networks target the following \glspl{kpi}: (i) very high throughput, in the order of 1 Gbps or more, to enable virtual reality applications and high-quality video streaming; (ii) ultra-low latency, possibly smaller than 1 ms on the wireless link, to support autonomous control applications; (iii) ultra-high reliability; (iv) low energy consumption; and (v) high availability of robust connections~\cite{ngmn5g,boccardi2014five}. 

In order to meet these requirements, a new approach in the design of the network is required, and new paradigms have recently emerged~\cite{boccardi2014five}. First, the densification of the network will increase the spatial reuse and, combined with the usage of mmWave frequencies, the available throughput. On the other hand, this will introduce new challenges related to mobility management~\cite{rangan2017potentials}. Second, with \gls{mec}, the content will be brought closer to the final users, in order to decrease the end-to-end latency~\cite{boccardi2014five}. Third, a higher level of automation will be introduced in cellular networks, relying on \gls{ml} techniques and \gls{sdn}, in order to manage the increased complexity of 5G networks.

The usage of \gls{ml} and \gls{ai} techniques to perform autonomous operations in cellular networks has been widely studied in recent years, with use cases that range from optimization of video flows~\cite{zorzi2015cobanets} to energy-efficient networks~\cite{li2017intelligent} and resource allocation~\cite{chinchali2018cellular}. This trend is coupled with the application of big-data analytics that leverage the huge amount of monitoring data generated in mobile networks to provide more insights on the behavior of networks at scale~\cite{he2016bigdata}. In the domain of mobile networks, these two technological components can empower costs savings, but also new applications, as we will show in this paper. However, despite the importance of this topic, the state of the art lacks considerations on how it is possible to effectively deploy machine learning algorithms and intelligence in cellular networks, and an evaluation of the gains of a data-driven approach with real large-scale network datasets.

% \vspace{-.2cm}
\subsection{Contributions}
\label{sec:contr}
% \vspace{-.1cm}

\begin{table*}
  \caption{Relevant literature on machine learning, \gls{mec} and edge controllers in cellular networks and novel contributions of this paper.}
  \label{table:related_work}
\footnotesize
  \centering
  \def\tabularxcolumn#1{m{#1}}
    \renewcommand{\arraystretch}{1}% Tighter
  % \begin{tabularx}{0.95\textwidth}{@{}X|X|X@{}}
  \begin{tabularx}{\textwidth}{@{}>{\hsize=0.5\hsize}X>{\hsize=0.5\hsize}X>{\hsize=1.3\hsize}X@{}}
  \toprule
  Topic & Relevant References & Contribution of this paper \\ \midrule
  Application of \gls{ml} in cellular networks & \cite{bui2017survey,pejovic2015anticipatory,jiang2017machine,he0216big,imran2014challenges,winstein2013tcp,gadaleta2017ddash} & Novel network-level architecture, integrated with \gls{3gpp} 5G specifications, and evaluation of its performance gains based on real network dataset. \\  \midrule
  Mobility prediction in cellular networks & \cite{becker2013human,becker2011tale,dong2013modeling} & Cluster-based approach to capture spatial correlation \\ \midrule
  \acrlong{mec} & \cite{tran2017collaborative,bastug2014living,boccardi2014five,darwish2018fog,rehman2017rededge} & \gls{mec}-based architecture used for \gls{ml} for network control and applications \\ \midrule
  \gls{sdn} in cellular networks & \cite{peng2016fog,xranwp,chen2015software,li2012towards,checko2015cloud,gudipati2013softran} & \gls{ml}-driven edge-\gls{sdn} controllers integrated in the \gls{ml} architecture\\ \bottomrule
  \end{tabularx}
  % \vspace{-.6cm}
\end{table*}

To address these limitations, in this paper we propose a data-driven control architecture for the practical implementation of \gls{ml} techniques in 5G cellular networks, and evaluate the gains that this architecture can introduce in some data-driven applications, using real data collected from hundreds of base stations of a major U.S. carrier in the San Francisco and Mountain View areas for more than a month. In particular, the main contributions of this paper are:
\begin{itemize}
	\item the design of a scalable and efficient multi-layer edge-based control architecture to deploy big-data and \gls{ml} applications in 5G systems. We propose to exploit controllers implemented in \gls{mec} and cloud facilities to collect the data generated by the network, run analytics and extract relevant metrics, that can be fed to intelligent algorithms to control the network itself and provide new services to the users. The \gls{ran} controllers, deployed at the edge, are associated with a cluster of base stations, and are thus responsible not only for \gls{ran} control, as proposed in~\cite{xranwp}, but also for running the data collection and \gls{ml} infrastructure. The network controller, placed in the operator's cloud, orchestrates the operations of the \gls{ran} controllers. 
	We characterize this architecture with respect to the latest 5G \gls{ran} specifications for \gls{3gpp} \gls{nr}, the 5G standard for cellular networks~\cite{38300}, and provide insights on how the controllers can interface with an \gls{nr} deployment, following the approach of an emerging open RAN initiative contributed by multiple operators and vendors~\cite{xranwp}. 

	\item the demonstration of the gains that data-driven techniques enabled by the proposed architecture can yield in network applications, leveraging a real world dataset on two use cases. 
	In the first, big data analytics are used to control the association between the base stations and the \gls{ran} controllers. We propose a \textit{dynamic clustering} method where base stations and controllers are grouped according to the day-to-day user mobility patterns, which are collected and processed by the \gls{ml} infrastructure. With respect to a static algorithm, based on the position of the base stations, the data-driven algorithm manages to decrease the number of inter-controller interactions and thus reduce the control plane latency.
	In the second example, we test different machine learning techniques (i.e., the \acrlong{brr}, the \acrlong{gpr} and the \acrlong{rfr}) for the \textit{prediction} of the number of users in the base stations of the network. We show that, thanks to the proposed \gls{ml} edge-based architecture, which makes it possible to exploit the spatial correlation of the  users, it is possible to increase the prediction accuracy with respect to that of decentralized schemes, with a reduction of the prediction error by up to 53\%.
\end{itemize} 

To the best of our knowledge, this is the first exhaustive contribution in which a practical control-plane \gls{ml} architecture, that can be applied on top of \gls{5g} \gls{nr} cellular networks, is evaluated using a real network dataset, showing promising results that indicate that new user services and optimization techniques based on machine learning in cellular networks are possible.

%  and compare an approach in which each base station predicts its number of users based only on local information compared to a strategy in which the controller predicts a vector with the number of users in all its base stations. In this third contribution, we show that it is possible to  on average,
% In the second part of the paper, in order to show why the controller-based architecture can be beneficial for machine learning applications, we present a use case which requires predicting the number of users in each base station at different time instants in the future. This application is a service that the network operator could offer to its customers that want to drive between two locations: given multiple routes available, which is the one the user should prefer in order to maximize its \gls{qos} in the network? We measure the \gls{qos} with different \glspl{kpi}, which can be computed as a function of the number of users attached to the base stations. 

% \vspace{-.4cm}
\subsection{Related Work}
\label{sec:rel_work}
% \vspace{-.2cm}

In the following paragraphs we will discuss the literature relevant to the scope of this paper, which is also summarized in Table~\ref{table:related_work}, and highlight the main differences we introduce with respect to the state of the art.

\begin{table*}[b]
  \centering
  \begin{tabular}[b]{@{}llll@{}}
  \toprule
      & Location & Time interval & Number of eNBs \\\midrule
    Campaign 1 & San Francisco & 01/31/2017 $-$ 02/26/2017, every day from 3 P.M. to 8 P.M. & 472 \\
    Campaign 2 & Palo Alto, Mountain View & 06/22/2018 $-$ 07/15/2018, whole day & 178 \\ 
    \bottomrule
  \end{tabular}
  \caption{Anonymized datasets used in this paper.}
  \label{table:meas}
\end{table*}

\begin{figure*}[t]
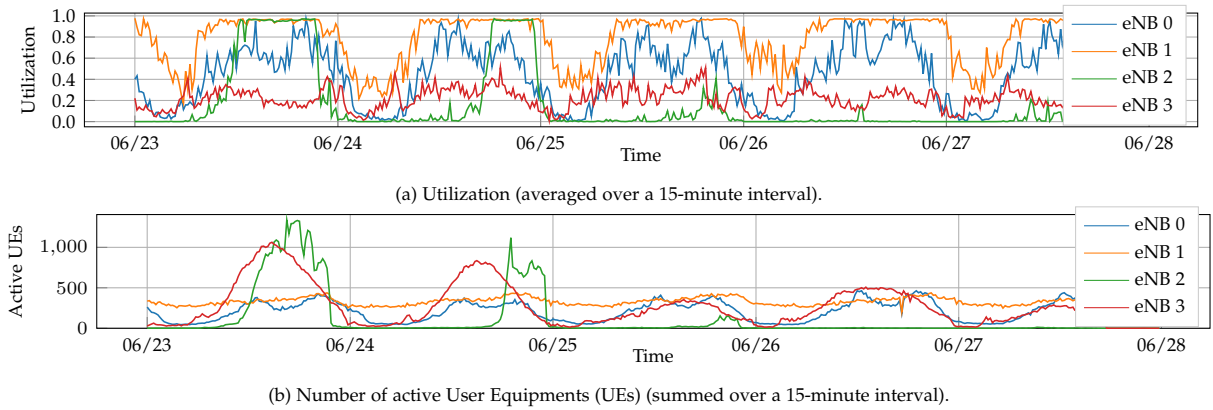

\centering
\begin{subfigure}[t]{\textwidth}
\centering
  \setlength\fwidth{.9\textwidth}
  \setlength\fheight{0.17\textwidth}
  \input{figures/utilquarter_hour_ue_timeseries_20180624.tex}
  \caption{Utilization (averaged over a 15-minute interval).}
  \label{fig:util}
\end{subfigure}
\begin{subfigure}[t]{\textwidth}
\centering
  \setlength\fwidth{.9\textwidth}
  \setlength\fheight{0.17\textwidth}
  \input{figures/numUequarter_hour_ue_timeseries_20180624.tex}
  \caption{Number of active \glspl{ue} (summed over a 15-minute interval).}
  \label{fig:numUe}
\end{subfigure}
% \begin{subfigure}[t]{\textwidth}
% \centering
% \setlength\abovecaptionskip{-.2cm}
%   \setlength\belowcaptionskip{-.3cm}
%   \setlength\fwidth{.9\textwidth}
%   \setlength\fheight{0.19\textwidth}
%   \input{figures/numHoInquarter_hour_ue_timeseries_20180624.tex}
%   \caption{Number of incoming handovers (summed over a 15-minute interval).}
%   \label{fig:numHo}
% \end{subfigure}
\caption{Example of timeseries from the traces collected for 4 \glspl{enb} in the Palo Alto dataset over 5 days.}
\label{fig:datasetmetrics}
\end{figure*}

\paragraph*{\textbf{\gls{ml} in cellular networks}}
The application of \gls{ml} techniques to cellular networks is a topic that has gained a lot of attention recently, thanks to the revived importance of \gls{ml} and \gls{ai} throughout all facets of the industry. 
%The paper~\cite{klaine2017survey} surveys algorithms and applications of \gls{ml} in 4G self-organizing networks. 
The surveys in~\cite{bui2017survey,pejovic2015anticipatory} present some recent results on how it is possible to apply regression techniques to mobile and cellular scenarios in order to optimize the network performance. The paper~\cite{jiang2017machine} gives an overview of how machine learning can play a role in next-generation 5G cellular networks, and lists relevant \gls{ml} techniques and algorithms. The usage of big-data-driven analytics for 5G is considered in~\cite{he0216big,imran2014challenges}, with a discussion of how data-driven approaches can empower self-organizing networks. However, none of these papers provides results based on real operators datasets at large scale that show the actual gains of data-driven and machine learning based approaches. Moreover, while practical implementations of machine learning algorithms for networks indeed exist for host-based applications (e.g., TCP~\cite{winstein2013tcp}, video streaming~\cite{gadaleta2017ddash}), or base-station-based use cases (e.g., scheduling~\cite{bui2018data}), the literature still lacks a discussion and an analysis of how it is possible to practically deploy the algorithms, collect real-time data and process it to enable new services in large-scale commercial networks. 
%The paper~\cite{malandrino2018cellular} discusses the role of network traces in 5G, but does not consider the real time collection and processing of the traces.

Furthermore, several papers report results on the prediction of mobility patterns of users in cellular networks. The authors of~\cite{becker2013human,becker2011tale} use network traces to study human mobility patterns, with the goal to infer large-scale patterns and understand city dynamics. The paper~\cite{dong2013modeling} proposes to use a leap graph to model the mobility pattern of single users. Other works focus on the prediction of the traffic generated by single base stations~\cite{xu2017high,sivakumar2011prediction}, or by groups of base stations~\cite{qiu2018spatio}, and do not consider the mobility patterns. With respect to the state of the art, in this paper we focus on the prediction of the number of users associated to a base station, in order to provide innovative services to the users themselves, and propose a novel cluster-based approach to improve the prediction accuracy, evaluating the performance of different algorithms on a real large-scale dataset.

\paragraph*{\textbf{\gls{mec} and controllers in cellular networks}}
The role of \gls{mec} has also been discussed in the context of 5G networks, e.g., to perform coordination~\cite{tran2017collaborative} and caching~\cite{bastug2014living}, and to offer low-latency content and control applications to the end users~\cite{boccardi2014five}. \gls{mec} is indeed considered a key element in the deployment of future autonomous driving vehicles, for which very short control loops will be needed~\cite{pandi2016joint}. A few papers consider specific cases for the application of machine learning and big data techniques at the edge, for example for intelligent transportation systems~\cite{darwish2018fog}, or the processing of data collected by internet-of-things devices~\cite{rehman2017rededge}, but, to the best of our knowledge, the usage of \gls{mec} to run data collection and machine learning algorithms for the prediction and optimization in 5G cellular networks has not been discussed in detail yet.

The edge has also been proposed for hosting controllers in cellular networks~\cite{peng2016fog,xranwp,chen2015software}. As the \gls{sdn} paradigm has become popular in wired networks~\cite{jain2013b4}, several software-defined approaches for the \gls{ran} have been described in the literature~\cite{li2012towards,checko2015cloud,gudipati2013softran}, and the telecom industry is moving towards open-controllers-based architectures for the deployment of 5G networks~\cite{xranwp}. With respect to existing studies, in this paper we propose to exploit the \gls{ran} controllers as proxies for the data collection in the \gls{ran} and the enforcement of machine learning algorithm-based policies. This approach has been considered in a wired-network context~\cite{cui2016when}, but this is the first paper that studies it in a 5G cellular network.

% controllers

% \vspace{-.8cm}
\subsection{Paper Structure}
% \vspace{-.1cm}

The remainder of the paper is organized as follows. In Sec.~\ref{sec:data} we present the real network data that will be used throughout the paper, and in Sec.~\ref{sec:ctrl} we describe the proposed architecture. In Sec.~\ref{sec:cluster} we provide details on the first application, i.e., the autonomous data-driven clustering of base stations. Results on the second application, i.e., the prediction accuracy for the number of users in the cells, are given in Sec.~\ref{sec:pred}, together with possible use cases. Finally, in Sec.~\ref{sec:concl} we conclude the paper.

% controllers

% \vspace{-.1cm}
\section{The Dataset}
\label{sec:data}
% \vspace{-.2cm}

This section describes the data that will be used in the evaluations in the remainder of the paper. The traces we exploit are based on the monitoring logs generated by 650 base stations of a national U.S. operator in two different areas, i.e., San Francisco and Palo Alto/Mountain View, for more than 600000 \glspl{ue} per day, properly anonymized during the collection phase. The base stations in the dataset belongs to a 4G LTE-A deployment, which represents the most advanced cellular technology commercially deployed at a large scale. Even if 5G \gls{nr} networks will have more advanced characteristics than \gls{lte} ones, this dataset can be seen as representative of an initial 5G deployment at sub-6 GHz frequencies in a dense urban scenario~\cite{8417851}. We consider two separate measurement campaigns, conducted in February 2017 in the San Francisco area and in June and July 2018 in the Palo Alto and Mountain View areas. Table~\ref{table:meas} summarizes the most relevant details of each measurement campaign. 

Given the sensitivity of this kind of data, we adopted standard procedures to ensure that individuals' privacy was not compromised during the data collection and the analysis. In particular, the records were anonymized by hashing the \glspl{ue}' \glspl{imsi}, which is the unique identifier that can be associated to a single customer in these traces. Moreover, for our analysis, we only used anonymized metrics that are based on aggregated usage at multiple layers: first, we consider users' data for each single cell (a cell is mapped to a sector and carrier frequency), and, then, aggregate the cells associated to the same base station (i.e., with the RF equipment in the same physical location). In this way, no user can be singled out by the results we present.

\begin{figure*}[t]
  \centering 
  \includegraphics[width=.85\textwidth]{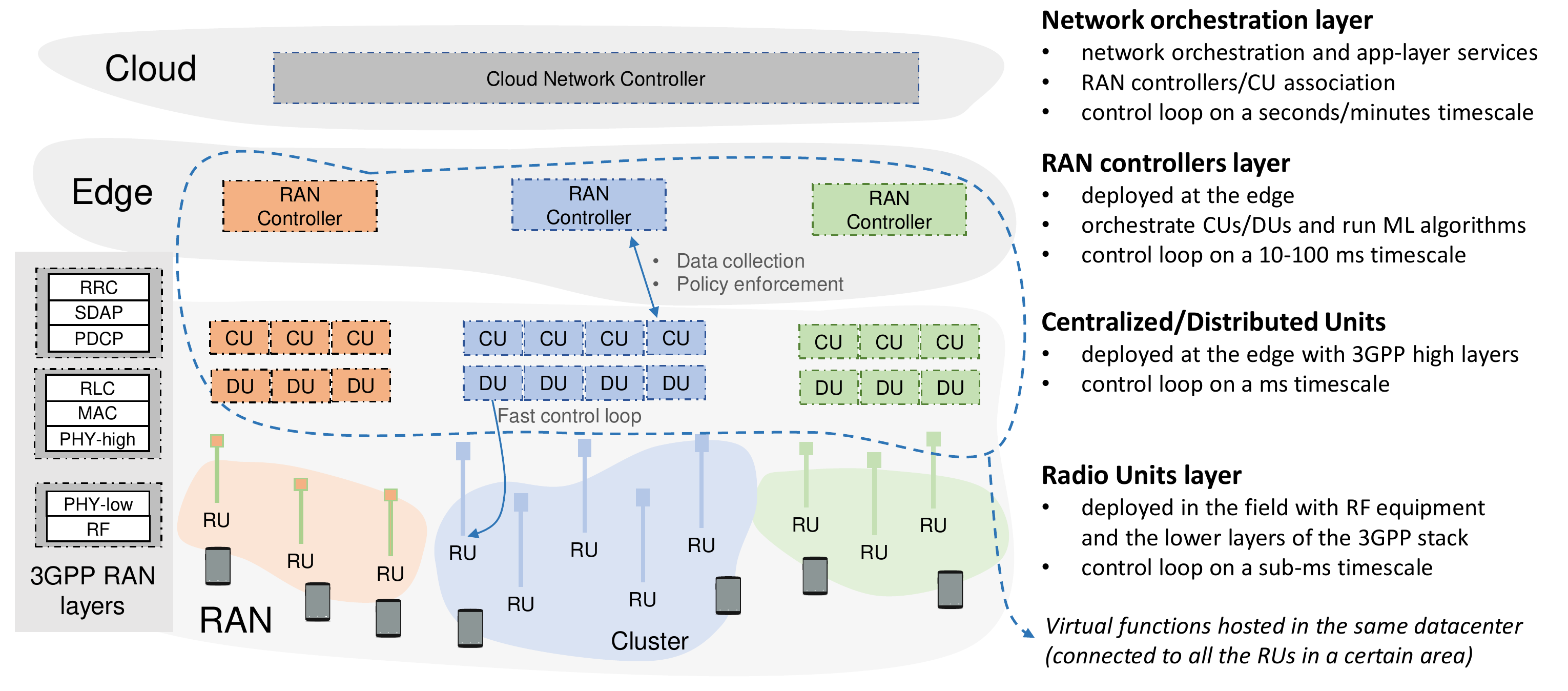}
  \caption{Proposed controller architecture for \gls{ran} control and machine learning at the edge.}
  \label{fig:ranCtrl}
\end{figure*}

The traces used in this paper record a set of standardized events in LTE \glspl{enb}, mainly related to the mobility of users. The raw data is further processed to construct time series of different quantities of interest in each \gls{enb} at different time scales (from minutes to weeks):
(i) the utilization of the \gls{enb}, which is represented by the ratio of used and available \glspl{prb};
(ii) the number of incoming and outgoing handovers, for both X2 and S1 handover events~\cite{36300};
and (iii) the number of active \glspl{ue}, obtained from context setup and release events. The measurement framework we used also offered the possibility of logging other events and extract other metrics, for example related to the latency experienced by the users, link statistics (e.g., error probability), or different estimates of the user and cell throughput. The events associated to these quantities, however, are reported less regularly and less frequently than those we consider, therefore they do not represent a reliable source for the estimation of the network performance. With respect to other publicly available datasets~\cite{dandelion}, this presents a more precise characterization of the mobility dynamics in the network and a finer granularity in the collected data.

Fig.~\ref{fig:datasetmetrics} shows an example of different timeseries for 4 \glspl{enb} in the Mountain View/Palo Alto area, with a time step of 15 minutes. It can be seen that, even though daily patterns can be identified, each \gls{enb} presents characteristic differences with the others.

\section{\gls{ran} Controllers as Enablers of\\Machine-Learning Applications at the Edge}
\label{sec:ctrl}
% \begin{itemize}
% 	\item Argue that the current network architecture can hardly scale to provide these services - need to collect data, run algorithms, etc
% 	\item Moreover, the 5G architecture is going to be disaggregated, with CU/DU split. Also mention the xRAN split and the usage of smart network controllers.
% 	\item These controllers can also be used to run analytics and intelligent services at the edge.
% 	\item Network data can also be used to perform the clustering, show gain and dynamics (how often do we need to update, show the quality of clutering using prev day/prev week statistics).
% \end{itemize}

The past and current generations of cellular networks were not designed to deploy machine learning and artificial intelligence algorithms at scale. The main reason is that there are no standardized interfaces that network operators can exploit to collect data from the base stations and the equipments of different vendors, and/or to modify the behavior of the network according to custom policies. Indeed, despite the \gls{son} capabilities embedded in the \gls{lte} standard~\cite{36300}, the deployment of autonomous networks is not widespread, and \gls{lte} \glspl{enb} are usually self-contained appliances to which the telecom operators have restricted access. Therefore, the control plane is usually decentralized, and the exchange of information among \glspl{enb} is limited~\cite{xranwp}. Accordingly, practical machine learning solutions that can deployed in a 4G \gls{lte} network are generally limited to \gls{son} parameters optimization for a few \glspl{enb}, generally with offline training and/or optimization, thus without real-time insights, or to the application of intelligent algorithms to the data that is collected in each single \gls{enb}, for example to predict the channel gain~\cite{chiariotti2017learning}, perform smart handovers~\cite{ali2016machine} or scheduling~\cite{chinchali2018cellular,bui2018data}. 

In order to make network management and operation more efficient, new design paradigms have emerged in the 5G domain. The main trend is related to the disaggregation of the base station (which in \gls{3gpp} \gls{nr} networks is the \gls{gnb}). The \gls{3gpp} has proposed different splits of the \gls{gnb} protocol stack~\cite{38300}, so that it will be possible to deploy a different \gls{ran} architecture, with the lower layers in \glspl{du} on poles and towers, and the higher layers in \glspl{cu} which can be hosted in a datacenter. The pooling of \glspl{cu} can enable more sophisticated orchestration operations, and energy savings~\cite{checko2015cloud}. On the other hand, the \glspl{du} that are deployed in the \gls{ran} are simpler and possibly smaller than 4G full-fledged base stations.

The second trend is related to the deployment in the wireless \gls{ran} of \gls{sdn} solutions based on open and smart network controllers~\cite{poularakis2017one}, which have already been adopted with success in large wired backbone networks~\cite{jain2013b4}. Along this line, the O-RAN Alliance, a consortium of network operators and equipment vendors, is standardizing controller interfaces between the \glspl{cu} and new custom \gls{ran} controllers that can be implemented and deployed by the telecom operators themselves. As mentioned in~\cite{xranwp}, an architecture with a split between the distributed hardware that performs data-plane-related functions and a more centralized software-based control plane can enable more advanced control procedures, thanks to the centralized view and the context awareness, and thus this approach is quickly becoming a de facto standard for the deployment of 5G cellular networks.

% \vspace{-.4cm}
\subsection{Proposed Architecture}
\label{sec:arch}
% \vspace{-.1cm}

In this paper, we propose to exploit the new design paradigms for the 5G \gls{ran} to make it possible to practically deploy intelligence in cellular networks, without the constraints and limitations previously described for 4G \gls{lte} deployments. As shown in Fig.~\ref{fig:ranCtrl}, our architecture leverages the different layers of controllers to aggregate and process the network data using machine learning and \gls{ai} techniques, with a multi-layer semi-distributed point of view that strikes a balance between the decentralized 4G approach and a completely centralized approach, which would be infeasible due to the amount of data to be processed. Notice that the proposed architecture applies to the control plane, and does not affect the routing of data packets.

% While this overlay is given by the integration of several state-of-the-art components, in this paper we propose for the first time to exploit it to, as we will discuss in the following paragraphs.

In the following paragraphs, we will introduce the proposed architecture and describe how it can be integrated in the \gls{nr} and O-RAN Alliance designs, following the \gls{mec} paradigm. Moreover, we will discuss the costs and the technical challenges associated to the deployment of the proposed architecture. In Sec.~\ref{sec:cluster} and Sec.~\ref{sec:pred} we will describe two \gls{ml}-based applications for networks, showing that the usage of the proposed architecture makes it possible to improve the performance with respect to decentralized, 4G-based approaches.

\subsubsection{\textbf{Integration with \gls{3gpp} networks}}

The proposed architecture exploits a multi-layer overlay that is compliant with \gls{3gpp} \gls{nr} networks, as reported in Fig.~\ref{fig:ranCtrl}. The overlay is composed by three main elements:
\begin{itemize}
	\item the \textit{\gls{ran}}, which is deployed to provide cellular service to the users, and includes the \gls{3gpp} \gls{nr} \glspl{cu}, \glspl{du} and \glspl{ru}. The \gls{ran} handles the data plane of the users, i.e., the user traffic is forwarded from or to the core network and the public Internet from the \glspl{cu}~\cite{38300}.
	\item the \textit{\gls{ran} controllers}, which control and coordinate the \gls{ran} elements, as proposed in~\cite{xranwp}. Each \gls{ran} controller is associated to a cluster of \glspl{gnb}, and is deployed in \gls{mec}, to minimize the communication latency with the \gls{ran}. Some of the control-plane processes are assigned to the \gls{ran} controllers, which can benefit from the cluster-based overview. For example, as proposed in~\cite{xranwp}, the \gls{ran} controllers can manage \gls{ue}-level connectivity, by coordinating handover decisions and performing load balancing, or can enforce \gls{qos} policies. Moreover, the \gls{ran} controllers can be deployed in the same edge datacenters that host the \gls{cu} for a certain area, to minimize the \gls{cu}-controller latency and to guarantee interconnectivity across the different controller domains, following the trends for cloud- and edge-based deployment of 5G networks~\cite{oranwg6}. 
	\item the \textit{Cloud Network Controller}, that orchestrates the \gls{ran} controllers (e.g., to establish the \gls{ran} controllers/\glspl{gnb} association) and provides application-layer services, and can be deployed in a remote cloud facility.
\end{itemize}

A multi-layer controller architecture combines the benefits of the scalability of a distributed approach with the performance gain given by a partially-centralized view of the network. Each layer implements control functionalities with different latency constraints, allowing the network to scale: the \glspl{du} schedule over-the-air transmissions on a sub-ms basis, the \gls{ran} controllers may decide upon users' association on a time scale of tens of milliseconds, and, finally, the Cloud Network Controller can operate on multiple-second (or even longer) intervals, for example to update the association between \glspl{gnb} and \gls{ran} controllers. At each additional layer, it is possible to support a larger number of devices (e.g., a \gls{du} controls tens of \glspl{ue} at most, while the \gls{ran} controller can be designed to handle hundreds of \glspl{ue}), and, given the more relaxed constraints on the decision time scale, it is possible to implement more refined and complex decision policies, based on machine learning algorithms enabled by the larger amount of data given by the clustered and/or centralized views. 
% For example, in the case of a handover, in a traditional architecture the decision is performed locally, and the \gls{gnb} signals to the neighbor identified as target that a \gls{ue} may handover, without actually being aware of the load of the neighbor, which may refuse to complete the procedure. In a controller-based architecture, instead, the \gls{gnb} can directly signal the information related to the users' channel quality to the controller, which has a global view of the attached \glspl{gnb} and can perform improved \glspl{gnb}-users associations. 

\subsubsection{\textbf{\gls{ran} Controllers, Machine Learning and Data Collection}}

While the \gls{ran} controllers are generally deployed to perform the aforementioned control plane task, we propose to leverage them to implement machine learning techniques at the edge of the network. A network operator can indeed use the proposed overlay to manage the data collection from the distributed \glspl{gnb} and enforce policies based on the learning applied to this data. Notice that, for some metrics, the controllers would not need explicit signaling for the data collection: for example, if a controller manages the \glspl{ue} sessions, as proposed in~\cite{xranwp}, then it is already aware of the number of users connected to each \gls{gnb} it controls.

The position of the \gls{ran} controllers in the overlay network strikes a balance between the breadth of their point of view, the amount of data they need to collect and process, and the number of the user sessions they can handle. 
In general, as the number of base stations associated to a controller grows (and, consequently, the number of controllers decreases, up to a single controller), it is possible to perform more refined optimizations, given that the knowledge of the state of the network is more complete. However, there is a limit to how much the data collection can be centralized. Indeed, if the operator is interested in running \textit{real-time} data-driven algorithms, for example to decide upon the association of \glspl{ue} and \glspl{gnb}, then we argue that a completely centralized architecture does not scale because of (i) the amount of data (for example, related to channel measurements) that needs to be collected and (ii) the collection and processing delay. In this regards, we observed that it is not possible to perform a real-time collection and processing of a subset of the monitoring data streamed from the Palo Alto/Mountain View network (178 base stations) in a single virtual machine with 8 x86 CPUs at 2.1 GHz. Moreover, controllers distributed in multiple datacenters at the network edge minimize the delay experienced by control messages exchanged with the \glspl{gnb}. 
On the other hand, a completely distributed approach (as in a 4G \gls{lte} network) cannot exploit \textit{any} centralized view and/or enforce coordinated policies, as previously mentioned, and, as we will show in Sec.~\ref{sec:pred} with real network data, does not perform as well as the controller-based architecture for the accurate prediction of the number of users in the network. 

% The \gls{ran} controllers 
% \red{More on how we can use the controllers (control + applications/user services)}

\subsubsection{\textbf{Technical Challenges}}

The usage of \gls{ran} controllers, however, introduces new technical challenges. First, new standard interfaces and signaling between the \glspl{gnb} and the controllers will need to be defined.\footnote{This effort is being pursued, among others, by the O-RAN Alliance~\cite{xranwp}}
For example, in a completely distributed architecture (e.g., \gls{lte}), for a handover there is a message exchange between neighboring base stations, and, then, the core network~\cite{36300}, while, if controllers are used, the \glspl{gnb} can interface directly with their controller to exploit its global view. Once the actual specifications for \gls{ran} controllers will be completed, it will be possible to also evaluate the signaling difference among these different architectures. 

Another interesting problem is related to the association of controllers and \glspl{gnb}. This issue has already been studied for \gls{sdn} controllers in wired networks~\cite{heller2012controller}, but wireless cellular networks have characteristics that introduce new dimensions to this problem, mainly related to the higher level of mobility of the endpoints of such networks, i.e., the \glspl{ue}.\footnote{Notice that in this paper we consider a control-plane \gls{gnb}-controller association, i.e., the controller is not involved in the processing of data-plane packets and low-level scheduling, which is what is instead usually considered in the design of controllers for interference coordination problems~\cite{deb2014algorithms}.} If the \gls{ran} controllers are used to manage user sessions and mobility events, then they will need to maintain a consistent state for each user associated to the \glspl{gnb} they control. Given that cellular users often move through the area covered by the cellular networks, it becomes of paramount importance to minimize the number of times a user performs a handover between two base stations controlled by different controllers. In this case, indeed, the two controllers would need to synchronize and share the user's state, and this would increase the control plane latency, as also observed in case of inter-controller communications in wired \gls{sdn} networks~\cite{zhang2016role}. Therefore, in the following section, we will describe a practical data-driven method to perform the association between \glspl{gnb} and controllers, testing the proposed algorithm on the San Francisco and the Mountain View/Palo Alto datasets.

\section{Big-data Driven \gls{ran} Controller Association}
\label{sec:cluster}
In the remainder of this paper we introduce our second major contribution, i.e., we describe two applications related to network control and optimization that show the advantages of using the proposed controller-based architecture described in Fig.~\ref{fig:ranCtrl}.
In particular, in this Section, we illustrate a data-driven approach for the control-plane association of \gls{ran} controllers and \glspl{gnb}. The algorithm we designed aims at minimizing the number of interactions between \glspl{gnb} belonging to different \gls{ran} controllers (since any controller that is added in the control loop severely impacts the control plane latency), and enables a dynamic allocation of the base stations to the different controllers. Moreover, it is based on the real data that the network itself can collect, thus it represents another example of how it is possible to exploit real-time analytics to self-optimize the performance.

% In the following paragraphs we will describe how it is possible to use the network data the controller themselves are collecting to dynamically decide upon the allocation of the base station to the different controllers. In particular, given the considerations on the inter-controller control plane latency introduced in Sec.~\ref{sec:arch}, in the architecture we propose it is important to minimize the number of interactions between \glspl{gnb} belonging to different controllers, since any controller that is added in the control loop severely impacts the control plane latency.

\vspace{-.3cm}
\subsection{Proposed Algorithm}
Our method is based on a semi-supervised constrained clustering on a graph weighted according to the transition probabilities among base stations. The algorithm is summarized with the pseudocode in Alg.~\ref{alg:cluster}. The input is represented by the timeseries of X2 and S1 handovers for all the $N_g$ \glspl{gnb} in the set $\mathcal{B}$, each tagged with the timestamp of the event and the pair $<source, destination>$ \glspl{gnb}, and by the time step $T_c$ to be considered for the computation of the transition probability matrices (e.g., fifteen minutes or a day). Moreover, the network operator can tune the number of \gls{ran} controllers $N_c$ according to the availability of computational resources and the number of base stations and related \glspl{ue} that each controller can support. 

\begin{algorithm}[t]
  %\sffamily
  \footnotesize
  % \setstretch{1}
  \caption{Network-data-driven \gls{ran} Controller Association Algorithm}\label{alg:cluster}
  \begin{algorithmic}[1]
    % \STATE An updated CRT is available at time $t$.
    \FOR {every time step $T_c$}
      \STATE \textbf{distributed data collection step (performed in each \gls{ran} controller):}
      \begin{ALC@g}
        
        \FOR {every \gls{ran} controller $p \in \{0, \dots, N_c - 1\}$ with associated \glspl{gnb} set $\mathcal{B}_{p}$}
          \FOR {every \gls{gnb} $i \in \mathcal{B}_{p}$}
            \STATE compute the number of handovers $N^{\rm ho}_{i, j} \forall  j \in \mathcal{B}$
          \ENDFOR
          \STATE report the statistics on the number of handovers to the Cloud Network Controller
        \ENDFOR
      \end{ALC@g}
      \STATE \textbf{clustering and association step (performed in the Cloud Network Controller):}
        \begin{ALC@g}
         \STATE compute the transition probability matrix $H$ based on the handovers between every pair of \glspl{gnb}
           \STATE define weighted graph $G = (V,E)$ with weight $W(G)_{i, j} = H_{i, j} + H_{j, i}$
           \STATE perform spectral clustering with constrained K means on $G$ to identify $N_c$ clusters
           \STATE apply the new association policy for the next time step
        \end{ALC@g}
    \ENDFOR
  \end{algorithmic}
\end{algorithm}

\begin{algorithm}[t]
  %\sffamily
  \footnotesize
  % \setstretch{1}
  \caption{Graph spectral clustering algorithm with constrained K means}\label{alg:spectral}
  \begin{algorithmic}[1]
    \STATE \textbf{input:} graph $G = (V,E)$ with weights $W(G)$
    \STATE compute the degree matrix $D_{i,i} = \sum_{j = 1}^{N_g} W(G)_{i, j}$
    \STATE compute the normalized Laplacian of $G$ as $L = I - D^{-1}W(G)$
    \STATE create the matrix $U \in \mathbb{R}^{N_g \times N_c}$ with the eigenvectors of $L$ associated to the $N_c$ smallest eigenvalues as columns
    \STATE apply constrained K means on the rows of $U$ to get $N_c$ clusters
  \end{algorithmic}
\end{algorithm}

\begin{figure*}[t]
    \centering
    \begin{subfigure}[t]{0.48\textwidth}
        \centering
            \setlength\belowcaptionskip{0.1cm}
        \includegraphics[width=.76\textwidth]{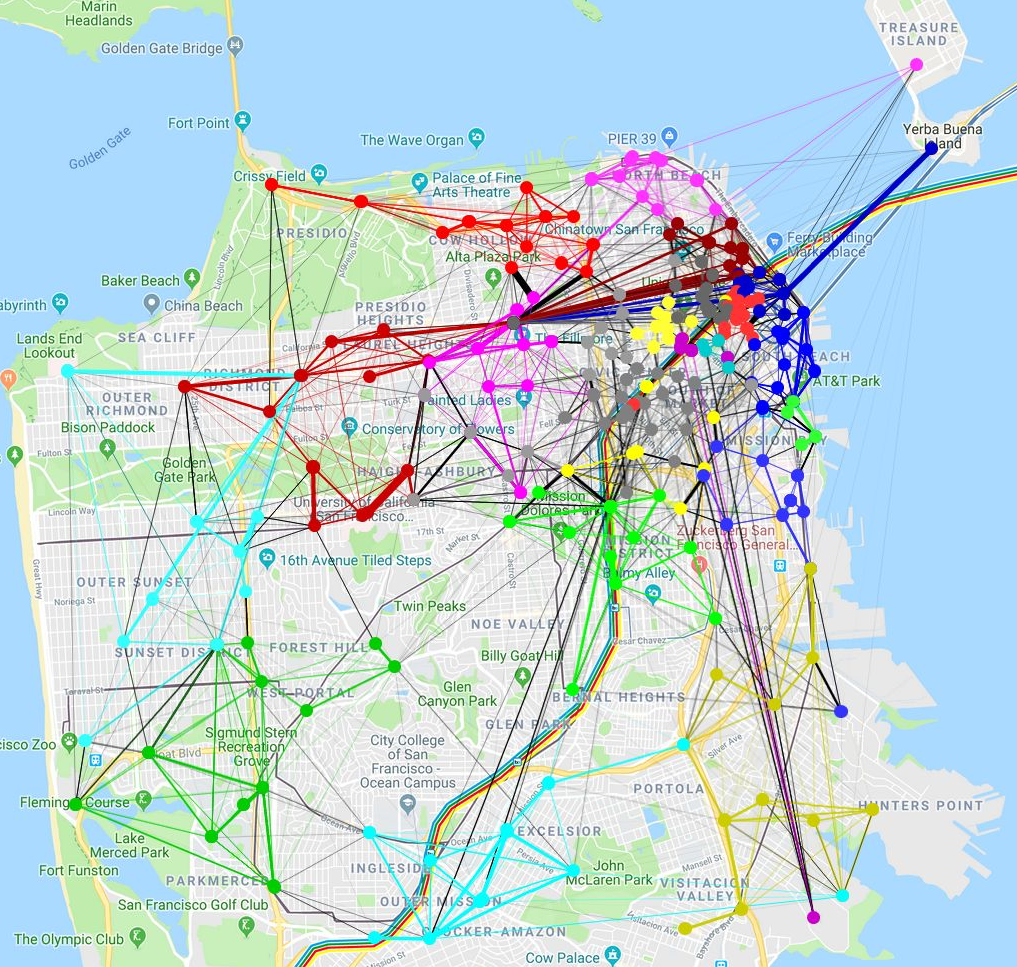}
        \caption{Clustering with Alg.~\ref{alg:cluster} in San Francisco.}
        \label{fig:speccl}
    \end{subfigure}\hfill
    \begin{subfigure}[t]{0.48\textwidth}
        \centering
            \setlength\belowcaptionskip{0.1cm}
        \includegraphics[width=.76\textwidth]{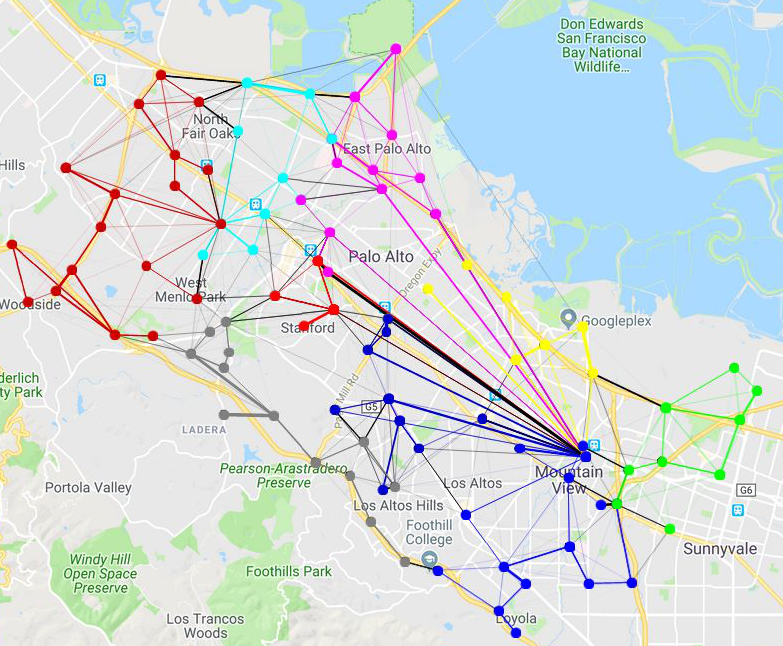}
        \caption{Clustering with Alg.~\ref{alg:cluster} in Mountain View.}
        \label{fig:specclmv}
    \end{subfigure}
    \begin{subfigure}[t]{0.48\textwidth}
        \centering
            \setlength\belowcaptionskip{0.1cm}
        \includegraphics[width=.76\textwidth]{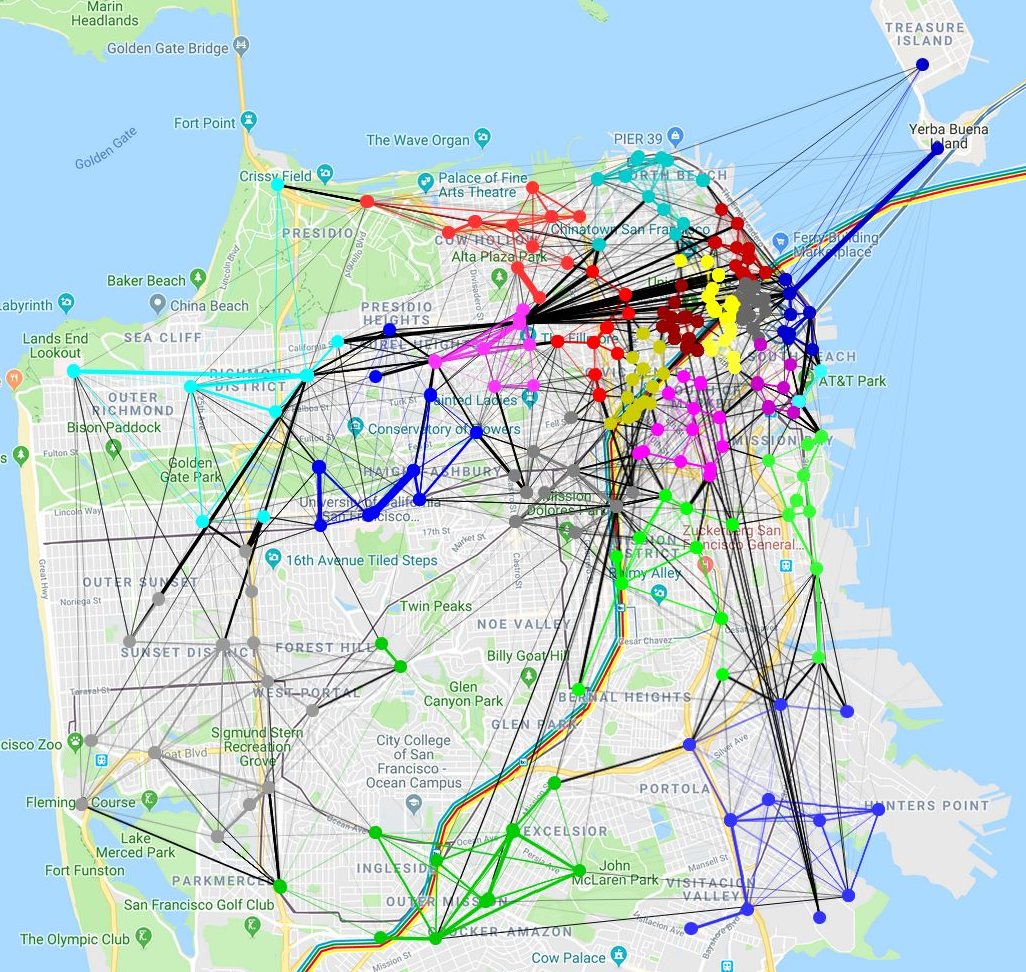}
        \caption{Clustering with the positions of the \glspl{gnb} in San Francisco.}
        \label{fig:geocl}
    \end{subfigure}\hfill
    \begin{subfigure}[t]{0.48\textwidth}
        \centering
            \setlength\belowcaptionskip{0.1cm}
        \includegraphics[width=.76\textwidth]{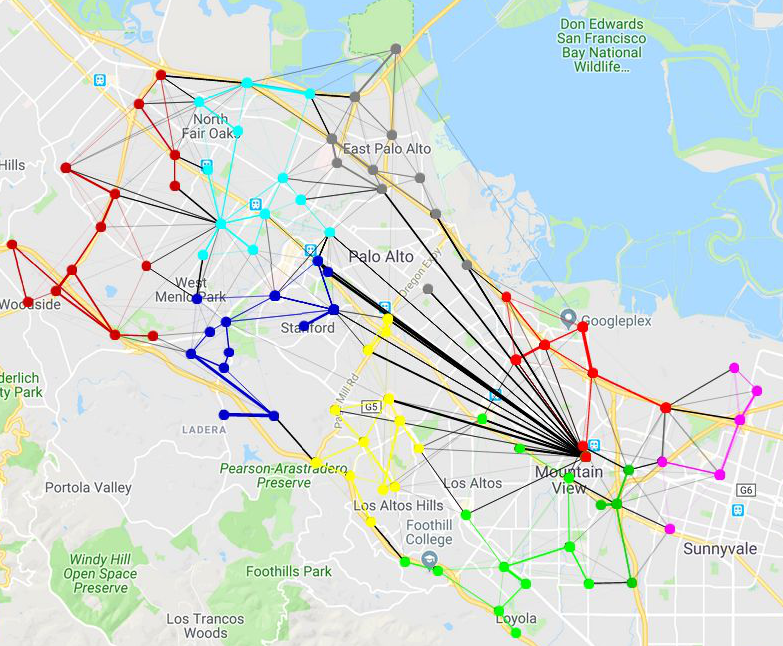}
        \caption{Clustering with the positions of the \glspl{gnb} in Mountain View.}
        \label{fig:geoclmv}
    \end{subfigure}
    \caption{Network-data- and position-based clusters in San Francisco, using data from 2017/02/01 with $T_c = 24$ hours and $N_c = 22$, and Mountain View/Palo Alto, with data from 2018/06/28 with $T_c = 24$ hours and $N_c = 10$. The colored dots represent the base stations, with different colors associated to different clusters. The lines connecting the dots represent the weights in the graph $G$ of the edge between the two \glspl{gnb}, with a thicker line representing a larger weight, i.e., sum of transition probabilities between the \glspl{gnb}. Finally, lines with the same color as the dots represent edges between vertices in the same cluster, and vice versa for black lines.}
    \label{fig:cl}
\end{figure*}

Every $T_c$, each \gls{ran} controller $p \in \{0, \dots, N_c - 1\}$, which has collected the timeseries of events for its \gls{gnb} $i$ in the set of controlled \glspl{gnb} $\mathcal{B}_{p}$, will process this data to extract the number of handovers $N^{\rm ho}_{i, j}, \forall i \in \mathcal{B}_p, \forall  j \in \mathcal{B}$, and will report this information to the Cloud Network Controller described in Sec.~\ref{sec:arch}. The Cloud Network Controller then aggregates the statistics from each \gls{ran} controller and builds a complete transition probability matrix $H$, where entry $(i, j)$ is 
\begin{equation}
	H_{i, j} = \begin{cases}
    \frac{N^{\rm ho}_{i, j}}{\sum_{j = 1}^{N_g} N^{\rm ho}_{i, j}} & \mbox{if}\; \sum_{j = 1}^{N_g} N^{\rm ho}_{i, j} \ne 0, \\
    0 & \mbox{otherwise.}
  \end{cases}
\end{equation}
Then, consider the fully-connected undirected graph $G = (V, E)$, where $V = \mathcal{B}$ is the set of $N_g$ vertices, and $E$ is the set of edges that represent possible transitions among the \glspl{gnb}. Each edge $e_{i, j}$ is weighted by the sum of the transition probabilities between \glspl{gnb} $i$ and $j$, i.e., $W(G)_{i, j} = H_{i, j} + H_{j, i}$, with $W(G)$ the weight matrix, to account for all the possible transitions (and thus interactions, and, possibly, message exchanges and state synchronizations) between the two \glspl{gnb}. In order to identify the set of \gls{gnb}-to-controllers associations that minimize the inter-controller communications, the proposed algorithm clusters the undirected graph $G$ to identify the groups of \glspl{gnb} in which the intra-cluster interactions (i.e., handovers and transfer of user sessions) are more frequent than inter-cluster ones.

We tested and considered different approaches for the clustering~\cite{schaeffer2007graph,nascimento2011spectral}, which, in this case, has to satisfy two constraints: (i) the number of clusters should be an input of the algorithm, to match the number of available controllers\footnote{Notice that in this case finding the optimal solution to the clustering problem is NP-hard, thus identifying the optimal solution is not feasible in large scale networks~\cite{blum2016foundations}.}; and (ii) the size of the clusters (i.e., number of \glspl{gnb} per cluster) should be balanced, to avoid overloading certain controllers while under-utilizing others. The first constraint rules out popular unsupervised graph clustering techniques based on community detection algorithms, which are also generally applied to directed graphs~\cite{malliaros2013clustering}. Therefore, we propose to use a variant of standard spectral clustering techniques for graphs~\cite{von2007tutorial}, which relies on a constrained version of K-means to balance the size of the clusters. Alg.~\ref{alg:spectral} lists the main steps of the procedure. 

Consider the degree matrix $D \in \mathbb{R}^{N_g \times N_g}$, i.e., a diagonal matrix with an entry $D_{i,i} = \sum_{j = 1}^{N_g} W(G)_{i, j}$ for each \gls{gnb} $i \in {1, \dots, N_g}$. Then, it is possible to compute the normalized graph Laplacian as $L = I - D^{-1}W(G)$ and extract the eigenvectors associated to the $N_c$ smallest eigenvalues, i.e., as many eigenvalues as the number of clusters to identify. The result is a matrix $U \in \mathbb{R}^{N_g \times N_c}$ with the eigenvectors as columns. 
Each row of this matrix, which corresponds to a specific \gls{gnb}, can be considered as a point in $\mathbb{R}^{N_c}$, and can be clustered using K means~\cite{von2007tutorial}. Standard K means, however, does not generate balanced clusters. Therefore, we replace this last step with a constrained K means algorithm, which modifies the standard K means by adding constraints on the minimum and maximum size of the clusters during the cluster assignment step. In this way, the cluster assignment problem can be formulated as a linear programming problem~\cite{bradley2000constrained}. 
%Notice that  \red{Should I say more on this?}
The final result is a set of $N_c$ clusters, and the Cloud Network Controller can apply the clustering policy to assign the \glspl{gnb} to the respective \gls{ran} controllers.

\subsection{Evaluation with Real Data}

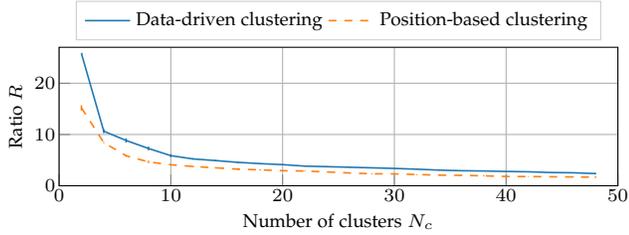
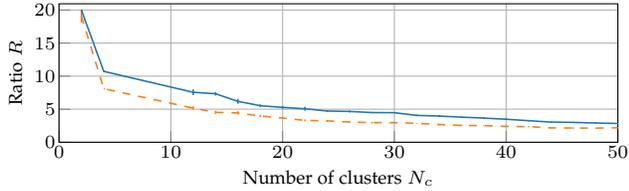
\begin{figure}[t]
    \centering 
    \begin{subfigure}[t]{0.495\textwidth}
      \centering
      \setlength\belowcaptionskip{.2cm}
      \setlength\abovecaptionskip{.1cm}
      \setlength\fwidth{\textwidth}
      \setlength\fheight{0.38\textwidth}
      % This file was created by matplotlib2tikz v0.6.17.
\begin{tikzpicture}

\definecolor{color1}{rgb}{1,0.498039215686275,0.0549019607843137}
\definecolor{color0}{rgb}{0.12156862745098,0.466666666666667,0.705882352941177}
\pgfplotsset{scaled x ticks=false}
\pgfplotsset{every tick label/.append style={font=\scriptsize}}
\begin{axis}[
xlabel={Number of clusters $N_c$},
ylabel={Ratio $R$},
xmin=0, xmax=50,
ymin=0, ymax=27.0203455812251,
ylabel shift=-2pt,
yticklabel shift=-2pt,
xlabel shift=-2pt,
xticklabel shift=-2pt,
width=\fwidth,
height=\fheight,
xlabel style={font=\scriptsize},
ylabel style={font=\scriptsize},
tick align=inside,
tick pos=left,
xmajorgrids,
x grid style={lightgray!92.02614379084967!black},
ymajorgrids,
y grid style={lightgray!92.02614379084967!black},
legend entries={{Data-driven clustering},{Position-based clustering}},
legend style={font=\scriptsize,at={(0.5, 1.05)}, anchor=south,draw=white!80.0!black},
legend cell align={left},
legend columns=2,
]
% \addlegendimage{no markers, color0}
% \addlegendimage{no markers, color1}
\path [draw=color0, semithick] (axis cs:2,25.806115)
--(axis cs:2,25.806115);

\path [draw=color0, semithick] (axis cs:4,10.256525280801)
--(axis cs:4,11.031664519199);

\path [draw=color0, semithick] (axis cs:6,8.41878377485386)
--(axis cs:6,9.25213702514614);

\path [draw=color0, semithick] (axis cs:8,6.92081514645295)
--(axis cs:8,7.63333525354705);

\path [draw=color0, semithick] (axis cs:10,5.57472965555468)
--(axis cs:10,6.17186934444532);

\path [draw=color0, semithick] (axis cs:12,5.08819958955109)
--(axis cs:12,5.37966513772164);

\path [draw=color0, semithick] (axis cs:14,4.7574226400206)
--(axis cs:14,5.0731283599794);

\path [draw=color0, semithick] (axis cs:16,4.45367069998261)
--(axis cs:16,4.68184570001739);

\path [draw=color0, semithick] (axis cs:18,4.24851160387345)
--(axis cs:18,4.38785519612655);

\path [draw=color0, semithick] (axis cs:20,3.99790681529109)
--(axis cs:20,4.24192343470891);

\path [draw=color0, semithick] (axis cs:22,3.76163346928346)
--(axis cs:22,3.89979753071654);

\path [draw=color0, semithick] (axis cs:24,3.63442841269518)
--(axis cs:24,3.81244478730482);

\path [draw=color0, semithick] (axis cs:26,3.55619317840622)
--(axis cs:26,3.66716842159378);

\path [draw=color0, semithick] (axis cs:28,3.40557833199978)
--(axis cs:28,3.59035966800022);

\path [draw=color0, semithick] (axis cs:30,3.31878542121018)
--(axis cs:30,3.44447997878982);

\path [draw=color0, semithick] (axis cs:32,3.18225349527236)
--(axis cs:32,3.28246290472764);

\path [draw=color0, semithick] (axis cs:34,2.97302835748401)
--(axis cs:34,3.13007564251599);

\path [draw=color0, semithick] (axis cs:36,2.87623435435115)
--(axis cs:36,2.99306284564885);

\path [draw=color0, semithick] (axis cs:38,2.82862034153602)
--(axis cs:38,2.93419651560684);

\path [draw=color0, semithick] (axis cs:40,2.75496136126046)
--(axis cs:40,2.83482103873954);

\path [draw=color0, semithick] (axis cs:42,2.67584109017875)
--(axis cs:42,2.77379710982125);

\path [draw=color0, semithick] (axis cs:44,2.54880343506494)
--(axis cs:44,2.61668996493506);

\path [draw=color0, semithick] (axis cs:46,2.49405891536477)
--(axis cs:46,2.57613808463523);

\path [draw=color0, semithick] (axis cs:48,2.25339272764323)
--(axis cs:48,2.50592527235677);

\path [draw=color1, semithick] (axis cs:2,14.7020856414484)
--(axis cs:2,15.7036283585516);

\path [draw=color1, semithick] (axis cs:4,8.32144253875506)
--(axis cs:4,8.51608266124494);

\path [draw=color1, semithick] (axis cs:6,5.67070525895381)
--(axis cs:6,6.03505214104619);

\path [draw=color1, semithick] (axis cs:8,4.43720937976776)
--(axis cs:8,4.87147882023224);

\path [draw=color1, semithick] (axis cs:10,4.0093080691004)
--(axis cs:10,4.1756681308996);

\path [draw=color1, semithick] (axis cs:12,3.62690219225252)
--(axis cs:12,3.8943616259293);

\path [draw=color1, semithick] (axis cs:14,3.36698764309321)
--(axis cs:14,3.64312955690679);

\path [draw=color1, semithick] (axis cs:16,3.06301474955745)
--(axis cs:16,3.42261485044255);

\path [draw=color1, semithick] (axis cs:18,2.99406281072659)
--(axis cs:18,3.21453638927341);

\path [draw=color1, semithick] (axis cs:20,2.8239838240004)
--(axis cs:20,3.0440269259996);

\path [draw=color1, semithick] (axis cs:22,2.80233020236611)
--(axis cs:22,2.94211339763389);

\path [draw=color1, semithick] (axis cs:24,2.62699095400457)
--(axis cs:24,2.76517304599543);

\path [draw=color1, semithick] (axis cs:26,2.39455935080393)
--(axis cs:26,2.60581284919607);

\path [draw=color1, semithick] (axis cs:28,2.25896128053589)
--(axis cs:28,2.42499811946411);

\path [draw=color1, semithick] (axis cs:30,2.24034876965954)
--(axis cs:30,2.39356843034046);

\path [draw=color1, semithick] (axis cs:32,2.09693969160208)
--(axis cs:32,2.28728970839792);

\path [draw=color1, semithick] (axis cs:34,2.00693003997356)
--(axis cs:34,2.13151136002644);

\path [draw=color1, semithick] (axis cs:36,1.94872127798089)
--(axis cs:36,2.12093352201911);

\path [draw=color1, semithick] (axis cs:38,1.83506896611778)
--(axis cs:38,2.04356731959651);

\path [draw=color1, semithick] (axis cs:40,1.76422956642113)
--(axis cs:40,1.85410863357887);

\path [draw=color1, semithick] (axis cs:42,1.74818978022789)
--(axis cs:42,1.84473801977211);

\path [draw=color1, semithick] (axis cs:44,1.70539875114124)
--(axis cs:44,1.80450644885876);

\path [draw=color1, semithick] (axis cs:46,1.6423335937867)
--(axis cs:46,1.7998044062133);

\path [draw=color1, semithick] (axis cs:48,1.5215033754972)
--(axis cs:48,1.81130329116947);

\addplot [semithick, color0]
table {%
2 25.806115
4 10.6440949
6 8.8354604
8 7.2770752
10 5.8732995
12 5.23393236363636
14 4.9152755
16 4.5677582
18 4.3181834
20 4.119915125
22 3.8307155
24 3.7234366
26 3.6116808
28 3.497969
30 3.3816327
32 3.2323582
34 3.051552
36 2.9346486
38 2.88140842857143
40 2.7948912
42 2.7248191
44 2.5827467
46 2.5350985
48 2.379659
};
\addplot [semithick, color1, dashed]
table {%
2 15.202857
4 8.4187626
6 5.8528787
8 4.6543441
10 4.0924881
12 3.76063190909091
14 3.5050586
16 3.2428148
18 3.1042996
20 2.934005375
22 2.8722218
24 2.696082
26 2.5001861
28 2.3419797
30 2.3169586
32 2.1921147
34 2.0692207
36 2.0348274
38 1.93931814285714
40 1.8091691
42 1.7964639
44 1.7549526
46 1.721069
48 1.66640333333333
};

\end{axis}

\end{tikzpicture}
      \caption{San Francisco scenario, 2017/02/02.}
      \label{fig:ratioSf}
    \end{subfigure}
    \begin{subfigure}[t]{0.495\textwidth}
      \centering
      \setlength\fwidth{\textwidth}
      \setlength\fheight{0.38\textwidth}
      % This file was created by matplotlib2tikz v0.6.17.
\begin{tikzpicture}

\definecolor{color1}{rgb}{1,0.498039215686275,0.0549019607843137}
\definecolor{color0}{rgb}{0.12156862745098,0.466666666666667,0.705882352941177}
\pgfplotsset{scaled x ticks=false}
\pgfplotsset{every tick label/.append style={font=\scriptsize}}

\begin{axis}[
xlabel={Number of clusters $N_c$},
ylabel={Ratio $R$},
xmin=0, xmax=50,
ymin=0, ymax=20.9473493511828,
ylabel shift=-2pt,
yticklabel shift=-2pt,
xlabel shift=-2pt,
xticklabel shift=-2pt,
width=\fwidth,
height=\fheight,
xlabel style={font=\scriptsize},
ylabel style={font=\scriptsize},
tick align=inside,
tick pos=left,
xmajorgrids,
x grid style={lightgray!92.02614379084967!black},
ymajorgrids,
y grid style={lightgray!92.02614379084967!black},
legend style={draw=white!80.0!black},
legend cell align={left}
]
\path [draw=color0, semithick] (axis cs:2,20.000982)
--(axis cs:2,20.000982);

\path [draw=color0, semithick] (axis cs:4,10.719815)
--(axis cs:4,10.719815);

\path [draw=color0, semithick] (axis cs:12,7.1279005100788)
--(axis cs:12,7.98868688992121);

\path [draw=color0, semithick] (axis cs:14,7.06831100084905)
--(axis cs:14,7.60006039915096);

\path [draw=color0, semithick] (axis cs:16,5.82418195343968)
--(axis cs:16,6.54896164656032);

\path [draw=color0, semithick] (axis cs:18,5.34534162673025)
--(axis cs:18,5.71515717326975);

\path [draw=color0, semithick] (axis cs:20,5.06858215846423)
--(axis cs:20,5.47360484153577);

\path [draw=color0, semithick] (axis cs:22,4.76893113246097)
--(axis cs:22,5.35199686753903);

\path [draw=color0, semithick] (axis cs:24,4.58988317230899)
--(axis cs:24,4.85841182769101);

\path [draw=color0, semithick] (axis cs:26,4.51790510139952)
--(axis cs:26,4.79745649860048);

\path [draw=color0, semithick] (axis cs:28,4.38997643698918)
--(axis cs:28,4.58793996301082);

\path [draw=color0, semithick] (axis cs:30,4.37239254237141)
--(axis cs:30,4.56686205762859);

\path [draw=color0, semithick] (axis cs:32,3.94762344995692)
--(axis cs:32,4.17176915004308);

\path [draw=color0, semithick] (axis cs:34,3.82862285189734)
--(axis cs:34,4.09544834810266);

\path [draw=color0, semithick] (axis cs:36,3.70626742146354)
--(axis cs:36,3.89159417853646);

\path [draw=color0, semithick] (axis cs:38,3.54149199847435)
--(axis cs:38,3.76327840152566);

\path [draw=color0, semithick] (axis cs:40,3.35846255724726)
--(axis cs:40,3.60192764275274);

\path [draw=color0, semithick] (axis cs:42,3.1832546676314)
--(axis cs:42,3.3424261323686);

\path [draw=color0, semithick] (axis cs:44,2.94978728591676)
--(axis cs:44,3.14412771408324);

\path [draw=color0, semithick] (axis cs:46,2.91678713434921)
--(axis cs:46,3.08812586565079);

\path [draw=color0, semithick] (axis cs:48,2.79890904637827)
--(axis cs:48,3.01822075362173);

\path [draw=color0, semithick] (axis cs:50,2.74308547368388)
--(axis cs:50,2.93011872631612);

\path [draw=color1, semithick] (axis cs:2,18.1000398318048)
--(axis cs:2,20.0470341681952);

\path [draw=color1, semithick] (axis cs:4,8.097644)
--(axis cs:4,8.097644);

\path [draw=color1, semithick] (axis cs:12,4.88773507481717)
--(axis cs:12,5.42334552518283);

\path [draw=color1, semithick] (axis cs:14,4.28497725573782)
--(axis cs:14,4.75957994426218);

\path [draw=color1, semithick] (axis cs:16,4.18741236757169)
--(axis cs:16,4.68245643242831);

\path [draw=color1, semithick] (axis cs:18,3.80174966874619)
--(axis cs:18,4.13209593125381);

\path [draw=color1, semithick] (axis cs:20,3.54422891429003)
--(axis cs:20,3.77379788570997);

\path [draw=color1, semithick] (axis cs:22,3.16643489459646)
--(axis cs:22,3.44265730540354);

\path [draw=color1, semithick] (axis cs:24,3.0931870982714)
--(axis cs:24,3.3757345017286);

\path [draw=color1, semithick] (axis cs:26,2.9570295629382)
--(axis cs:26,3.1458994370618);

\path [draw=color1, semithick] (axis cs:28,2.8463449621525)
--(axis cs:28,3.0930654378475);

\path [draw=color1, semithick] (axis cs:30,2.85966001275706)
--(axis cs:30,3.05372758724294);

\path [draw=color1, semithick] (axis cs:32,2.79373168496297)
--(axis cs:32,2.91456291503703);

\path [draw=color1, semithick] (axis cs:34,2.58256355786722)
--(axis cs:34,2.78913724213278);

\path [draw=color1, semithick] (axis cs:36,2.42214059409549)
--(axis cs:36,2.6583132059045);

\path [draw=color1, semithick] (axis cs:38,2.44999071760107)
--(axis cs:38,2.57945028239893);

\path [draw=color1, semithick] (axis cs:40,2.30376793976118)
--(axis cs:40,2.48091166023882);

\path [draw=color1, semithick] (axis cs:42,2.2612557561654)
--(axis cs:42,2.45116164383461);

\path [draw=color1, semithick] (axis cs:44,2.09702605000307)
--(axis cs:44,2.23946794999693);

\path [draw=color1, semithick] (axis cs:46,2.04073050844277)
--(axis cs:46,2.25794609155723);

\path [draw=color1, semithick] (axis cs:48,2.09097732357416)
--(axis cs:48,2.18029487642585);

\path [draw=color1, semithick] (axis cs:50,2.13272178456431)
--(axis cs:50,2.24126901543569);

\addplot [semithick, color0, forget plot]
table {%
2 20.000982
4 10.719815
12 7.5582937
14 7.3341857
16 6.1865718
18 5.5302494
20 5.2710935
22 5.060464
24 4.7241475
26 4.6576808
28 4.4889582
30 4.4696273
32 4.0596963
34 3.9620356
36 3.7989308
38 3.6523852
40 3.4801951
42 3.2628404
44 3.0469575
46 3.0024565
48 2.9085649
50 2.8366021
};
\addplot [semithick, color1, dashed, forget plot]
table {%
2 19.073537
4 8.097644
12 5.1555403
14 4.5222786
16 4.4349344
18 3.9669228
20 3.6590134
22 3.3045461
24 3.2344608
26 3.0514645
28 2.9697052
30 2.9566938
32 2.8541473
34 2.6858504
36 2.5402269
38 2.5147205
40 2.3923398
42 2.3562087
44 2.168247
46 2.1493383
48 2.1356361
50 2.1869954
};

\end{axis}

\end{tikzpicture}
      \caption{Mountain View/Palo Alto scenario, 2018/06/28.}
      \label{fig:ratioMv}
    \end{subfigure}
    \caption{Ratio $R$ between intra- and inter-cluster handovers as a function of the number of clusters $N_c$, with clustering based on daily updates.}
    \label{fig:numClusters}
\end{figure}

We compare the proposed network-data-based strategy (whose results are reported in Fig.~\ref{fig:speccl} for the San Francisco area and Fig.~\ref{fig:specclmv} for the Mountain View area) with a baseline, in which the constrained K means is directly applied to the latitude and longitude of the \glspl{gnb} (shown in Figs.~\ref{fig:geocl} and~\ref{fig:geoclmv}, respectively). Indeed, several approaches have been proposed in the literature to cluster, for example, remote radio heads and \glspl{bbu} into \gls{bbu} pools, according to different targets~\cite{cluster1,cluster2,aliu2013survey}. However, none of these focuses on the minimization of the control plane latency, but rather on data-plane issues, such as the minimization of interference or coordinated multipoint transmissions. Therefore, as a baseline, we consider the basic clustering approach based on the geographical position of the base stations. This method is static, and can be applied in networks that do not rely on data-driven approaches for configuration purposes, for example because the operator does not collect and/or make use of real-time network analytics. In the absence of this kind of data, we argue that geographic clustering is an approach in line with the goal to minimize inter-controller interactions, given that users are expected to move among neighboring base stations, which the geographical clustering will group under the same \gls{ran} controller.

\begin{figure*}[t]
    \centering
    \begin{subfigure}[t]{.49\textwidth}
      \centering
      \setlength\abovecaptionskip{-.2cm}
      \setlength\fwidth{\textwidth}
      \setlength\fheight{0.4\textwidth}
      % This file was created by matplotlib2tikz v0.6.17.
\begin{tikzpicture}

\definecolor{color1}{rgb}{1,0.498039215686275,0.0549019607843137}
\definecolor{color0}{rgb}{0.12156862745098,0.466666666666667,0.705882352941177}
\definecolor{color3}{rgb}{0.83921568627451,0.152941176470588,0.156862745098039}
\definecolor{color2}{rgb}{0.172549019607843,0.627450980392157,0.172549019607843}
\definecolor{color5}{rgb}{0.549019607843137,0.337254901960784,0.294117647058824}
\definecolor{color4}{rgb}{0.580392156862745,0.403921568627451,0.741176470588235}
\pgfplotsset{scaled x ticks=false}
\pgfplotsset{every tick label/.append style={font=\scriptsize}}

\begin{axis}[
xlabel={Time},
ylabel={Number of handovers},
xmin=-0.9, xmax=18.9,
xtick={3, 7, 11, 15},
xticklabels={16:00, 17:00, 18:00, 19:00},
ymin=47611.2404529324, ymax=440065.545956787,
ylabel shift=-2pt,
yticklabel shift=-2pt,
xlabel shift=-2pt,
xticklabel shift=-2pt,
width=\fwidth,
height=\fheight,
xlabel style={font=\scriptsize},
ylabel style={font=\scriptsize},
tick align=inside,
tick pos=left,
xmajorgrids,
x grid style={lightgray!92.02614379084967!black},
ymajorgrids,
y grid style={lightgray!92.02614379084967!black},
legend style={font=\scriptsize,at={(1.2,1.2)}, anchor=south, draw=white!80.0!black},
legend columns=2,
legend cell align={left},
legend entries={{Intra cluster HO - 15 min updates},{Inter cluster HO - 15 min updates},{Intra cluster HO - daily updates},{Inter cluster HO - daily updates},{Intra cluster HO - static (geo clustering)},{Inter cluster HO - static (geo clustering)}}
]
% \addlegendimage{no markers, color0}
% \addlegendimage{no markers, color1}
% \addlegendimage{no markers, color2}
% \addlegendimage{no markers, color3}
% \addlegendimage{no markers, color4}
% \addlegendimage{no markers, color5}
\addplot [semithick, color0, dashed]
table {%
0 348825.912277604
1 367454.917996629
2 363570.784099137
3 368675.085260036
4 390379.62459657
5 383751.407751082
6 380849.329840351
7 406753.189902161
8 419255.557651564
9 421257.347775163
10 411618.480875597
11 402023.591664037
12 381488.746125695
13 368371.866306317
14 338093.834445555
15 321943.160571693
16 303357.216850653
17 270543.97209392
18 254977.927478711
};
\addplot [semithick, color1, dashed]
table {%
0 78182.0877223961
1 86267.0820033712
2 86779.2159008631
3 92573.9147399643
4 94488.3754034301
5 97433.5922489178
6 102522.67015965
7 117622.810097839
8 115478.442348436
9 121921.652224837
10 112548.519124403
11 109563.408335963
12 110595.253874305
13 97700.133693683
14 97510.165554445
15 86037.8394283067
16 79471.7831493465
17 79168.0279060797
18 65450.0725212894
};
\addplot [semithick, color2]
table {%
0 344588.463986213
1 365901.906870291
2 361394.050951375
3 364663.064176538
4 385669.481978828
5 377182.200226656
6 379779.344955736
7 409258.66036721
8 416737.57530998
9 422226.71388843
10 408123.581755196
11 398233.010849068
12 384724.452858
13 364804.321236953
14 342088.61526396
15 322446.742545183
16 301294.297406395
17 276689.658149653
18 254435.679999328
};
\addplot [semithick, color3]
table {%
0 82419.5360137868
1 87820.0931297093
2 88955.9490486255
3 96585.9358234618
4 99198.5180211722
5 104002.799773344
6 103592.655044264
7 115117.33963279
8 117996.42469002
9 120952.28611157
10 116043.418244804
11 113353.989150932
12 107359.547142
13 101267.678763047
14 93515.3847360404
15 85534.2574548174
16 81534.7025936045
17 73022.3418503475
18 65992.3200006717
};
\addplot [semithick, color4, dashdotted]
table {%
0 326751.641443689
1 345571.593143435
2 342128.015372112
3 344322.593786827
4 366666.221483727
5 357754.792766276
6 359333.243981498
7 387568.282056037
8 393729.892127237
9 397403.251681373
10 384829.360727103
11 374700.67569684
12 361434.780937388
13 342839.057421769
14 318636.280792962
15 298541.696647644
16 280758.964684413
17 257643.189236384
18 236798.115556812
};
\addplot [semithick, color5, dashdotted]
table {%
0 100256.358556311
1 108150.406856565
2 108221.984627888
3 116926.406213173
4 118201.778516272
5 123430.207233724
6 124038.756018502
7 136807.717943963
8 141004.107872763
9 145775.748318627
10 139337.639272897
11 136886.32430316
12 130649.219062612
13 123232.942578231
14 116967.719207038
15 109439.303352356
16 102070.035315587
17 92068.8107636156
18 83629.8844431883
};
% \path [draw=black, fill opacity=0] (axis cs:0,47611.2404529324)
% --(axis cs:0,440065.545956787);

% \path [draw=black, fill opacity=0] (axis cs:1,47611.2404529324)
% --(axis cs:1,440065.545956787);

% \path [draw=black, fill opacity=0] (axis cs:-0.9,0)
% --(axis cs:18.9,0);

% \path [draw=black, fill opacity=0] (axis cs:-0.9,1)
% --(axis cs:18.9,1);

\end{axis}

\end{tikzpicture}
      \caption{Number of intra- and inter-cluster handovers for 2017/02/02 in San Francisco, $N_c=22$.}
      \label{fig:hoTsSf}
    \end{subfigure}\hfill
    \begin{subfigure}[t]{.49\textwidth}
      \centering
      \setlength\fwidth{\textwidth}
      \setlength\fheight{0.4\textwidth}
      \input{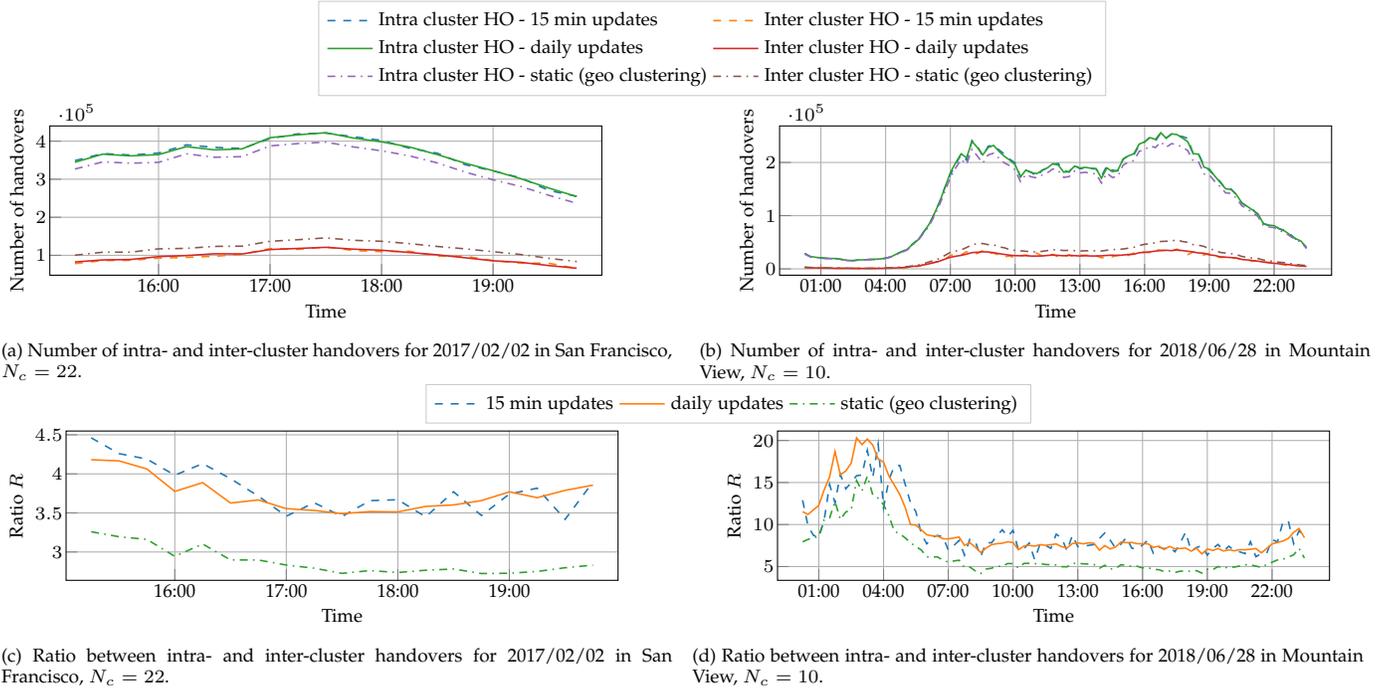}
      \caption{Number of intra- and inter-cluster handovers for 2018/06/28 in Mountain View, $N_c=10$.}
      \label{fig:hoTsMv}
    \end{subfigure}
    \begin{subfigure}[t]{.49\textwidth}
      \centering
      \setlength\abovecaptionskip{-.2cm}
      \setlength\fwidth{\textwidth}
      \setlength\fheight{0.4\textwidth}
      % This file was created by matplotlib2tikz v0.6.17.
\begin{tikzpicture}

\definecolor{color1}{rgb}{1,0.498039215686275,0.0549019607843137}
\definecolor{color0}{rgb}{0.12156862745098,0.466666666666667,0.705882352941177}
\definecolor{color2}{rgb}{0.172549019607843,0.627450980392157,0.172549019607843}
\pgfplotsset{scaled x ticks=false}
\pgfplotsset{every tick label/.append style={font=\scriptsize}}
\begin{axis}[
xlabel={Time},
ylabel={Ratio $R$},
xmin=-0.9, xmax=18.9,
xtick={3, 7, 11, 15},
xticklabels={16:00, 17:00, 18:00, 19:00},
ymin=2.63725997734406, ymax=4.54859004199861,
ylabel shift=-2pt,
yticklabel shift=-2pt,
xlabel shift=-2pt,
xticklabel shift=-2pt,
width=\fwidth,
height=\fheight,
xlabel style={font=\scriptsize},
ylabel style={font=\scriptsize},
tick align=inside,
tick pos=left,
xmajorgrids,
x grid style={lightgray!92.02614379084967!black},
ymajorgrids,
y grid style={lightgray!92.02614379084967!black},
legend style={font=\scriptsize,at={(1.2, 1.05)}, anchor=south,draw=white!80.0!black},
legend entries={{15 min updates},{daily updates},{static (geo clustering)}},
legend columns=3,
legend cell align={left}
]
% \addlegendimage{no markers, color0}
% \addlegendimage{no markers, color1}
% \addlegendimage{no markers, color2}
\addplot [semithick, color0, dashed]
table {%
0 4.46171140269613
1 4.25950327127408
2 4.18960669700544
3 3.98249427277248
4 4.13150954209758
5 3.93859447130612
6 3.71478161120157
7 3.45811488064111
8 3.63059588547734
9 3.45514795844728
10 3.65725363672376
11 3.66932352479653
12 3.44941335872575
13 3.77043359491467
14 3.46726756664955
15 3.74187872116386
16 3.81716887213381
17 3.41733878245493
18 3.89576233694427
};
\addplot [semithick, color1]
table {%
0 4.18090759366289
1 4.16649417952515
2 4.06261812522317
3 3.77552964691529
4 3.88785527921408
5 3.6266542924677
6 3.66608370828475
7 3.55514348813737
8 3.53178137731514
9 3.49085352135433
10 3.51699034661512
11 3.51318038149339
12 3.58351411774437
13 3.60237664862987
14 3.65809985415288
15 3.76979647851052
16 3.69528909559092
17 3.7891096223222
18 3.855534704595
};
\addplot [semithick, color2, dashdotted]
table {%
0 3.25916127564281
1 3.19528703763224
2 3.16135410516164
3 2.94478043872381
4 3.10203641676381
5 2.89843791713678
6 2.89694330639609
7 2.83294164891184
8 2.79232923116343
9 2.7261273309519
10 2.76184785916607
11 2.73731271260521
12 2.76645190480761
13 2.78204066420086
14 2.72413861664654
15 2.7279202946536
16 2.75065021596539
17 2.79837642193377
18 2.83150116891139
};
% \path [draw=black, fill opacity=0] (axis cs:0,2.63725997734406)
% --(axis cs:0,4.54859004199861);

% \path [draw=black, fill opacity=0] (axis cs:1,2.63725997734406)
% --(axis cs:1,4.54859004199861);

% \path [draw=black, fill opacity=0] (axis cs:-0.9,0)
% --(axis cs:18.9,0);

% \path [draw=black, fill opacity=0] (axis cs:-0.9,1)
% --(axis cs:18.9,1);

\end{axis}

\end{tikzpicture}
      \caption{Ratio between intra- and inter-cluster handovers for 2017/02/02 in San Francisco, $N_c=22$.}
      \label{fig:hoRatioSf}
    \end{subfigure}\hfill
    \begin{subfigure}[t]{.49\textwidth}
      \centering
      \setlength\fwidth{\textwidth}
      \setlength\fheight{0.4\textwidth}
      % This file was created by matplotlib2tikz v0.6.17.
\begin{tikzpicture}

\definecolor{color1}{rgb}{1,0.498039215686275,0.0549019607843137}
\definecolor{color0}{rgb}{0.12156862745098,0.466666666666667,0.705882352941177}
\definecolor{color2}{rgb}{0.172549019607843,0.627450980392157,0.172549019607843}
\pgfplotsset{scaled x ticks=false}
\pgfplotsset{every tick label/.append style={font=\scriptsize}}
\begin{axis}[
xlabel={Time},
ylabel={Ratio $R$},
xmin=-4.65, xmax=97.65,
ymin=3.36313584910291, ymax=21.1269966536852,
xtick={3, 15, 27, 39, 51, 63, 75, 87},
xticklabels={01:00, 04:00, 07:00, 10:00, 13:00, 16:00, 19:00, 22:00},
ylabel shift=-2pt,
yticklabel shift=-2pt,
xlabel shift=-2pt,
xticklabel shift=-2pt,
width=\fwidth,
height=\fheight,
xlabel style={font=\scriptsize},
ylabel style={font=\scriptsize},
tick align=inside,
tick pos=left,
xmajorgrids,
x grid style={lightgray!92.02614379084967!black},
ymajorgrids,
y grid style={lightgray!92.02614379084967!black},
legend style={font=\scriptsize,draw=white!80.0!black},
legend cell align={left}
]
% \addlegendimage{no markers, color0}
% \addlegendimage{no markers, color1}
% \addlegendimage{no markers, color2}
\addplot [semithick, color0, dashed]
table {%
0 12.8938249666815
1 9.8302376087422
2 8.72023747754668
3 8.46980987183977
4 10.55311339527
5 14.7678818713477
6 12.5802387267905
7 15.9681603706343
8 14.2435110612219
9 14.9848672785909
10 15.813415036108
11 15.8967956845812
12 18.8216560509554
13 15.589519650655
14 19.5705959443269
15 12.0066707095209
16 15.3653585926928
17 16.9323262839879
18 17.0353264424964
19 14.8127820748098
20 12.5786386028764
21 11.5227791921294
22 8.88126345303256
23 7.70761086259179
24 8.48115872806153
25 8.69846741947527
26 8.55791968582241
27 6.9662409804957
28 7.48849091421848
29 8.10520385534376
30 6.35333987124926
31 8.26347373765301
32 7.99255032014101
33 6.17002528743768
34 7.85376268173779
35 7.82475045038736
36 8.6269729356545
37 9.38184283731681
38 8.57667283831667
39 9.30330203417634
40 6.96789987635096
41 7.13631091241625
42 7.797956787114
43 5.81690867664098
44 7.6929149930748
45 7.52840971056647
46 7.16747604751763
47 8.85094040654555
48 8.41337795482862
49 6.86720705447182
50 8.7414115827405
51 7.36835274492833
52 7.40761065897175
53 7.52114314072793
54 7.61643312091797
55 8.34466826674694
56 8.99803484830386
57 8.14201011234243
58 7.18831547221285
59 7.67204042254346
60 8.45991236535985
61 7.18986550605007
62 7.08996327725355
63 7.58119607728966
64 8.14024081175271
65 6.59016297699967
66 7.32638483810585
67 7.1142497645288
68 7.35099673154939
69 6.76071932385731
70 7.84423568020734
71 9.11475926160257
72 6.82053815262844
73 7.27461142075884
74 7.47457972366943
75 6.35871734145969
76 7.16887192669989
77 7.62989961795185
78 6.86522715683518
79 7.42291745763603
80 6.89282675877416
81 6.75444624667856
82 6.53168884844326
83 7.95488736354899
84 6.1732481376873
85 6.53418695649177
86 6.56584548320358
87 8.22550383064365
88 7.91559104145953
89 10.2803690949284
90 10.4594958741646
91 7.72306442644287
92 9.25158055538867
93 8.6022866050931
};
\addplot [semithick, color1]
table {%
0 11.5300480769231
1 11.2028555812963
2 11.7488791319595
3 12.2806066009051
4 14.1432044952728
5 15.6555225447978
6 18.6913461538462
7 15.9191301824505
8 16.3772365460398
9 17.2915044404798
10 20.3195484352951
11 19.5082897940705
12 20.1700680272109
13 19.4467168998924
14 17.8230755255868
15 17.426116838488
16 15.7623007623008
17 14.6881194661028
18 13.5912988250238
19 12.0453926544181
20 9.99373612381942
21 9.93196465632755
22 9.36512772949191
23 8.78279165007066
24 8.65429568090495
25 8.63061406301565
26 8.31241773948592
27 8.29264916813196
28 8.40081386274078
29 8.51210593714495
30 7.52465053647255
31 7.8288579579767
32 7.32305948973832
33 6.68152457888688
34 7.19283136631265
35 7.60196076948202
36 7.72300579164844
37 7.77296281976619
38 7.94719745290001
39 7.85732938963261
40 7.01051164243745
41 7.43484534381778
42 7.52525883393154
43 7.42639529765203
44 7.60703494761197
45 7.52944126822993
46 7.60943894107066
47 7.69185241507842
48 7.37289432289239
49 7.20856443862311
50 7.74408666399026
51 7.81798354762874
52 7.70334413036996
53 7.81661645371136
54 7.76867360001956
55 6.94795987085456
56 7.49729167220129
57 7.09562374555608
58 7.33288244137647
59 7.91769833062049
60 7.84153982358435
61 7.87299336741403
62 7.71947512751897
63 7.7124372144714
64 7.7116089866763
65 7.28156693489018
66 7.4281037705933
67 7.02541173031919
68 7.40109620295257
69 7.11046076654432
70 7.15113936421234
71 7.20653612363035
72 6.83109146129253
73 7.1848707980029
74 6.50239590063147
75 7.10816520374471
76 6.91617018778223
77 7.15021523774556
78 6.8521625548557
79 6.99740743065433
80 6.86961470455155
81 7.01774614968293
82 6.99391933889017
83 7.06166740583262
84 7.14386063017557
85 6.63152874833625
86 7.14266375250117
87 7.7017369516119
88 7.85590766133465
89 8.03995524805292
90 8.33359429890399
91 9.06724335341491
92 9.5337918726202
93 8.43936495554355
};
\addplot [semithick, color2, dashdotted]
table {%
0 7.93826807659331
1 8.23271737156402
2 8.42318624416696
3 8.58096086258805
4 10.5758746045317
5 10.77356217279
6 12.7442953020134
7 10.5890540948424
8 11.5023047247984
9 11.7216770944257
10 15.2283755170952
11 14.0001655666182
12 15.7612208258528
13 13.5332823259373
14 13.1290999989465
15 12.3549190535492
16 11.0338308457711
17 9.18900388006554
18 8.89654305919518
19 8.4236443763173
20 7.83741111203723
21 7.36807338552201
22 6.97513780590738
23 6.15329809681221
24 6.15070371792214
25 6.14725338922137
26 5.765595633215
27 5.52620700098387
28 5.66245314604467
29 5.75234029389464
30 4.89978340575864
31 4.94363297155546
32 4.48961744245221
33 4.17058406749301
34 4.70162174388903
35 4.79906526804676
36 5.03899548137086
37 5.01898308353906
38 5.36469837314681
39 5.34312426040224
40 4.83921498599679
41 5.37251040184158
42 5.38755099337829
43 5.33975240068259
44 5.3603227490318
45 5.24492807747174
46 5.22344850840828
47 5.19440474858167
48 4.9885741790987
49 4.96810830111758
50 5.36331674396092
51 5.39716694536366
52 5.32681668058393
53 5.29959330966625
54 5.25090036681082
55 4.80336475756444
56 5.15527279888641
57 4.71186963497125
58 4.75790689431668
59 5.21306899921057
60 5.15812865735926
61 5.02228350555668
62 5.03844557164393
63 4.83369834736694
64 4.88176770613748
65 4.65536862778623
66 4.63444815672593
67 4.40289051106864
68 4.47124828224523
69 4.29830384063748
70 4.4509389181675
71 4.44815725384511
72 4.4542558613094
73 4.51544217827941
74 4.224140316163
75 4.70130718673338
76 4.63254966175112
77 5.00194248785953
78 4.80948687392861
79 4.9535262754692
80 4.93456592818342
81 5.06587482713296
82 5.21509577840578
83 5.20253066911936
84 5.05833254732269
85 4.89541806749012
86 5.23763350198691
87 5.5177084240732
88 5.78814642999679
89 5.85677108021879
90 6.13397167285064
91 6.39188349023239
92 7.16810004593197
93 6.01161949933227
};
% \path [draw=black, fill opacity=0] (axis cs:0,3.36313584910291)
% --(axis cs:0,21.1269966536852);

% \path [draw=black, fill opacity=0] (axis cs:1,3.36313584910291)
% --(axis cs:1,21.1269966536852);

% \path [draw=black, fill opacity=0] (axis cs:-4.65,0)
% --(axis cs:97.65,0);

% \path [draw=black, fill opacity=0] (axis cs:-4.65,1)
% --(axis cs:97.65,1);

\end{axis}

\end{tikzpicture}
      \caption{Ratio between intra- and inter-cluster handovers for 2018/06/28 in Mountain View, $N_c=10$.}
      \label{fig:hoRatioMv}
    \end{subfigure}
    \setlength\belowcaptionskip{-.3cm}
    \caption{Number of intra- and inter-cluster handovers (and relative ratio $R$) with different clustering strategies, in different deployments (i.e., San Francisco, with 472 base stations, and Mountain View/Palo Alto, with 178).}
    \label{fig:hots}
\end{figure*}

Fig.~\ref{fig:speccl} reports an example of the clustering applied to the $N_g = 472$ San Francisco base stations, with $N_c = 22$ clusters and $T_c = 24$ hours, i.e., with one clustering update per day, using the data collected in the previous day. The size of the clusters is constrained in $\{0.8 N_g/N_c, \dots, 1.2 N_g/N_c\}$. 
By comparing Figs.~\ref{fig:speccl} and~\ref{fig:geocl}, it can be seen that network-based clustering maintains a proximity criterion (i.e., base stations which are close together are generally clustered together), but this is not as strict as in the geographical one, as it strives to match the users' mobility with the clusters. Consider for example the base station at the bottom right of the figures: it serves an area close to U.S. Route 101, and public transportation stations, thus there are a lot of handovers happening directly from base stations in the downtown area to that \gls{gnb}. Consequently, the network-based approach clusters it with the purple cluster in the city center, reducing the number of inter-controller handovers with respect to the position-based strategy, which associates it to the other base stations at the bottom of the map. In general, it can be seen that in Fig.~\ref{fig:geocl} there are more large black lines connecting the \glspl{gnb}, meaning that base stations with a high level of interactions are placed under different controllers in different clusters. Another example of this can be seen in the comparison between Figs.~\ref{fig:specclmv} and~\ref{fig:geoclmv} for the transitions along the Caltrain railway line that crosses the map on the diagonal. In Fig.~\ref{fig:specclmv}, most of the lines along the railway are colored, showing that intra-cluster handovers happen between the interested base stations, and vice versa in Fig.~\ref{fig:geoclmv}. 

\subsubsection{Reduction of inter-cluster handovers}

The effectiveness of the data-driven approach is eventually highlighted by the reduction in inter-controller handovers. As mentioned in Sec.~\ref{sec:ctrl}, intra-controller handovers can be managed locally, by the controller which is in common to the source and target base stations. Inter-controller handoffs, instead, require the coordination and synchronization of the two controllers, thus increasing the control plane latency to at least twice that of handovers related to a single controller. The actual overhead on the latency introduced by inter-controller communications will depend on signaling specifications that have not been developed yet, and on the controller implementation and processing capabilities, as mentioned in Sec.~\ref{sec:ctrl}, but the need to avoid inter-controller synchronization is valid in any case. Therefore, we report as metrics the number of intra- and inter-controller handovers and their ratio as a function of the number of controllers\footnote{The number of controllers an operator will need to deploy on a network will depend on the capacity of the controllers themselves and the signaling they will need to support.} (and thus clusters) $N_c$ and the time interval between two consecutive updates $T_c$.

In Fig.~\ref{fig:numClusters} we present the ratio $R$ between intra- and inter-cluster handovers by considering $T_c = 24$ hours as fixed, and changing the number of clusters $N_c$. For each value of $N_c$, we run multiple times the clustering algorithms, to average the behavior of K means and provide confidence intervals. It can be seen that the gain of the network-data-based solution over the position-based one is almost constant, especially as the number of clusters grows, with an average increase of the ratio $R$ of 45.38\% for the San Francisco case and 42.62\% for the Mountain View/Palo Alto scenario. The behavior in the two scenarios with $N_c = 2$, however, is different: while in the San Francisco case $N_c=2$ yields the largest difference for the value of $R$ between the network-data- and the location-based clustering, in the Mountain View context it corresponds to the minimum difference. This is due to the difference in the geography of the two areas, as shown in Fig.~\ref{fig:cl}: the San Francisco dataset covers a much larger number of base stations than the other one, and the mobility patterns of the users are less regular, thus the clustering based on the network data can find a better solution than that based on location.

Moreover, in Figs.~\ref{fig:hoTsSf} and~\ref{fig:hoTsMv}, we report the number of handovers for the two configurations shown in Fig.~\ref{fig:cl}, with $T_c = 24$ hours, and for a more dynamic solution based on more frequent updates (i.e., $T_c = 15$ minutes). Moreover, Figs.~\ref{fig:hoRatioSf} and~\ref{fig:hoRatioMv} also plot the ratio between the intra- and inter-cluster handovers.
Notice that the number of handovers reported in Fig.~\ref{fig:hoTsSf} refers to the events happened on February 2nd, while the clustering is based on the data from the previous day. For the 15-minute update case, the clustering is updated every 15 minutes to reflect the statistics from the previous 15 minutes. However, as Fig.~\ref{fig:hoTsSf} shows, updating the clusters with a daily periodicity, using data from the previous day, does not result in significantly degraded performance with respect to the 15-minute updates case. Notice also that a cluster update has some cost in terms of control signaling between the \glspl{gnb} and the controllers. Moreover, the daily-based update builds the graph and the clustering according to a more robust statistics, i.e., based on the transitions for the whole day. This is particularly evident if we consider the example in Figs.~\ref{fig:hoTsMv} and~\ref{fig:hoRatioMv}, which report the same metrics but for a whole day in the Mountain View/Palo Alto area and $N_c=10$ clusters. As it can be seen, at night, when the number of handovers is low, the clustering with update step $T_c = 15$ minutes exhibits a very high variation in the ratio between intra- and inter-cluster handovers, and in some cases has a performance which is similar to that of the geographic case, while the curve for the daily-based update shows a more stable behavior and better performance.

To summarize, we showed that the data-driven clustering based on the proposed architecture (i) adapts to the mobility of users, in different scenarios, thus reducing the inter-controller interactions and, consequently, the control plane latency, and (ii) can be updated on a daily basis without significant performance loss with respect to a more dynamic solution.

\begin{figure}
  \centering
  \setlength\fwidth{0.9\columnwidth}
  \setlength\fheight{0.4\columnwidth}
  \input{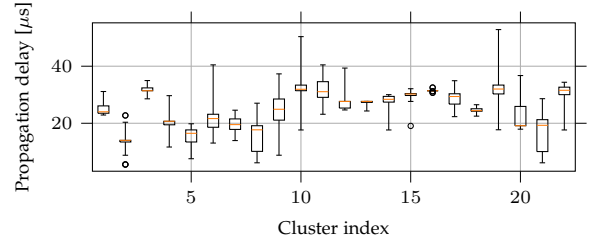}
\caption{Propagation delay between the location of a \gls{ru} (or \gls{gnb}, if not split) and a possible datacenter location at the center of the San Francisco area, and the clusters identified using the data-driven clustering.}
\label{fig:lat}
\end{figure}

Notice that the control plane latency is not significantly impacted by the higher geographical displacement that the data-driven clustering algorithm introduces (as shown in Fig.~\ref{fig:cl}, some locations of the \gls{ru} of a \gls{gnb} may be spaced by several kilometers). Indeed, the communication delay between the \glspl{cu} and the controller is minimized by co-locating the \glspl{cu} and controllers in the same datacenter at the network edge, as mentioned in Sec.~\ref{sec:arch}. Moreover, even in the case in which the \glspl{gnb} were deployed (for example, in the San Francisco area) with \gls{ru}, \gls{du} and \gls{cu} in the same location, the aforementioned delay would be much smaller than the timescale at which the controllers operate. In Fig.~\ref{fig:lat}, we consider the propagation delay on fiber optic cables from the position of each node to that of a possible datacenter in the center of the San Francisco area (e.g., which could represent the location at which fibers to the deployed units would be terminated). As can be seen, for this specific region, the values are all smaller than 53 $\mu$s, with an average of 25 $\mu$s, and are thus much smaller than the timescale at which the \gls{ran} controllers operate (i.e., tens of milliseconds).

\section{Predicting Network KPIs Using Controllers}
\label{sec:pred}

In this section, we present an additional application of the \gls{ml} architecture presented in Sec.~\ref{sec:ctrl}, in which the point of view of the \gls{ran} controllers is exploited to predict the number of users attached to each base station of the cellular network. This metric can be used to forecast useful \glspl{kpi} such as the user throughput, the outage duration and the overall network load. In the following paragraphs, we will first discuss the quality of the prediction with several machine learning algorithms by considering a single cluster among those presented in Fig.~\ref{fig:speccl} for San Francisco. 
%, and then will extend the discussion to all the clusters, using the most promising approaches identified for the first cluster.
The main comparison will be between the accuracy of the prediction with (i) methods that only use local information, i.e., in which each base station is a separate entity (as in 4G) and has available only its own data for the training of the machine learning algorithm, and (ii) techniques that exploit the architecture described in Sec.~\ref{sec:ctrl} to collect and process data, and thus for which it is possible to perform predictions based on the joint history of multiple base stations associated to each controller.
Then, we will extend the analysis to all the clusters, using the most promising approaches identified for the test-cluster, showing how a cluster-based approach reduces the prediction error with respect to a local-based approach. Finally, we will describe some prediction-based applications for network automation and new user services.

\subsection{Data Preprocessing}
The performance analysis presented in this Section is based on the San Francisco dataset. We sampled the number of users in each base station with a time step $T_s = 5$ minutes, and divided the dataset into a training set (which will be used for k-fold cross validation) and a test set. The training set is based on the interval from January 31st to February 20th, while the test set goes from February 21st to February 26th. In the following, the notation $N_{\alpha}^{\beta}$ indicates the number of elements of type $\alpha$ (e.g., $\alpha = u$ for users, $\alpha = bs$ for base stations) associated to the entity $\beta$. For example, $N_u^b$ indicates the number of users in base stations $b$, while $N_{bs}^d$ the number of base stations in set $d$.

\begin{table*}[t]
  \centering
  \begin{tabular}{ll}
    \toprule
    Regression method & Hyperparameters \\\midrule
    \acrlong{brr}~\cite{mackay1992bayesian,shi2016bayesian} & $\alpha \in \{10^{-6}, 10^{-3}, 1, 10, 100\}$, $\lambda \in \{10^{-6}, 10^{-3}, 1, 10, 100\}$ \\
    \acrlong{rfr}~\cite{breiman2001random,douglass2015high} & Number of trees $N_{rf} \in \{1000, 5000, 10000\}$ \\
    \acrlong{gpr}~\cite{rasmussen2004gaussian} & $\alpha \in \{10^{-6}, 10^{-4}, 10^{-2}, 0.1\}$, $\sigma_k \in \{0.001, 0.01\}$ \\ 
    \bottomrule
  \end{tabular}
  \caption{Values of the hyperparameters of the different regressors for the k-fold cross-validation.}
  \label{table:hyperparams}
\end{table*}

Let $\mathcal{B}$ be the set of base stations in San Francisco. For base station $b \in \mathcal{B}$, consider a multi-step ahead prediction of the number of users $N_u^b(t + L)$ at times $t + 1, \dots, t + L$ (where $L \ge 1$ is the \textit{look-ahead} step of the prediction), given the real-time data before time $t$. The features we identified are (i) the past $W$ samples of the number of users (where $W$ is the window of the history used for the prediction), i.e., $N_u^b(t + \tau), \tau \in [-W + 1, 0]$; (ii) an integer $h(t) \in \{0, \dots, 4\}$ that represents the hour of the day (from 3 P.M. to 8 P.M.); and (iii) a boolean $\omega(t)$ that indicates whether the selected day is a weekday. We also tested the cell utilization and the number of handovers as possible features, however they showed small correlation with the prediction target. For each day, given the discontinuities of the collected data, we discard the first $W$ samples, thus the actual size of the training ($M_{tr}$) and test ($M_{te}$) sets depends on the value of $W$.

For the local-based prediction, in which each base station predicts the future number of users based on the knowledge of its own data, the training and test set are composed by the feature matrix $\mathbf{X} \in \mathbb{R}^{M_i, 3W}, i \in \{tr, te\}$, in which each row is a vector 
\begin{multline}
  [N_u^b(t - W + 1), h(t - W + 1), \omega(t - W + 1) \dots,\\N_u^b(t), h(t), \omega(t)],
\end{multline}
and by the target vector $\mathbf{y} \in \mathbb{R}^{M_i, 1}, i \in \{tr, te\}$. For the cluster-based method, instead, the goal is to predict the vector of the numbers of users for all the base stations in the cluster. 
Therefore, for the set $\mathcal{C}_d = \{k_d, \dots, j_d\}  \subset \mathcal{B}$ with the $N^{d}_{bs}$ base stations of cluster $d$, each row of the target matrix $\mathbf{Y} \in \mathbb{R}^{M_i, N^{d}_{bs}}, i \in \{tr, te\}$ is composed by a vector 
\begin{equation}
  [N_u^{k_d}(t + L), \dots, N_u^{j_d}(t + L)].
\end{equation}
Each row of the feature matrix $\mathbf{X} \in \mathbb{R}^{M_i, (N^{d}_{bs} + 2)W}, i \in \{tr, te\}$ is instead a vector 
\begin{multline}
  [N_u^{k_d}(t - W + 1), \dots, \\ N_u^{j_d}(t - W + 1), h(t - W + 1), \omega(t - W + 1), \dots, \\ N_u^{k_d}(t), \dots, N_u^{j_d}(t), h(t), \omega(t)].
\end{multline}
The values of the numbers of users in the training and test sets are transformed with the function $\log(1+x)$ and scaled so that each feature assumes values between 0 and 1. The scaling is fitted on the training set, and then applied also to the test set. For the evaluation of the performance of the different methods and prediction algorithms, we use the \gls{rmse}, defined for a single base station $b$ as 
\begin{equation}
\sigma_b = \sqrt{\frac{1}{M_{te}} \sum_{t = 1}^{M_{te}} (y_b(t) - \hat{y}_b(t))^2},
\end{equation}
with $y_b$ the time series of the real values for the number of users for base station $b$, and $\hat{y}_b$ the predicted one.

\subsection{Algorithm Comparison}
We tested several machine learning algorithms tailored for prediction, i.e., the \gls{brr} for the local-based prediction, and the \gls{gpr} and \gls{rfr} for both the local- and the cluster-based predictions, using the implementations from the popular open-source library scikit-learn~\cite{scikit-learn}.\footnote{An approach based on neural networks was also considered, but, due to the reduced size of the training set, underperformed with respect to the other regression methods.} For each of these methods, we considered different values of $W \in \{1, \dots, 10\}$ and predicted at different future steps $L \in \{1, \dots, 9\}$, i.e., over a time horizon of 45 minutes. 3-fold cross-validation was performed for each method and value of $L$ and $W$ to identify the best hyperparameters, among those summarized in Table~\ref{table:hyperparams}. The split in each fold is done using the \texttt{TimeSeriesSplit} of scikit-learn, i.e., without shuffling, and with increasing indices in each split, to maintain the temporal relation among consecutive samples. We also considered an \gls{arma} predictor for the local-based case, using the implementation in~\cite{seabold2010statsmodels}.

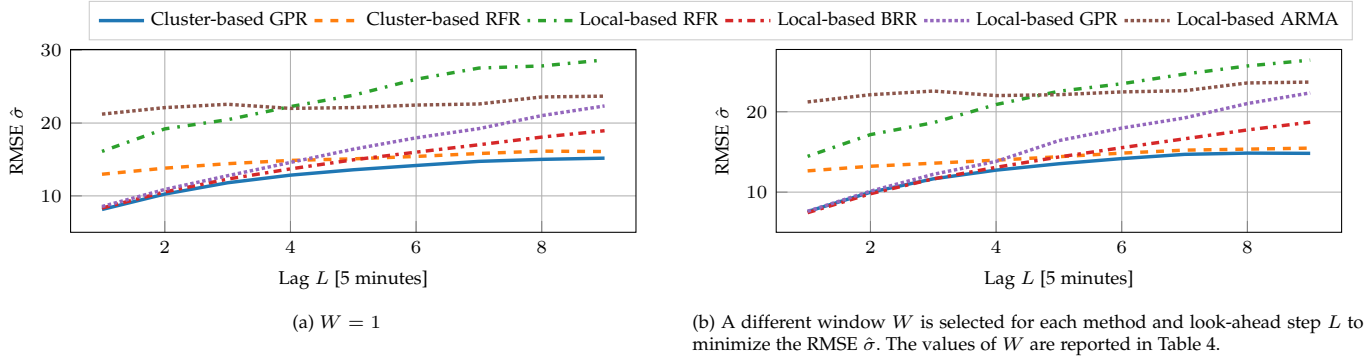
\begin{figure*}[t]
\centering
  \begin{subfigure}[t]{0.49\textwidth}
    \centering
    \setlength\abovecaptionskip{-.2cm}
    \setlength\fwidth{1.02\textwidth}
    \setlength\fheight{.45\textwidth}
    % This file was created by matplotlib2tikz v0.6.17.
\begin{tikzpicture}

\definecolor{color1}{rgb}{1,0.498039215686275,0.0549019607843137}
\definecolor{color0}{rgb}{0.12156862745098,0.466666666666667,0.705882352941177}
\definecolor{color3}{rgb}{0.83921568627451,0.152941176470588,0.156862745098039}
\definecolor{color2}{rgb}{0.172549019607843,0.627450980392157,0.172549019607843}
\definecolor{color5}{rgb}{0.549019607843137,0.337254901960784,0.294117647058824}
\definecolor{color4}{rgb}{0.580392156862745,0.403921568627451,0.741176470588235}
\pgfplotsset{every tick label/.append style={font=\scriptsize}}
\pgfplotsset{scaled x ticks=false}
\begin{axis}[
xlabel={Lag $L$ [5 minutes]},
ylabel={RMSE $\hat{\sigma}$},
xmin=0.5, xmax=9.5,
xlabel style={font=\scriptsize},
ylabel style={font=\scriptsize},
ymin=5, ymax=30.1002686495289,
width=\fwidth,
height=\fheight,
tick align=inside,
tick pos=left,
xmajorgrids,
x grid style={lightgray!92.02614379084967!black},
ymajorgrids,
y grid style={lightgray!92.02614379084967!black},
legend style={font=\scriptsize,at={(1.15,1.05)}, anchor=south, draw=white!80.0!black},
legend columns=6,
legend entries={{Cluster-based \gls{gpr}},{Cluster-based \gls{rfr}},{Local-based \gls{rfr}},{Local-based \gls{brr}},{Local-based \gls{gpr}}, {Local-based \gls{arma}}},
legend cell align={left}
]
% \addlegendimage{no markers, color0}
% \addlegendimage{no markers, color1}
% \addlegendimage{no markers, color2}
% \addlegendimage{no markers, color3}
% \addlegendimage{no markers, color4}
% \addlegendimage{no markers, color5}
\addplot [thick, line width=1.3pt, color0]
table {%
1 8.13295311317822
2 10.2491557288124
3 11.8027169962765
4 12.8486202568259
5 13.5854330150954
6 14.1736838475352
7 14.7390800328519
8 15.014733748166
9 15.1792145653684
};
\addplot [thick, line width=1.3pt, color1, dashed]
table {%
1 12.9729767499125
2 13.8078626367641
3 14.4216932843194
4 14.872387440232
5 15.0862738200385
6 15.4174965897589
7 15.8182394018721
8 16.1438430960707
9 16.0873769394958
};
\addplot [thick, line width=1.3pt, color2, dash pattern=on 1pt off 3pt on 3pt off 3pt]
table {%
1 16.1085817306441
2 19.2008863595974
3 20.458700657538
4 22.2591594226348
5 23.816060813441
6 26.0124475512306
7 27.5508520888724
8 27.8410475303921
9 28.6669225233609
};
\addplot [thick, line width=1.3pt, color3, dashdotted]
table {%
1 8.32742262223218
2 10.5447439932311
3 12.308023620399
4 13.7071527577455
5 14.9509847173459
6 16.002065203405
7 17.0204002336048
8 18.0723949420066
9 18.9457603772524
};
\addplot [thick, line width=1.3pt, color4, densely dotted]
table {%
1 8.53897801978611
2 10.8828718835585
3 12.7705277673578
4 14.5931768956141
5 16.4110936257917
6 17.9708278931006
7 19.2358655074419
8 21.0235629981515
9 22.3488455680325
};

\addplot [thick, line width=1.3pt, color5, densely dotted]
table {%
1 21.22469961
2 22.12063682
3 22.57701663
4 22.01329843
5 22.12540525
6 22.47114782
7 22.61955288
8 23.58539471
9 23.69325246
};

% LSTM: do not report
% \addplot [semithick, color5]
% table {%
% 1 12.3764471268485
% 2 14.1685025898326
% 3 15.6939866869951
% 4 16.4127404224143
% 5 16.32861407292
% 6 16.9053309354027
% 7 17.3974527093857
% 8 17.9927821264473
% 9 18.4700756903833
% 10 0
% };

\end{axis}

\end{tikzpicture}
    \caption{$W = 1$}
    \label{fig:rmseCluster0sameWin}
  \end{subfigure}\hfill
  \begin{subfigure}[t]{0.49\textwidth}
    \centering
    \setlength\fwidth{1.02\textwidth}
    \setlength\fheight{.45\textwidth}
    % This file was created by matplotlib2tikz v0.6.17.
\begin{tikzpicture}

\definecolor{color1}{rgb}{1,0.498039215686275,0.0549019607843137}
\definecolor{color0}{rgb}{0.12156862745098,0.466666666666667,0.705882352941177}
\definecolor{color3}{rgb}{0.83921568627451,0.152941176470588,0.156862745098039}
\definecolor{color2}{rgb}{0.172549019607843,0.627450980392157,0.172549019607843}
\definecolor{color5}{rgb}{0.549019607843137,0.337254901960784,0.294117647058824}
\definecolor{color4}{rgb}{0.580392156862745,0.403921568627451,0.741176470588235}

\pgfplotsset{every tick label/.append style={font=\scriptsize}}
\pgfplotsset{scaled x ticks=false}
\begin{axis}[
xlabel={Lag $L$ [5 minutes]},
ylabel={RMSE $\hat{\sigma}$},
xmin=0.5, xmax=9.5,
xlabel style={font=\scriptsize},
ylabel style={font=\scriptsize},
ymin=5, ymax=27.750482737001,
width=\fwidth,
height=\fheight,
tick align=inside,
tick pos=left,
xmajorgrids,
x grid style={lightgray!92.02614379084967!black},
ymajorgrids,
y grid style={lightgray!92.02614379084967!black},
legend style={font=\scriptsize,at={(0.97,0.03)}, anchor=south east, draw=white!80.0!black},
legend cell align={left}
]
% \addlegendimage{no markers, color0}
% \addlegendimage{no markers, color1}
% \addlegendimage{no markers, color2}
% \addlegendimage{no markers, color3}
% \addlegendimage{no markers, color4}
% \addlegendimage{no markers, color5}
\addplot [thick, line width=1.3pt, color0]
table {%
1 7.58411043177041
2 9.98106747593326
3 11.6512745769582
4 12.7360648753582
5 13.5307617433895
6 14.1736838475352
7 14.7003414294417
8 14.8555202095695
9 14.8255002115967
};
\addplot [thick, line width=1.3pt, color1, dashed]
table {%
1 12.644043200937
2 13.2204870972954
3 13.5984517438214
4 13.9745843247023
5 14.4074179728453
6 14.8551047685713
7 15.2236120341599
8 15.3485994738299
9 15.4765796317052
};
\addplot [thick, line width=1.3pt, color2, dash pattern=on 1pt off 3pt on 3pt off 3pt]
table {%
1 14.4529933262429
2 17.1568941700643
3 18.6391186061374
4 20.9004568244193
5 22.5539814124731
6 23.4831210561152
7 24.6919572232756
8 25.7009612801131
9 26.4290311780962
};
\addplot [thick, line width=1.3pt, color3, dashdotted]
table {%
1 7.44118893930456
2 9.79746864764284
3 11.6461836240763
4 13.0882298429861
5 14.3730289254162
6 15.528718021315
7 16.6354187520533
8 17.7382733236912
9 18.7035975133551
};
\addplot [thick, line width=1.3pt, color4, densely dotted]
table {%
1 7.61518575981789
2 10.118625261603
3 12.214575821443
4 13.8215163147264
5 16.4110936257917
6 17.9708278931006
7 19.2358655074419
8 21.0235629981515
9 22.3488455680325
};

\addplot [thick, line width=1.3pt, color5, densely dotted]
table {%
1 21.22469961
2 22.12063682
3 22.57701663
4 22.01329843
5 22.12540525
6 22.47114782
7 22.61955288
8 23.58539471
9 23.69325246
};

% no lstm
% \addplot [semithick, color5]
% table {%
% 1 10.2450102076666
% 2 12.0791325361793
% 3 13.2976308699348
% 4 14.6687941978375
% 5 14.972812628417
% 6 15.2809707515347
% 7 15.7568376842915
% 8 16.8255051174152
% 9 17.6983851639752
% 10 0
% };

\end{axis}

\end{tikzpicture}
    \caption{A different window $W$ is selected for each method and look-ahead step $L$ to minimize the \gls{rmse} $\hat{\sigma}$. The values of $W$ are reported in Table~\ref{table:windowValues}.}
    \label{fig:rmseCluster0diffWin}
  \end{subfigure}
  \setlength\belowcaptionskip{-.3cm}
  \caption{\gls{rmse} $\hat{\sigma}$ for different local- and cluster-based prediction methods, as a function of the look-ahead step $L$, and for different windows $W$.}
  \label{fig:rmseCluster0}
\end{figure*}

The \gls{brr} (which is used for urban traffic prediction in~\cite{shi2016bayesian}) combines the Bayesian probabilistic approach and the ridge $L_2$ regularization~\cite{mackay1992bayesian}. The Bayesian framework makes it possible to adapt to the data, and only needs the tuning of the parameters $\alpha$ and $\lambda$ of the Gamma priors. However, it does not generalize to multi-output prediction, thus we applied this method only to the local-based scenario.

The \gls{rfr} (used in~\cite{douglass2015high} for population prediction) is a classic ensemble method that trains $N_{rf}$ regression trees from bootstrap samples of the training set and averages their output for the prediction~\cite{breiman2001random}. The only hyperparameters to be tuned are (i) the number of trees $N_{rf}$, for which a higher value implies better generalization properties, but also longer training time; and (ii) the number of random features to sample when splitting the nodes to build additional tree branches, which is set to be equal to the number of features for regression problems. It supports prediction of scalars and vectors, thus we tested it with both the local- and the cluster-based approaches.

The \gls{gpr} is a regressor that fits a Gaussian Process to the observed data~\cite{rasmussen2004gaussian}. The prior has a zero mean, and the covariance matrix described by a kernel. In this case, we chose a kernel in the form
\begin{equation}
  k(x_i, x_j) = \sigma_k^2 + x_i \cdot x_j + \left(1 + \frac{d(x_i, x_j)^2}{2\alpha l^2}\right)^{-\alpha} + \delta_{x_i x_j},
\end{equation}
i.e., the sum of a dot product kernel, that can model non-stationary trends, a rational quadratic kernel with $l=1$ and $\alpha=1$, and a white kernel, that explains the noisy part of the signal. The \gls{gpr} can be used for both single-output and multi-output regressions.

Finally, an \gls{arma} model predicts future values of a certain sequence. The model is based on two polynomials, of order $p$ and $q$. The first is an autoregressive term, which models the stochastic process as a weighted sum of the past history (up to $p$ steps in the past) and a random noise term. The second is a moving average model, which models the randomness in past samples with $q$ white noise terms, added to the average of the process. We selected $p = 4$ and $q = 2$. A first-order differentiation was applied to remove trend from the time series.

Notice that, as we use a real dataset, it is not possible to estimate the theoretical lower bound of the prediction, as it is given by the variance of the unknown distribution of the stochastic process underlying the data we consider.

\subsection{Performance analysis for a sample cluster}

For the comparison between the aforementioned regressors, we consider the cluster $d=0$ with $N^0_d = 22$ base stations in the San Francisco area. We assume that the cluster is stable throughout the training and testing period. In a real deployment, when the base station association to the available controllers changes, a re-training will be needed, together with additional signaling between the controllers, to share the data related to the base stations whose association was updated. 

\begin{figure}[t]
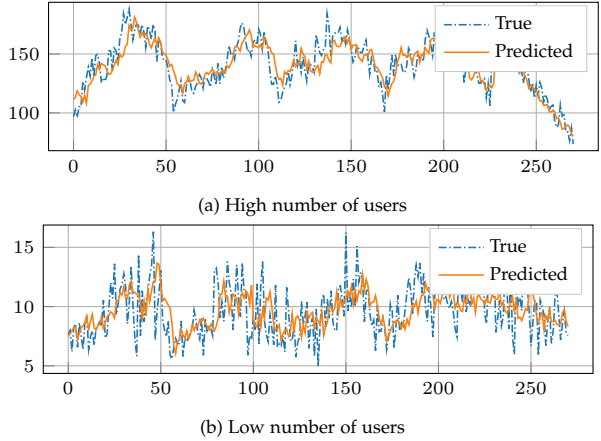

  \centering
  \begin{subfigure}[t]{\columnwidth}
    \centering
      \setlength\belowcaptionskip{.1cm}
    \setlength\fwidth{\columnwidth}
    \setlength\fheight{.4\textwidth}
    \input{figures/0_l_3_w_3000_0.tex}
    \caption{High number of users}
    \label{fig:highNumUe}
  \end{subfigure} 
  \begin{subfigure}[t]{\columnwidth}
    \centering
    \setlength\fwidth{\columnwidth}
    \setlength\fheight{.4\textwidth}
    \input{figures/0_l_3_w_3000_1.tex}
    \caption{Low number of users}
    \label{fig:lowNumUe}
  \end{subfigure} 
  \setlength\belowcaptionskip{-.5cm}
  \caption{Example of predicted vs true time series, for $L=3$ (i.e., 15 minutes ahead), $W=3$ and the cluster-based \gls{gpr} on two base stations for cluster 0.}
  \label{fig:tsExample}
\end{figure}

\begin{figure}[t]
  \centering
  \begin{subfigure}[t]{\columnwidth}
    \setlength\fwidth{\textwidth}
    \setlength\fheight{0.45\textwidth}
    % This file was created by matplotlib2tikz v0.6.17.
\begin{tikzpicture}

\definecolor{color1}{rgb}{0.953,0,0.129}%
\definecolor{color3}{rgb}{1,0.808,0}%
% \definecolor{mycolor5}{rgb}{0.204,0.847,0}%
\definecolor{color0}{rgb}{0.157,0.655,0}%
\definecolor{color2}{rgb}{0.196,0.086,0.69}%500,0.93300}%

\pgfplotsset{every tick label/.append style={font=\scriptsize}}

\begin{axis}[
ybar,
xlabel={Lag $L$ [5 minutes]},
ylabel={RMSE $\hat{\sigma}$},
xlabel style={font=\scriptsize},
ylabel style={font=\scriptsize},
xmin=0.5,
xmax=5.5,
ymin=5, ymax=20,
width=\fwidth,
height=\fheight,
tick align=inside,
tick pos=left,
xtick=data,
xmajorgrids,
x grid style={lightgray!92.02614379084967!black},
ymajorgrids,
y grid style={lightgray!92.02614379084967!black},
legend columns=2,
legend cell align={left},
legend style={at={(0.01,0.8)}, font=\scriptsize, anchor=south west, draw=white!80.0!black}
]

\addplot [bar width=0.12, semithick, fill=color0]
table {%
1 9.5391051474811
2 12.2645872255432
3 14.8782902953242
4 16.8159355202722
5 19.2357141052311
%6 20.6665796880898
};
\addlegendentry{25 hours};

\addplot [bar width=0.12, semithick, fill=color1]
table {%
1 8.46391252089279
2 11.2167140690967
3 13.4569022221144
4 14.9956064542077
5 16.6355845612118
%6 17.8798124559697
};
\addlegendentry{50 hours};

\addplot [bar width=0.12, semithick, fill=color2]
table {%
1 8.16347679662205
2 10.8149728303017
3 12.9549052872079
4 14.3245407020426
5 15.5385567531121
%6 16.4787052703422
};
\addlegendentry{75 hours};

\addplot [bar width=0.12, semithick, fill=color3, postaction={pattern=crosshatch dots}]
table {%
1 7.58411043193974
2 10.0132885934511
3 11.6811373441736
4 12.8041765413189
5 13.6708583009388
%6 14.2951313780908
};
\addlegendentry{Full training (100 hours)};

\end{axis}

\end{tikzpicture}
    \caption{\gls{rmse} $\hat{\sigma}$ of the cluster-based \gls{gpr} on cluster 0 when varying the amount of data used for training, at different future time steps $L$.}
    \label{fig:partial}
  \end{subfigure}\hfill
  \begin{subfigure}[t]{\columnwidth}
    \setlength\fwidth{\textwidth}
    \setlength\fheight{0.45\textwidth}
    \input{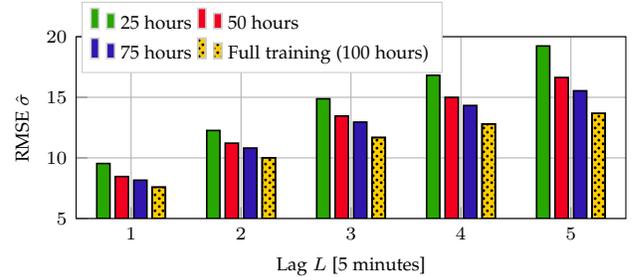}
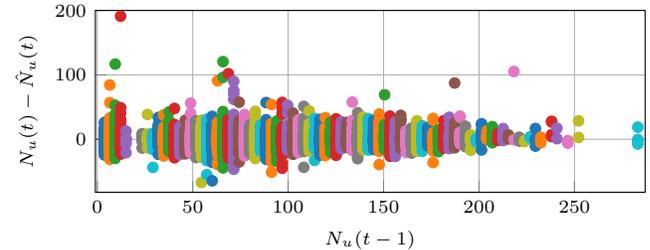
    % \caption{$L = 2$, $W = 2$}
    \caption{Residual error $N_u(t) - \hat{N}_u(t)$, where $N_u(t)$ is the true value of the number of users at time $t$, and $\hat{N}_u(t)$ is the predicted one, as a function of the true value of the number of users $N_u(t-1)$ at time $t-1$. $L=2$.}
    \label{fig:residual}
  \end{subfigure}
  \setlength\belowcaptionskip{-.5cm}
  \caption{Additional results on the prediction accuracy for cluster 0 with the cluster-based \gls{gpr}, $W=2$.}
  \label{fig:moreResults}  
\end{figure}

\begin{table}[b]
  \centering
  \begin{tabular}{llllllllll}
    \toprule
    Look-ahead step $L$ & 1 & 2 & 3 & 4 & 5 & 6 & 7 & 8 & 9 \\ \midrule 
    \gls{brr} & 6 & 6 & 4 & 4 & 3 & 3 & 3 & 2 & 2 \\
    cluster-\gls{gpr} & 3 & 2 & 2 & 2 & 2 & 1 & 6 & 5 & 4 \\ \bottomrule
  \end{tabular}
  \caption{Values of $W$ for the plot in Fig.~\ref{fig:rmseCluster0diffWin} for the \gls{brr} and the cluster-based \gls{gpr}}
  \label{table:windowValues}
\end{table}

\begin{figure*}[t]
  \centering
  \begin{subfigure}[t]{0.16\textwidth}
    \centering  
    \setlength\belowcaptionskip{.1cm}
    \setlength\abovecaptionskip{0cm}
    \setlength\fwidth{1.25\textwidth}
    \setlength\fheight{0.85\textwidth}
    % This file was created by matplotlib2tikz v0.6.17.
\begin{tikzpicture}

\definecolor{color1}{rgb}{1,0.498039215686275,0.0549019607843137}
\definecolor{color0}{rgb}{0.12156862745098,0.466666666666667,0.705882352941177}

\pgfplotsset{every tick label/.append style={font=\tiny}}
\begin{axis}[
xlabel={Lag $L$ [5 minutes]},
ylabel={RMSE $\hat{\sigma}$},
xmin=0.6, xmax=9.4,
ymin=6.62006592743373, ymax=15.3978022382902,
width=\fwidth,
height=\fheight,
xlabel style={font=\tiny},
ylabel style={font=\tiny},
tick align=inside,
tick pos=left,
xmajorgrids,
x grid style={lightgray!92.02614379084967!black},
ymajorgrids,
y grid style={lightgray!92.02614379084967!black},
legend cell align={left},
ylabel shift=-5pt,
yticklabel shift=-3pt,
xlabel shift=-2pt,
xticklabel shift=-2pt,legend style={at={(0.03,0.97)}, anchor=north west, draw=white!80.0!black}
]
%\addlegendimage{no markers,  color0}
%\addlegendimage{no markers,  color1}
\addplot [semithick, color0, dashed]
table {%
1 7.01905394156357
2 8.47119008432215
3 9.53387991587303
4 10.3221806932308
5 10.8468177967779
6 11.2274362790595
7 11.4516452286612
8 11.6067946689476
9 11.6754596070706
};
\addplot [semithick, color1]
table {%
1 7.55437634547429
2 9.19083018094716
3 10.2925292507154
4 11.2995394761323
5 12.2522068599733
6 13.0534208643394
7 13.7899086365793
8 14.440232007271
9 14.9988142241603
};

\end{axis}

\end{tikzpicture}
    \caption{Cluster 1}
    % \label{fig:cluster1}
  \end{subfigure}\hfill
  \begin{subfigure}[t]{0.16\textwidth}
    \centering  
    \setlength\belowcaptionskip{.1cm}
    \setlength\abovecaptionskip{0cm}
    \setlength\fwidth{1.25\textwidth}
    \setlength\fheight{0.85\textwidth}
    % This file was created by matplotlib2tikz v0.6.17.
\begin{tikzpicture}

\definecolor{color1}{rgb}{1,0.498039215686275,0.0549019607843137}
\definecolor{color0}{rgb}{0.12156862745098,0.466666666666667,0.705882352941177}

\pgfplotsset{every tick label/.append style={font=\tiny}}
\begin{axis}[
xlabel={Lag $L$ [5 minutes]},
ylabel={RMSE $\hat{\sigma}$},
xmin=0.6, xmax=9.4,
ymin=3.24422217391589, ymax=11.8345414223563,
width=\fwidth,
height=\fheight,
xlabel style={font=\tiny},
ylabel style={font=\tiny},
tick align=inside,
tick pos=left,
xmajorgrids,
x grid style={lightgray!92.02614379084967!black},
ymajorgrids,
y grid style={lightgray!92.02614379084967!black},
legend cell align={left},
ylabel shift=-5pt,
yticklabel shift=-3pt,
xlabel shift=-2pt,
xticklabel shift=-2pt,legend style={at={(0.03,0.97)}, anchor=north west, draw=white!80.0!black}
]
%\addlegendimage{no markers,  color0}
%\addlegendimage{no markers,  color1}
\addplot [semithick, color0, dashed]
table {%
1 3.63469123066318
2 4.61551391825844
3 5.31234884270068
4 5.79634187741774
5 6.20606561913891
6 6.48468573820226
7 6.7089867536876
8 6.97284677817527
9 7.39681956307643
};
\addplot [semithick, color1]
table {%
1 5.11921572183331
2 6.4346890570691
3 7.4565259038608
4 8.23479667103188
5 8.98851621835124
6 9.71586187684679
7 10.3488068067091
8 10.8831760406468
9 11.444072365609
};

\end{axis}

\end{tikzpicture}
    \caption{Cluster 2}
    % \label{fig:cluster1}
  \end{subfigure}\hfill
  \begin{subfigure}[t]{0.16\textwidth}
    \centering  
    \setlength\abovecaptionskip{-.4cm}
    \setlength\fwidth{1.25\textwidth}
    \setlength\fheight{0.85\textwidth}
    % This file was created by matplotlib2tikz v0.6.17.
\begin{tikzpicture}

\definecolor{color1}{rgb}{1,0.498039215686275,0.0549019607843137}
\definecolor{color0}{rgb}{0.12156862745098,0.466666666666667,0.705882352941177}

\pgfplotsset{every tick label/.append style={font=\tiny}}
\begin{axis}[
xlabel={Lag $L$ [5 minutes]},
ylabel={RMSE $\hat{\sigma}$},
xmin=0.6, xmax=9.4,
ymin=15.0528110017702, ymax=29.2395836143206,
width=\fwidth,
height=\fheight,
xlabel style={font=\tiny},
ylabel style={font=\tiny},
tick align=inside,
tick pos=left,
xmajorgrids,
x grid style={lightgray!92.02614379084967!black},
ymajorgrids,
y grid style={lightgray!92.02614379084967!black},
legend entries={{Cluster-based \gls{gpr}},{Local-based \gls{brr}}},
legend cell align={left},
legend columns=2,
ylabel shift=-5pt,
yticklabel shift=-3pt,
xlabel shift=-2pt,
xticklabel shift=-2pt,
legend style={at={(1.3,1.05)}, anchor=south, draw=white!80.0!black,font=\tiny}
]
%\addlegendimage{no markers,  color0}
%\addlegendimage{no markers,  color1}
\addplot [semithick, color0, dashed]
table {%
1 15.6976643023407
2 18.1992872941832
3 20.2364567621757
4 21.1292676904307
5 21.6693453422311
6 22.5036565830335
7 22.8664855058005
8 23.7706828228559
9 23.6571307027204
};
\addplot [semithick, color1]
table {%
1 18.3640313853548
2 19.6801526234426
3 20.7670863640074
4 22.2521750627845
5 23.5363652595682
6 24.8198856667718
7 26.0673556292008
8 27.3194688638417
9 28.5947303137501
};

\end{axis}

\end{tikzpicture}
    \caption{Cluster 3}
    % \label{fig:cluster1}
  \end{subfigure}\hfill
  \begin{subfigure}[t]{0.16\textwidth}
    \centering  
    \setlength\belowcaptionskip{.1cm}
    \setlength\abovecaptionskip{0cm}
    \setlength\fwidth{1.25\textwidth}
    \setlength\fheight{0.85\textwidth}
    % This file was created by matplotlib2tikz v0.6.17.
\begin{tikzpicture}

\definecolor{color1}{rgb}{1,0.498039215686275,0.0549019607843137}
\definecolor{color0}{rgb}{0.12156862745098,0.466666666666667,0.705882352941177}

\pgfplotsset{every tick label/.append style={font=\tiny}}
\begin{axis}[
xlabel={Lag $L$ [5 minutes]},
ylabel={RMSE $\hat{\sigma}$},
xmin=0.6, xmax=9.4,
ymin=5.62973430516347, ymax=12.444684860937,
width=\fwidth,
height=\fheight,
xlabel style={font=\tiny},
ylabel style={font=\tiny},
tick align=inside,
tick pos=left,
xmajorgrids,
x grid style={lightgray!92.02614379084967!black},
ymajorgrids,
y grid style={lightgray!92.02614379084967!black},
legend cell align={left},
ylabel shift=-5pt,
yticklabel shift=-3pt,
xlabel shift=-2pt,
xticklabel shift=-2pt,legend style={at={(0.03,0.97)}, anchor=north west, draw=white!80.0!black}
]
%\addlegendimage{no markers,  color0}
%\addlegendimage{no markers,  color1}
\addplot [semithick, color0, dashed]
table {%
1 6.05249337498807
2 7.23989906837573
3 8.00874833168413
4 8.64827678102341
5 9.10625505679366
6 9.47913791301803
7 10.4666697869186
8 10.4260724476286
9 10.4338358531474
};
\addplot [semithick, color1]
table {%
1 5.93950478497136
2 7.15997016361002
3 8.05360320919011
4 8.82063761501487
5 9.49136250865239
6 10.1659751666848
7 10.8133387413308
8 11.4923400287689
9 12.1349143811291
};

\end{axis}

\end{tikzpicture}
    \caption{Cluster 4}
    % \label{fig:cluster1}
  \end{subfigure}\hfill
  \begin{subfigure}[t]{0.16\textwidth}
    \centering  
    \setlength\belowcaptionskip{.1cm}
    \setlength\abovecaptionskip{0cm}
    \setlength\fwidth{1.25\textwidth}
    \setlength\fheight{0.85\textwidth}
    % This file was created by matplotlib2tikz v0.6.17.
\begin{tikzpicture}

\definecolor{color1}{rgb}{1,0.498039215686275,0.0549019607843137}
\definecolor{color0}{rgb}{0.12156862745098,0.466666666666667,0.705882352941177}

\pgfplotsset{every tick label/.append style={font=\tiny}}
\begin{axis}[
xlabel={Lag $L$ [5 minutes]},
ylabel={RMSE $\hat{\sigma}$},
xmin=0.6, xmax=9.4,
ymin=9.84078480601725, ymax=24.0774805504972,
width=\fwidth,
height=\fheight,
xlabel style={font=\tiny},
ylabel style={font=\tiny},
tick align=inside,
tick pos=left,
xmajorgrids,
x grid style={lightgray!92.02614379084967!black},
ymajorgrids,
y grid style={lightgray!92.02614379084967!black},
legend cell align={left},
ylabel shift=-5pt,
yticklabel shift=-3pt,
xlabel shift=-2pt,
xticklabel shift=-2pt,legend style={at={(0.03,0.97)}, anchor=north west, draw=white!80.0!black}
]
%\addlegendimage{no markers,  color0}
%\addlegendimage{no markers,  color1}
\addplot [semithick, color0, dashed]
table {%
1 10.4879073398572
2 12.944502725817
3 14.5812225359933
4 15.5433300060572
5 16.3950154347517
6 17.0507702567931
7 17.8191449472743
8 18.2221544931662
9 18.9132349855718
};
\addplot [semithick, color1]
table {%
1 10.5727658086082
2 13.3448245272807
3 15.3544283573074
4 16.9981453037761
5 18.4639754259079
6 19.7315239431979
7 21.0289457097609
8 22.2328288055347
9 23.4303580166572
};
\end{axis}

\end{tikzpicture}
    \caption{Cluster 5}
    % \label{fig:cluster1}
  \end{subfigure}\hfill
  \begin{subfigure}[t]{0.16\textwidth}
    \centering  
    \setlength\belowcaptionskip{.1cm}
    \setlength\abovecaptionskip{0cm}
    \setlength\fwidth{1.25\textwidth}
    \setlength\fheight{0.85\textwidth}
    % This file was created by matplotlib2tikz v0.6.17.
\begin{tikzpicture}

\definecolor{color1}{rgb}{1,0.498039215686275,0.0549019607843137}
\definecolor{color0}{rgb}{0.12156862745098,0.466666666666667,0.705882352941177}

\pgfplotsset{every tick label/.append style={font=\tiny}}
\begin{axis}[
xlabel={Lag $L$ [5 minutes]},
ylabel={RMSE $\hat{\sigma}$},
xmin=0.6, xmax=9.4,
ymin=5.52605652914888, ymax=16.6354719731097,
width=\fwidth,
height=\fheight,
xlabel style={font=\tiny},
ylabel style={font=\tiny},
tick align=inside,
tick pos=left,
xmajorgrids,
x grid style={lightgray!92.02614379084967!black},
ymajorgrids,
y grid style={lightgray!92.02614379084967!black},
legend cell align={left},
ylabel shift=-5pt,
yticklabel shift=-3pt,
xlabel shift=-2pt,
xticklabel shift=-2pt,legend style={at={(0.03,0.97)}, anchor=north west, draw=white!80.0!black}
]
%\addlegendimage{no markers,  color0}
%\addlegendimage{no markers,  color1}
\addplot [semithick, color0, dashed]
table {%
1 6.03102995841983
2 7.9382612822622
3 9.30878226322458
4 10.338992702433
5 11.0661085695673
6 11.3511347103195
7 11.7516645135186
8 11.3916641107262
9 11.6243955309928
};
\addplot [semithick, color1]
table {%
1 6.23894360940412
2 8.38874153840343
3 10.0150508918577
4 11.4563322545496
5 12.5893651042189
6 13.6391183153377
7 14.5724899931572
8 15.3853761247848
9 16.1304985438388
};

\end{axis}

\end{tikzpicture}
    \caption{Cluster 6}
    % \label{fig:cluster1}
  \end{subfigure}\hfill
  \begin{subfigure}[t]{0.16\textwidth}
    \centering  
    \setlength\belowcaptionskip{.1cm}
    \setlength\abovecaptionskip{0cm}
    \setlength\fwidth{1.25\textwidth}
    \setlength\fheight{0.85\textwidth}
    % This file was created by matplotlib2tikz v0.6.17.
\begin{tikzpicture}

\definecolor{color1}{rgb}{1,0.498039215686275,0.0549019607843137}
\definecolor{color0}{rgb}{0.12156862745098,0.466666666666667,0.705882352941177}
\pgfplotsset{every tick label/.append style={font=\tiny}}

\pgfplotsset{every tick label/.append style={font=\tiny}}
\begin{axis}[
xlabel={Lag $L$ [5 minutes]},
ylabel={RMSE $\hat{\sigma}$},
xlabel style={font=\tiny},
ylabel style={font=\tiny},
ylabel shift=-5pt,
yticklabel shift=-3pt,
xlabel shift=-2pt,
xticklabel shift=-2pt,
xmin=0.6, xmax=9.4,
ymin=6.95338313388046, ymax=25.3331227962898,
width=\fwidth,
height=\fheight,
tick align=inside,
tick pos=left,
xmajorgrids,
x grid style={lightgray!92.02614379084967!black},
ymajorgrids,
y grid style={lightgray!92.02614379084967!black},
legend cell align={left},
legend style={at={(0.03,0.97)}, anchor=north west, draw=white!80.0!black}
]
% \addlegendimage{no markers, color0}
% \addlegendimage{no markers, color1}
\addplot [semithick, color0, dashed]
table {%
1 7.78882584580816
2 9.56084074510655
3 10.7609503316483
4 11.6387503544365
5 12.1088884084688
6 12.2059328611921
7 12.4985251755996
8 12.8469724255839
9 12.9495011920412
};
\addplot [semithick, color1]
table {%
1 8.12559115959759
2 10.6971147664826
3 12.8381742055978
4 14.8370875216752
5 16.7569986154741
6 18.5579751476293
7 20.4767528367644
8 22.5479271118924
9 24.4976800843621
};

\end{axis}

\end{tikzpicture}
    \caption{Cluster 7}
    % \label{fig:cluster1}
  \end{subfigure}\hfill
  \begin{subfigure}[t]{0.16\textwidth}
    \centering  
    \setlength\belowcaptionskip{.1cm}
    \setlength\abovecaptionskip{0cm}
    \setlength\fwidth{1.25\textwidth}
    \setlength\fheight{0.85\textwidth}
    % This file was created by matplotlib2tikz v0.6.17.
\begin{tikzpicture}

\definecolor{color1}{rgb}{1,0.498039215686275,0.0549019607843137}
\definecolor{color0}{rgb}{0.12156862745098,0.466666666666667,0.705882352941177}
\pgfplotsset{every tick label/.append style={font=\tiny}}

\pgfplotsset{every tick label/.append style={font=\tiny}}
\begin{axis}[
xlabel={Lag $L$ [5 minutes]},
ylabel={RMSE $\hat{\sigma}$},
xlabel style={font=\tiny},
ylabel style={font=\tiny},
ylabel shift=-5pt,
yticklabel shift=-3pt,
xlabel shift=-2pt,
xticklabel shift=-2pt,
xmin=0.6, xmax=9.4,
ymin=6.8526678823917, ymax=19.9106934969324,
width=\fwidth,
height=\fheight,
tick align=inside,
tick pos=left,
xmajorgrids,
x grid style={lightgray!92.02614379084967!black},
ymajorgrids,
y grid style={lightgray!92.02614379084967!black},
legend cell align={left},
legend style={at={(0.03,0.97)}, anchor=north west, draw=white!80.0!black}
]
% \addlegendimage{no markers, color0}
% \addlegendimage{no markers, color1}
\addplot [semithick, color0, dashed]
table {%
1 7.44621450123446
2 9.29922614169848
3 10.6656215291348
4 11.8289246931934
5 12.6331444913132
6 13.2441603237306
7 13.6606813993967
8 13.904538670399
9 14.1082045252145
};
\addplot [semithick, color1]
table {%
1 8.14732150793079
2 10.387595895948
3 12.222316170041
4 13.7815708644206
5 15.1009957866894
6 16.2690149447863
7 17.3363782918801
8 18.3658316996672
9 19.3171468780896
};

\end{axis}

\end{tikzpicture}
    \caption{Cluster 8}
    % \label{fig:cluster1}
  \end{subfigure}\hfill
  \begin{subfigure}[t]{0.16\textwidth}
    \centering  
    \setlength\belowcaptionskip{.1cm}
    \setlength\abovecaptionskip{0cm}
    \setlength\fwidth{1.25\textwidth}
    \setlength\fheight{0.85\textwidth}
    % This file was created by matplotlib2tikz v0.6.17.
\begin{tikzpicture}

\definecolor{color1}{rgb}{1,0.498039215686275,0.0549019607843137}
\definecolor{color0}{rgb}{0.12156862745098,0.466666666666667,0.705882352941177}

\pgfplotsset{every tick label/.append style={font=\tiny}}
\begin{axis}[
xlabel={Lag $L$ [5 minutes]},
ylabel={RMSE $\hat{\sigma}$},
xlabel style={font=\tiny},
ylabel style={font=\tiny},
ylabel shift=-5pt,
yticklabel shift=-3pt,
xlabel shift=-2pt,
xticklabel shift=-2pt,
xmin=0.6, xmax=9.4,
ymin=3.59946454136782, ymax=20.6567937322613,
width=\fwidth,
height=\fheight,
tick align=inside,
tick pos=left,
xmajorgrids,
x grid style={lightgray!92.02614379084967!black},
ymajorgrids,
y grid style={lightgray!92.02614379084967!black},
legend cell align={left},
legend style={at={(0.03,0.97)}, anchor=north west, draw=white!80.0!black}
]
% \addlegendimage{no markers, color0}
% \addlegendimage{no markers, color1}
\addplot [semithick, color0, dashed]
table {%
1 4.37479768640844
2 7.38014395806991
3 9.48981135576176
4 11.0399971564959
5 12.0601033805137
6 12.8206406468876
7 13.3558164663095
8 12.9985368980886
9 12.9401042573777
};
\addplot [semithick, color1]
table {%
1 4.76651317553699
2 8.24160071407812
3 10.6005886678932
4 12.7318070234025
5 14.4621397323348
6 16.0197696593525
7 17.4331666763734
8 18.871116990569
9 19.8814605872207
};

\end{axis}

\end{tikzpicture}
    \caption{Cluster 9}
    % \label{fig:cluster1}
  \end{subfigure}\hfill
  \begin{subfigure}[t]{0.16\textwidth}
    \centering  
    \setlength\belowcaptionskip{.1cm}
    \setlength\abovecaptionskip{0cm}
    \setlength\fwidth{1.25\textwidth}
    \setlength\fheight{0.85\textwidth}
    % This file was created by matplotlib2tikz v0.6.17.
\begin{tikzpicture}

\definecolor{color1}{rgb}{1,0.498039215686275,0.0549019607843137}
\definecolor{color0}{rgb}{0.12156862745098,0.466666666666667,0.705882352941177}

\pgfplotsset{every tick label/.append style={font=\tiny}}
\begin{axis}[
xlabel={Lag $L$ [5 minutes]},
ylabel={RMSE $\hat{\sigma}$},
xlabel style={font=\tiny},
ylabel style={font=\tiny},
ylabel shift=-5pt,
yticklabel shift=-3pt,
xlabel shift=-2pt,
xticklabel shift=-2pt,
xmin=0.6, xmax=9.4,
ymin=6.76671567904875, ymax=15.6419540087058,
width=\fwidth,
height=\fheight,
tick align=inside,
tick pos=left,
xmajorgrids,
x grid style={lightgray!92.02614379084967!black},
ymajorgrids,
y grid style={lightgray!92.02614379084967!black},
legend cell align={left},
legend style={at={(0.03,0.97)}, anchor=north west, draw=white!80.0!black}
]

\addplot [semithick, color1]
table {%
1 7.93203003255926
2 9.50000506376864
3 10.587061527222
4 11.5273992775804
5 12.4518408394599
6 13.4197414947215
7 14.6996185686723
8 14.6407868133652
9 15.2385340846305
};
\addplot [semithick, color0, dashed]
table {%
1 7.17013560312407
2 8.70366584703016
3 9.77888016645915
4 10.9401335966708
5 11.776524277872
6 12.5316958368859
7 13.2148010728093
8 13.8831830981097
9 14.5720763412505
};

\end{axis}

\end{tikzpicture}
    \caption{Cluster 10}
    % \label{fig:cluster1}
  \end{subfigure}\hfill
  \begin{subfigure}[t]{0.16\textwidth}
    \centering  
    \setlength\belowcaptionskip{.1cm}
    \setlength\abovecaptionskip{0cm}
    \setlength\fwidth{1.25\textwidth}
    \setlength\fheight{0.85\textwidth}
    % This file was created by matplotlib2tikz v0.6.17.
\begin{tikzpicture}

\definecolor{color1}{rgb}{1,0.498039215686275,0.0549019607843137}
\definecolor{color0}{rgb}{0.12156862745098,0.466666666666667,0.705882352941177}

\pgfplotsset{every tick label/.append style={font=\tiny}}
\begin{axis}[
xlabel={Lag $L$ [5 minutes]},
ylabel={RMSE $\hat{\sigma}$},
xmin=0.6, xmax=9.4,
ymin=2.20556914793487, ymax=10.9528482248374,
width=\fwidth,
height=\fheight,
xlabel style={font=\tiny},
ylabel style={font=\tiny},
tick align=inside,
tick pos=left,
xmajorgrids,
x grid style={lightgray!92.02614379084967!black},
ymajorgrids,
y grid style={lightgray!92.02614379084967!black},
legend cell align={left},
ylabel shift=-5pt,
yticklabel shift=-3pt,
xlabel shift=-2pt,
xticklabel shift=-2pt,legend style={at={(0.03,0.97)}, anchor=north west, draw=white!80.0!black}
]
%\addlegendimage{no markers,  color0}
%\addlegendimage{no markers,  color1}
\addplot [semithick, color0, dashed]
table {%
1 2.60317274233953
2 3.16794922243067
3 3.51081261774223
4 3.83469812566139
5 4.08933741036077
6 4.18632991568326
7 4.15515371429606
8 4.23021963462245
9 4.34815370010601
};
\addplot [semithick, color1]
table {%
1 4.617035396103
2 5.9689749747172
3 7.03242504534691
4 8.0407526400044
5 8.71303569713253
6 9.32859011781181
7 9.80552590657542
8 10.2206250555699
9 10.5552446304328
};

\end{axis}

\end{tikzpicture}
    \caption{Cluster 11}
    % \label{fig:cluster1}
  \end{subfigure}\hfill
  \begin{subfigure}[t]{0.16\textwidth}
    \centering  
    \setlength\belowcaptionskip{.1cm}
    \setlength\abovecaptionskip{0cm}
    \setlength\fwidth{1.25\textwidth}
    \setlength\fheight{0.85\textwidth}
    % This file was created by matplotlib2tikz v0.6.17.
\begin{tikzpicture}

\definecolor{color1}{rgb}{1,0.498039215686275,0.0549019607843137}
\definecolor{color0}{rgb}{0.12156862745098,0.466666666666667,0.705882352941177}

\pgfplotsset{every tick label/.append style={font=\tiny}}
\begin{axis}[
xlabel={Lag $L$ [5 minutes]},
ylabel={RMSE $\hat{\sigma}$},
xmin=0.6, xmax=9.4,
ymin=1.87429352287567, ymax=24.7479174746339,
width=\fwidth,
height=\fheight,
xlabel style={font=\tiny},
ylabel style={font=\tiny},
tick align=inside,
tick pos=left,
xmajorgrids,
x grid style={lightgray!92.02614379084967!black},
ymajorgrids,
y grid style={lightgray!92.02614379084967!black},
legend cell align={left},
ylabel shift=-5pt,
yticklabel shift=-3pt,
xlabel shift=-2pt,
xticklabel shift=-2pt,legend style={at={(0.03,0.97)}, anchor=north west, draw=white!80.0!black}
]
%\addlegendimage{no markers,  color0}
%\addlegendimage{no markers,  color1}
\addplot [semithick, color0, dashed]
table {%
1 2.91400370250104
2 4.00873110630603
3 4.82605165926214
4 4.97677610193741
5 5.34993122057525
6 5.37657496411583
7 5.41421872212744
8 5.5837797042194
9 5.84713960620749
};
\addplot [semithick, color1]
table {%
1 7.31630619209556
2 10.8482685146053
3 13.8553729273334
4 16.2626924381947
5 18.3569270104272
6 20.0498923742496
7 21.4845637776803
8 22.6753823842289
9 23.7082072950085
};

\end{axis}

\end{tikzpicture}
    \caption{Cluster 12}
    % \label{fig:cluster1}
  \end{subfigure}\hfill
  \begin{subfigure}[t]{0.16\textwidth}
    \centering  
    \setlength\belowcaptionskip{.1cm}
    \setlength\abovecaptionskip{0cm}
    \setlength\fwidth{1.25\textwidth}
    \setlength\fheight{0.85\textwidth}
    % This file was created by matplotlib2tikz v0.6.17.
\begin{tikzpicture}

\definecolor{color1}{rgb}{1,0.498039215686275,0.0549019607843137}
\definecolor{color0}{rgb}{0.12156862745098,0.466666666666667,0.705882352941177}

\pgfplotsset{every tick label/.append style={font=\tiny}}
\begin{axis}[
xlabel={Lag $L$ [5 minutes]},
ylabel={RMSE $\hat{\sigma}$},
xmin=0.6, xmax=9.4,
ymin=5.68536367717939, ymax=25.3196389247189,
width=\fwidth,
height=\fheight,
xlabel style={font=\tiny},
ylabel style={font=\tiny},
tick align=inside,
tick pos=left,
xmajorgrids,
x grid style={lightgray!92.02614379084967!black},
ymajorgrids,
y grid style={lightgray!92.02614379084967!black},
legend cell align={left},
ylabel shift=-5pt,
yticklabel shift=-3pt,
xlabel shift=-2pt,
xticklabel shift=-2pt,legend style={at={(0.03,0.97)}, anchor=north west, draw=white!80.0!black}
]
%\addlegendimage{no markers,  color0}
%\addlegendimage{no markers,  color1}
\addplot [semithick, color0, dashed]
table {%
1 6.57783073388573
2 10.0639888143963
3 13.9226091888152
4 14.8949033212384
5 17.2019528218381
6 16.3614763716705
7 18.1568020841849
8 18.7341401959908
9 16.8868432898926
};
\addplot [semithick, color1]
table {%
1 8.12449647377324
2 11.4740158810464
3 14.3127842399225
4 16.6840377927609
5 18.7126818218614
6 20.3424368808079
7 21.8955279321579
8 23.2545725987464
9 24.4271718680125
};

\end{axis}

\end{tikzpicture}
    \caption{Cluster 13}
    % \label{fig:cluster1}
  \end{subfigure}\hfill
  \begin{subfigure}[t]{0.16\textwidth}
    \centering  
    \setlength\belowcaptionskip{.1cm}
    \setlength\abovecaptionskip{0cm}
    \setlength\fwidth{1.25\textwidth}
    \setlength\fheight{0.85\textwidth}
    % This file was created by matplotlib2tikz v0.6.17.
\begin{tikzpicture}

\definecolor{color1}{rgb}{1,0.498039215686275,0.0549019607843137}
\definecolor{color0}{rgb}{0.12156862745098,0.466666666666667,0.705882352941177}

\pgfplotsset{every tick label/.append style={font=\tiny}}
\begin{axis}[
xlabel={Lag $L$ [5 minutes]},
ylabel={RMSE $\hat{\sigma}$},
xmin=0.6, xmax=9.4,
ymin=3.36259543967565, ymax=24.6534147319157,
width=\fwidth,
height=\fheight,
xlabel style={font=\tiny},
ylabel style={font=\tiny},
tick align=inside,
tick pos=left,
xmajorgrids,
x grid style={lightgray!92.02614379084967!black},
ymajorgrids,
y grid style={lightgray!92.02614379084967!black},
legend cell align={left},
ylabel shift=-5pt,
yticklabel shift=-3pt,
xlabel shift=-2pt,
xticklabel shift=-2pt,legend style={at={(0.03,0.97)}, anchor=north west, draw=white!80.0!black}
]
%\addlegendimage{no markers,  color0}
%\addlegendimage{no markers,  color1}
\addplot [semithick, color0, dashed]
table {%
1 4.33035995295929
2 7.60527831447842
3 9.11715988466185
4 8.95364356901316
5 9.57125906983605
6 10.0694233714827
7 10.4580913264275
8 10.5139268276154
9 10.3387536440248
};
\addplot [semithick, color1]
table {%
1 5.43649591196104
2 10.0376368481207
3 13.1743351233658
4 15.8945264132613
5 17.8394585184836
6 19.6331867083123
7 21.2344026314613
8 22.3975767596396
9 23.685650218632
};

\end{axis}

\end{tikzpicture}
    \caption{Cluster 14}
    % \label{fig:cluster1}
  \end{subfigure}\hfill
  \begin{subfigure}[t]{0.16\textwidth}
    \centering  
    \setlength\belowcaptionskip{.1cm}
    \setlength\abovecaptionskip{0cm}
    \setlength\fwidth{1.25\textwidth}
    \setlength\fheight{0.85\textwidth}
    % This file was created by matplotlib2tikz v0.6.17.
\begin{tikzpicture}

\definecolor{color1}{rgb}{1,0.498039215686275,0.0549019607843137}
\definecolor{color0}{rgb}{0.12156862745098,0.466666666666667,0.705882352941177}

\pgfplotsset{every tick label/.append style={font=\tiny}}
\begin{axis}[
xlabel={Lag $L$ [5 minutes]},
ylabel={RMSE $\hat{\sigma}$},
xmin=0.6, xmax=9.4,
ymin=1.04115463713453, ymax=3.31795410304463,
width=\fwidth,
height=\fheight,
xlabel style={font=\tiny},
ylabel style={font=\tiny},
tick align=inside,
tick pos=left,
xmajorgrids,
x grid style={lightgray!92.02614379084967!black},
ymajorgrids,
y grid style={lightgray!92.02614379084967!black},
legend cell align={left},
ylabel shift=-5pt,
yticklabel shift=-3pt,
xlabel shift=-2pt,
xticklabel shift=-2pt,legend style={at={(0.03,0.97)}, anchor=north west, draw=white!80.0!black}
]
%\addlegendimage{no markers,  color0}
%\addlegendimage{no markers,  color1}
\addplot [semithick, color0, dashed]
table {%
1 1.14464552194863
2 1.48340361092487
3 1.68118816729467
4 1.87971856251333
5 2.00196325649125
6 2.11664462651031
7 2.17758151234594
8 2.22431418110551
9 2.22755777666312
};
\addplot [semithick, color1]
table {%
1 1.41909289399871
2 1.78038677269489
3 2.04464452046641
4 2.28495351648176
5 2.48933531012762
6 2.69362272473556
7 2.88652323423153
8 3.05368150779645
9 3.21446321823053
};

\end{axis}

\end{tikzpicture}
    \caption{Cluster 15}
    % \label{fig:cluster1}
  \end{subfigure}\hfill
  \begin{subfigure}[t]{0.16\textwidth}
    \centering  
    \setlength\belowcaptionskip{.1cm}
    \setlength\abovecaptionskip{0cm}
    \setlength\fwidth{1.25\textwidth}
    \setlength\fheight{0.85\textwidth}
    % This file was created by matplotlib2tikz v0.6.17.
\begin{tikzpicture}

\definecolor{color1}{rgb}{1,0.498039215686275,0.0549019607843137}
\definecolor{color0}{rgb}{0.12156862745098,0.466666666666667,0.705882352941177}

\pgfplotsset{every tick label/.append style={font=\tiny}}
\begin{axis}[
xlabel={Lag $L$ [5 minutes]},
ylabel={RMSE $\hat{\sigma}$},
xmin=0.6, xmax=9.4,
ymin=4.93242952602717, ymax=15.9895164503649,
width=\fwidth,
height=\fheight,
xlabel style={font=\tiny},
ylabel style={font=\tiny},
tick align=inside,
tick pos=left,
xmajorgrids,
x grid style={lightgray!92.02614379084967!black},
ymajorgrids,
y grid style={lightgray!92.02614379084967!black},
legend cell align={left},
ylabel shift=-5pt,
yticklabel shift=-3pt,
xlabel shift=-2pt,
xticklabel shift=-2pt,legend style={at={(0.03,0.97)}, anchor=north west, draw=white!80.0!black}
]
%\addlegendimage{no markers,  color0}
%\addlegendimage{no markers,  color1}
\addplot [semithick, color0, dashed]
table {%
1 5.43502438622433
2 7.11458613597398
3 8.17663442756053
4 8.93098812872745
5 9.55520512377759
6 9.76185641517592
7 9.33148917718277
8 9.20876906448366
9 9.3287845391505
};
\addplot [semithick, color1]
table {%
1 5.91594550824663
2 7.97622364674966
3 9.54692984653486
4 10.8819631629819
5 11.9820747694062
6 12.9786080919779
7 13.8696357480839
8 14.6911547005391
9 15.4869215901677
};

\end{axis}

\end{tikzpicture}
    \caption{Cluster 16}
    % \label{fig:cluster1}
  \end{subfigure}\hfill
  \begin{subfigure}[t]{0.16\textwidth}
    \centering  
    \setlength\belowcaptionskip{.1cm}
    \setlength\abovecaptionskip{0cm}
    \setlength\fwidth{1.25\textwidth}
    \setlength\fheight{0.85\textwidth}
    % This file was created by matplotlib2tikz v0.6.17.
\begin{tikzpicture}

\definecolor{color1}{rgb}{1,0.498039215686275,0.0549019607843137}
\definecolor{color0}{rgb}{0.12156862745098,0.466666666666667,0.705882352941177}
\pgfplotsset{every tick label/.append style={font=\tiny}}
\pgfplotsset{every tick label/.append style={font=\tiny}}
\begin{axis}[
xlabel={Lag $L$ [5 minutes]},
ylabel={RMSE $\hat{\sigma}$},
xlabel style={font=\tiny},
ylabel style={font=\tiny},
ylabel shift=-5pt,
yticklabel shift=-3pt,
xlabel shift=-2pt,
xticklabel shift=-2pt,
xmin=0.6, xmax=9.4,
ymin=4.74427609721371, ymax=12.726180865432,
width=\fwidth,
height=\fheight,
tick align=inside,
tick pos=left,
xmajorgrids,
x grid style={lightgray!92.02614379084967!black},
ymajorgrids,
y grid style={lightgray!92.02614379084967!black},
legend cell align={left},
legend style={at={(0.03,0.97)}, anchor=north west, draw=white!80.0!black}
]
% \addlegendimage{no markers, color0}
% \addlegendimage{no markers, color1}
\addplot [semithick, color0, dashed]
table {%
1 5.10708995031455
2 6.36366722280376
3 7.15273597362418
4 7.79402226088025
5 8.25024347127471
6 8.50052149412227
7 8.68787017202916
8 8.88839602802637
9 9.09102074477982
};
\addplot [semithick, color1]
table {%
1 5.16925765008344
2 6.66109587459427
3 7.76068471832383
4 8.67774098755941
5 9.44654432388014
6 10.2056608672713
7 10.9257761421114
8 11.6653068537308
9 12.3633670123312
};

\end{axis}

\end{tikzpicture}
    \caption{Cluster 17}
    % \label{fig:cluster1}
  \end{subfigure}\hfill
  \begin{subfigure}[t]{0.16\textwidth}
    \centering  
    \setlength\belowcaptionskip{.1cm}
    \setlength\abovecaptionskip{0cm}
    \setlength\fwidth{1.25\textwidth}
    \setlength\fheight{0.85\textwidth}
    % This file was created by matplotlib2tikz v0.6.17.
\begin{tikzpicture}

\definecolor{color1}{rgb}{1,0.498039215686275,0.0549019607843137}
\definecolor{color0}{rgb}{0.12156862745098,0.466666666666667,0.705882352941177}
\pgfplotsset{every tick label/.append style={font=\tiny}}
\pgfplotsset{every tick label/.append style={font=\tiny}}
\begin{axis}[
xlabel={Lag $L$ [5 minutes]},
ylabel={RMSE $\hat{\sigma}$},
xlabel style={font=\tiny},
ylabel style={font=\tiny},
ylabel shift=-5pt,
yticklabel shift=-3pt,
xlabel shift=-2pt,
xticklabel shift=-2pt,
xmin=0.6, xmax=9.4,
ymin=6.50157046865135, ymax=20.1812642173424,
width=\fwidth,
height=\fheight,
tick align=inside,
tick pos=left,
xmajorgrids,
x grid style={lightgray!92.02614379084967!black},
ymajorgrids,
y grid style={lightgray!92.02614379084967!black},
legend cell align={left},
legend style={at={(0.03,0.97)}, anchor=north west, draw=white!80.0!black}
]
% \addlegendimage{no markers, color0}
% \addlegendimage{no markers, color1}
\addplot [semithick, color0, dashed]
table {%
1 7.12337472995548
2 9.53292188383557
3 11.0888743020479
4 12.202522340779
5 12.6541773338118
6 12.3111890537669
7 12.5461560107966
8 12.7750194359452
9 12.987307681551
};
\addplot [semithick, color1]
table {%
1 7.64507353443581
2 10.1938224236961
3 12.0592134284796
4 13.6653118186253
5 14.9826121697056
6 16.208210406923
7 17.3957638363182
8 18.5607900158896
9 19.5594599560382
};

\end{axis}

\end{tikzpicture}
    \caption{Cluster 18}
    % \label{fig:cluster1}
  \end{subfigure}\hfill
  \begin{subfigure}[t]{0.16\textwidth}
    \centering  
    \setlength\belowcaptionskip{.1cm}
    \setlength\abovecaptionskip{0cm}
    \setlength\fwidth{1.25\textwidth}
    \setlength\fheight{0.85\textwidth}
    % This file was created by matplotlib2tikz v0.6.17.
\begin{tikzpicture}

\definecolor{color1}{rgb}{1,0.498039215686275,0.0549019607843137}
\definecolor{color0}{rgb}{0.12156862745098,0.466666666666667,0.705882352941177}

\pgfplotsset{every tick label/.append style={font=\tiny}}
\begin{axis}[
xlabel={Lag $L$ [5 minutes]},
ylabel={RMSE $\hat{\sigma}$},
xlabel style={font=\tiny},
ylabel style={font=\tiny},
ylabel shift=-5pt,
yticklabel shift=-3pt,
xlabel shift=-2pt,
xticklabel shift=-2pt,
xmin=0.6, xmax=9.4,
ymin=7.59510173542651, ymax=21.1443825110806,
width=\fwidth,
height=\fheight,
tick align=inside,
tick pos=left,
xmajorgrids,
x grid style={lightgray!92.02614379084967!black},
ymajorgrids,
y grid style={lightgray!92.02614379084967!black},
legend cell align={left},
legend style={at={(0.03,0.97)}, anchor=north west, draw=white!80.0!black}
]

\addplot [semithick, color0, dashed]
table {%
1 8.21097813431988
2 10.2696090561427
3 11.5669566692588
4 12.366735832717
5 12.8792896998022
6 12.9880728495191
7 13.1411848587309
8 13.2477852754781
9 13.3847127890363
};
\addplot [semithick, color1]
table {%
1 8.22457534899973
2 10.5937859520454
3 12.3610022611298
4 14.0212168063307
5 15.4870564233978
6 16.8340081168741
7 18.14097983112
8 19.3522018580286
9 20.5285061121872
};

\end{axis}

\end{tikzpicture}
    \caption{Cluster 19}
    % \label{fig:cluster1}
  \end{subfigure}\hfill
  \begin{subfigure}[t]{0.16\textwidth}
    \centering  
    \setlength\belowcaptionskip{.1cm}
    \setlength\abovecaptionskip{0cm}
    \setlength\fwidth{1.25\textwidth}
    \setlength\fheight{0.85\textwidth}
    % This file was created by matplotlib2tikz v0.6.17.
\begin{tikzpicture}

\definecolor{color1}{rgb}{1,0.498039215686275,0.0549019607843137}
\definecolor{color0}{rgb}{0.12156862745098,0.466666666666667,0.705882352941177}

\pgfplotsset{every tick label/.append style={font=\tiny}}
\begin{axis}[
xlabel={Lag $L$ [5 minutes]},
ylabel={RMSE $\hat{\sigma}$},
xlabel style={font=\tiny},
ylabel style={font=\tiny},
ylabel shift=-5pt,
yticklabel shift=-3pt,
xlabel shift=-2pt,
xticklabel shift=-2pt,
xmin=0.6, xmax=9.4,
ymin=7.04320485183961, ymax=23.4505827720509,
width=\fwidth,
height=\fheight,
tick align=inside,
tick pos=left,
xmajorgrids,
x grid style={lightgray!92.02614379084967!black},
ymajorgrids,
y grid style={lightgray!92.02614379084967!black},
legend cell align={left},
legend style={at={(0.03,0.97)}, anchor=north west, draw=white!80.0!black}
]

\addplot [semithick, color1]
table {%
1 11.1794904170949
2 14.2050422802673
3 14.7144455864695
4 17.0400907307544
5 17.7828747630845
6 22.5396680761736
7 22.7047928665868
8 21.6762112871288
9 20.9839385959666
};
\addplot [semithick, color0, dashed]
table {%
1 7.78899475730376
2 9.54150144340876
3 10.9189328614128
4 12.0501843268286
5 13.2141775769773
6 14.3506468640314
7 15.3944392111337
8 16.395229597856
9 17.4662698171478
};

\end{axis}

\end{tikzpicture}
    \caption{Cluster 20}
    % \label{fig:cluster1}
  \end{subfigure}\hfill
    \begin{subfigure}[t]{0.16\textwidth}
    \centering  
    \setlength\belowcaptionskip{.1cm}
    \setlength\abovecaptionskip{0cm}
    \setlength\fwidth{1.25\textwidth}
    \setlength\fheight{0.85\textwidth}
    % This file was created by matplotlib2tikz v0.6.17.
\begin{tikzpicture}

\definecolor{color1}{rgb}{1,0.498039215686275,0.0549019607843137}
\definecolor{color0}{rgb}{0.12156862745098,0.466666666666667,0.705882352941177}

\pgfplotsset{every tick label/.append style={font=\tiny}}
\begin{axis}[
xlabel={Lag $L$ [5 minutes]},
ylabel={RMSE $\hat{\sigma}$},
xmin=0.6, xmax=9.4,
ymin=6.09918711448929, ymax=23.8315903985497,
width=\fwidth,
height=\fheight,
xlabel style={font=\tiny},
ylabel style={font=\tiny},
tick align=inside,
tick pos=left,
xmajorgrids,
x grid style={lightgray!92.02614379084967!black},
ymajorgrids,
y grid style={lightgray!92.02614379084967!black},
legend cell align={left},
ylabel shift=-5pt,
yticklabel shift=-3pt,
xlabel shift=-2pt,
xticklabel shift=-2pt,legend style={at={(0.03,0.97)}, anchor=north west, draw=white!80.0!black}
]
% \addlegendimage{no markers, color0}
% \addlegendimage{no markers, color1}
\addplot [semithick, color0, dashed]
table {%
1 6.90520544558294
2 10.8006134490972
3 13.3365258073745
4 14.4727100881015
5 15.5523723076944
6 17.1249402413108
7 16.977425731252
8 16.859967494956
9 16.6413200931726
};
\addplot [semithick, color1]
table {%
1 7.55155430502273
2 11.0954576557253
3 13.6235893574278
4 15.8168937131483
5 17.6429526069769
6 19.1746985998438
7 20.5971055731067
8 21.7820933478441
9 23.0255720674561
};

\end{axis}

\end{tikzpicture}
    \caption{Cluster 21}
    % \label{fig:cluster1}
  \end{subfigure}
  %   \begin{subfigure}[t]{0.16\textwidth}
  %   \centering  
  %   \setlength\belowcaptionskip{.1cm}
  %   \setlength\abovecaptionskip{-.1cm}
  %   \setlength\fwidth{1.25\textwidth}
  %   \setlength\fheight{1.1\textwidth}
  %   \input{figures/rmse_cluster/v2-cluster-22-rmse.tex}
  %   \caption{Cluster 22}
  %   % \label{fig:cluster1}
  % \end{subfigure}\hfill
  %\setlength\belowcaptionskip{-1cm}
  \caption{Cluster-based \gls{gpr} vs local-based \gls{brr} for the other clusters.}
  \label{fig:allClusters}
\end{figure*}
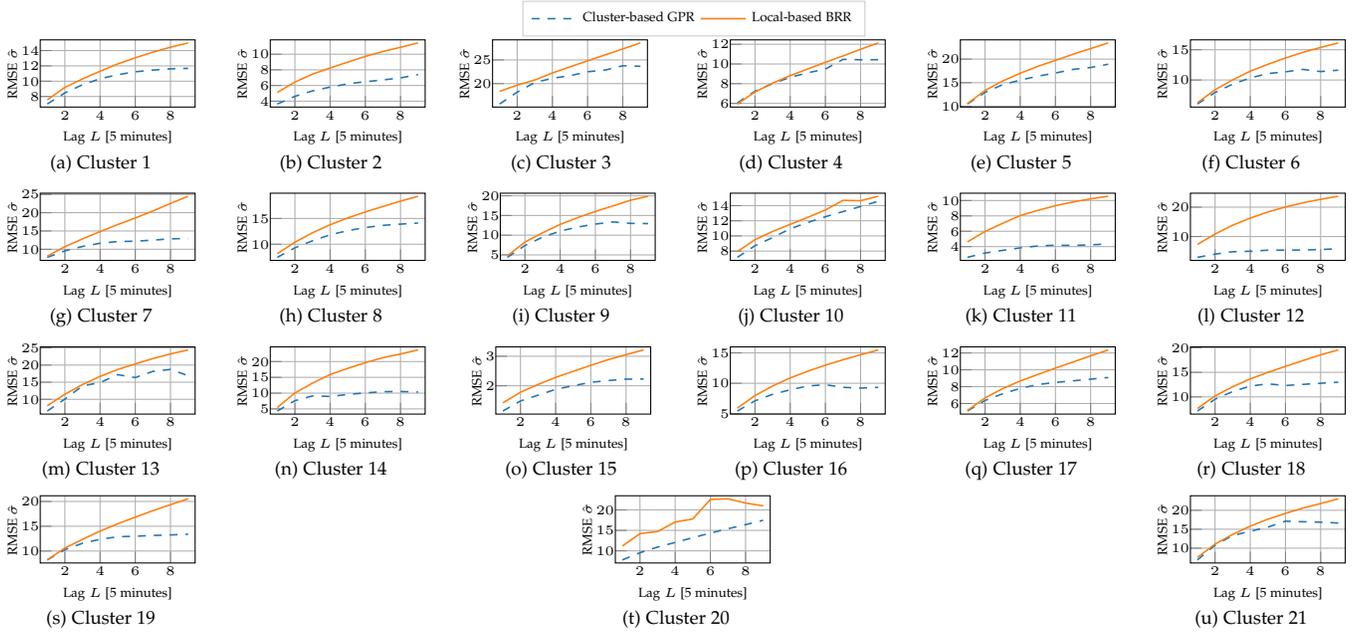

In order to compare the local- and the cluster-based methods, we report in Fig.~\ref{fig:rmseCluster0} the average \gls{rmse} $\hat{\sigma} = \mathbb{E}_{i \in \mathcal{C}_0}[\sigma_i]$ of the base stations in the set $\mathcal{C}_0$ associated to cluster $0$. As expected, the \gls{rmse} increases with the look-ahead step $L$. Among the local-based methods, the \gls{brr} gives the best results for all the values of the look-ahead step $L$, with a gain of up to 18\% and 55\% with respect to the \gls{gpr} and \gls{rfr} for $L=9$. The \gls{gpr}, instead, is the best among the cluster-based techniques, with an improvement up to 50\% from the \gls{rfr} (for $L = 1$). When comparing the local- and the cluster-based methods, the latter performs better, especially as the look-ahead step increases, since the curve of the \gls{rmse} for the cluster-based \gls{gpr} flattens around $\hat{\sigma} = 14.8$, while that for both the \gls{brr} and the local-based \gls{gpr} continues to increase. In this case, instead, for small values of $L$ the performance of local- and cluster-based methods is similar. Finally, the \gls{arma} model (with local information) underperforms all the others for small values of $L$, and is better than the local-based \gls{rfr} (with $W=1$) for $L \ge 4$.

Table~\ref{table:windowValues} reports the values of the window $W$ used in Fig.~\ref{fig:rmseCluster0diffWin} for the two best performing methods, the \gls{brr} and the \gls{gpr}. 
%When considering the impact of the window $W$ on the \gls{rmse}, 
By comparing Figs.~\ref{fig:rmseCluster0sameWin}, in which the window $W$ is fixed, and~\ref{fig:rmseCluster0diffWin}, where $W$ is selected for each step $L$ to yield the smallest \gls{rmse} $\hat{\sigma}$, it can be seen that the difference is minimal for the best performing methods (i.e., below 5\%), while it is more significant for the local-based \gls{rfr}. Moreover, the spatial dimension has more impact on the quality of the prediction than the temporal one. Indeed, while by changing $W$ the \gls{rmse} for the \gls{gpr} and \gls{brr} improves by up to 5\%, when introducing the multi-output prediction with the \gls{gpr} the \gls{rmse} decreases by up to 50\%. 
Differently from prior works in which the single user mobility is predicted~\cite{dong2013modeling}, we are indeed considering the number of users at a cell level, and, in this case, the possible transitions between neighboring cells are limited by the geography of the scenario, and by the available means of transport. Therefore, there exists a spatial correlation between the number of users in the neighboring base stations and the number of users in the considered base station at some time in the future, given that the mobility flows are constrained by the aforementioned factors.

Nonetheless, there exist still some limitations to the accuracy of the prediction of the number of users. Fig.~\ref{fig:tsExample} reports an example of the predicted (for $L = 3$, i.e., 15 minutes) and the true time series for two different base stations, with a high and low number of users. As it can be seen, the true time series have some daily patterns, but are also quite noisy. As a consequence, the predicted time series manage to track the daily pattern, but cannot predict the exact value of the number of users. This is more evident when the number of \glspl{ue} is low, as in Fig.~\ref{fig:lowNumUe}, which also exhibits smaller daily variations.

Finally, Fig.~\ref{fig:moreResults} reports additional results on the prediction performance of the cluster-based \gls{gpr}. In Fig.~\ref{fig:partial}, we compare the \gls{rmse} $\hat{\sigma}$ obtained on the testing dataset when using partial training datasets of different sizes, i.e., with 25, 50, 75 hours, or the complete training dataset (i.e., 100 hours). The \gls{rmse} monotonically decreases as the size of the training dataset increases, showing that there is room for improvement with a richer past history. Moreover, the difference is more marked when considering a higher prediction lag $L$, i.e., the full training dataset yields an \gls{rmse} which is 25\% smaller than the 25-hours dataset for $L=1$ and 40\% for $L=5$. 

Fig.~\ref{fig:residual} shows an example of residual analysis, which can help understand the limits of the cluster-based \gls{gpr} on the available San Francisco dataset. The y-axis reports the residual error $N_u(t) - \hat{N}_u(t)$, with $N_u(t)$ and $\hat{N}_u(t)$ the true and predicted number of users at time $t$, and the x-axis one of the features used in the prediction, i.e., the true number of users $N_u(t-1)$ at the previous time step $t-1$. Notice that the x-axis is quantized into 100 bins in order to improve the visualization of the residuals. It can be seen that the largest errors happen (infrequently) on the left part of the plot, i.e., when there is a sudden increase in the number of users in the base station, transitioning from a small $N_u(t-1)$ to a large $N_u(t)$. 

\subsection{Performance analysis for the other clusters}

Given the promising results of the cluster-based approach on the first cluster, we selected the best performing local- and cluster-based methods, i.e., respectively, the \gls{brr} and the \gls{gpr}, and performed the prediction on all the clusters reported in Fig.~\ref{fig:speccl}. The results are reported in Fig.~\ref{fig:allClusters} for each single cluster. The cluster-based method always outperforms the local-based one, and, in most cases, also exhibits a smaller \gls{rmse} for small values of the look-ahead step $L$, contrary to what happens for cluster 0. The reduction in the average \gls{rmse} over all the clusters $\mathbb{E}_{clusters}[\hat{\sigma}]$ is 18.3\% for $L=1$ (from $\mathbb{E}_{clusters}[\hat{\sigma}] = 7.24$ to $\mathbb{E}_{clusters}[\hat{\sigma}] = 6.11$) and increases up to 53\% for $L=9$ (from $\mathbb{E}_{clusters}[\hat{\sigma}] = 17.42$ to $\mathbb{E}_{clusters}[\hat{\sigma}] = 11.34$).

\subsection{Possible Applications}
\label{sec:route}

The results presented in Figs.~\ref{fig:rmseCluster0} and~\ref{fig:allClusters} show that the cluster-based method is more capable than local-based ones to capture the user dynamics in the cellular network. The prediction of the number of users in a base station can be used to optimize the performance of the network in a number of different ways: for example, it can enable predictive load-balancing, bearer pre-configuration, scaling of \gls{ran} resources, sleeping periods for base stations, and so on. We believe that the increase in the prediction accuracy that the cluster-based method yields can be beneficial to practically enable these anticipatory and prediction-based optimizations.

\begin{table*}[t]
  \centering
  \setlength\belowcaptionskip{-.3cm}
  \renewcommand{\arraystretch}{1}
  \footnotesize
  \begin{tabular}{l|cccc|cccc|cccc}
    \toprule
    & \multicolumn{4}{c|}{Feb. 23rd, 19:00} & \multicolumn{4}{c|}{Feb. 24th, 19:00} & \multicolumn{4}{c}{Feb. 24th, 19:20} \\\midrule
    Route & \cellcolor{blue!25} R1 & \cellcolor{red!25} R2 & \cellcolor{green!25} R3 & \cellcolor{purple!25} R4 & \cellcolor{blue!25} R1 & \cellcolor{red!25} R2 & \cellcolor{green!25} R3 & \cellcolor{purple!25} R4 & \cellcolor{blue!25} R1 & \cellcolor{red!25} R2 & \cellcolor{green!25} R3 & \cellcolor{purple!25} R4\\\midrule
    $\hat{S}$ [Mbit/s] &  1.93 & \cellcolor{red!25} 2.51 & 2.36 & \cellcolor{gray!25} 2.74 & 1.72 & 2.00 & \cellcolor{green!25}2.28 & \cellcolor{gray!25} 2.89 & 2.05 & \cellcolor{red!25}2.49 & 1.98 & \cellcolor{gray!25} 2.86 \\
    $D_{o, \max}$ [s] & \cellcolor{blue!25}133.47 & 157.8 & 172.5 & \cellcolor{gray!25} 171.2 & 152.4 & 157 & \cellcolor{green!25}148.8 & \cellcolor{gray!25} 169.1 & 152.1 & \cellcolor{red!25}123.7 & 172.5 & \cellcolor{gray!25}116.7 \\
    \bottomrule
  \end{tabular}
  \caption{Average throughput $\hat{S}$ and maximum outage duration $D_{o, \max}$ on the four itineraries from Fig.~\ref{fig:routesMap}, for different departure times in February 2017. For the three routes with a similar duration, the colored cells represent the best route for the metric of interest.}
  \label{table:metrics}
\end{table*}

\begin{figure}[t]
  \centering
  \setlength\belowcaptionskip{-.3cm}
  \includegraphics[width=.8\columnwidth]{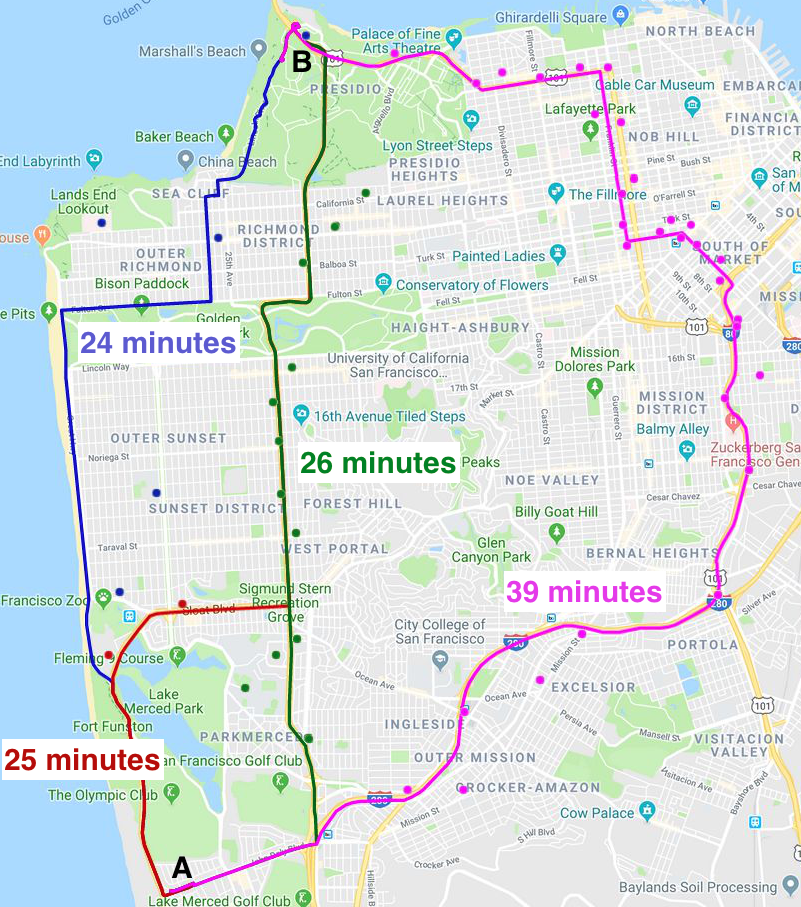}
  \caption{Map of the routes. The dots represent the visited base stations. Notice that, for route 2 (the red one), several base stations are shared with either the blue or the green routes.}
  \label{fig:routesMap}
\end{figure}

Moreover, network operators can exploit the prediction to offer novel services to the end users. For example, consider a vehicle that has to travel from point $A$ to point $B$ in an area covered by cellular service. While on the journey, the passengers may want to participate in a conference call, or, if not driving, surf the web or stream multimedia content. Therefore, given the choice of multiple routes with similar \glspl{eta}, the passengers may prefer to choose an itinerary with a slightly higher \gls{eta} but with a better network performance, because, for example, it crosses an area with a better coverage, or with fewer users. 
This becomes particularly relevant in view of the envisioned transition to an autonomous driving future, in which active driving might not be required and working or getting entertained in the car will become a common trend. In order to address this need, cellular network operators can exploit the architecture described in Sec.~\ref{sec:ctrl} and the prediction of the number of active users in the cells to offer anticipatory services to the end users and inform them on which is the best route for their journey.

Fig.~\ref{fig:routesMap} shows an example of three different routes in the San Francisco area, together with different metrics in Table~\ref{table:metrics}, which are computed from the predicted number of users, in different dates. It can be seen that the fastest route (i.e., route 1, in blue), is not always the one offering the best service in the three departure times considered. 
For the first three routes, which have a similar travel time, the best route changes at different departure times: for the throughput, on Feb. 23rd, 19:00, route 2 (red) is better than the others, while in the next day at the same time the best itinerary is route 3 (green). 
When considering also the longest route, which still leads from the origin to the destination, but takes 50\% more time than the shortest, it can be seen that it always offers the highest average throughput, but, in some cases, is one of the worst in terms of maximum outage duration.

This example shows that, according to the users' needs, it is possible to identify and select different routes that have different performance in terms of throughput and outage. Moreover, the routes are ranked differently according to various departure times. Therefore, simply applying the analytics given by the average statistics from the previous days may not yield reliable results in terms of routes ranking. This makes the case for adopting the medium-term prediction techniques described in this Section to forecast the expected value of the metrics in the time interval in which the user will travel, based on the actual network conditions for the same day.

\section{Conclusions}
\label{sec:concl}

Machine learning, software-defined networks and edge cloud will be key components of the next generation of cellular networks. In this paper we investigated how these three elements can be jointly used in the system design for 5G networks, providing insights and results based on a dataset collected from hundreds of base stations of a major U.S. cellular network in two different cities for more than a month. 

After reviewing the relevant state of the art, we investigated how it is possible to practically introduce machine learning and big-data-based policies in 5G cellular networks. We proposed an overlay architecture on top of \gls{3gpp} \gls{nr}, in which multiple layers of controllers with different functionalities are used to collect the data from the \gls{ran}, process it and use it to infer intelligent policies that can be applied to the cellular network. 

Next, we discussed a first application of the proposed architecture, i.e., a data-driven association algorithm between the \glspl{gnb} and the \gls{ran} controllers themselves. We described a clustering solution that limits the interactions among different controllers to minimize the need for inter-controller synchronization and reduce the control plane latency, and evaluated the performance of the proposed approach using data from a real network.

Then, we outlined a second possible application enabled by our architecture, providing an extensive set of results related to the prediction accuracy of the number of users in base stations, using one month of data collected from the San Francisco base stations. In particular, we showed how the usage of the cluster-based architecture proposed in this paper can reduce the prediction error. With respect to a solution in which each base station tries to perform the regression based solely on its own data, as realized by a completely distributed architecture (e.g., in \gls{lte}), the controller-based design makes it possible to aggregate data from multiple neighboring base stations, and to predict a vector with the number of users in the nodes associated to the controller. This captures the spatial correlation given by the mobility of users, and, especially when  increasing the temporal horizon of the prediction, reduces the \gls{rmse} by up to 53\%.
Finally, we also described some prediction-enabled use cases, either to control the network itself, or to offer innovative predictive services to network users, for example by recommending different driving itineraries to improve the user experience in the network. We illustrated a real example in the San Francisco area, showing how the fastest route does not necessarily yield the best throughput, or the minimum outage, and that the best itinerary according to these metrics (which we derive from the number of users in each base station) may differ according to the departure time, so that a prediction-based approach is useful.

We believe that this paper addresses for the first time several issues related to the practical deployment of machine learning techniques in 5G cellular networks, providing results and conclusions based on a real-network dataset. As future work, we will test different prediction algorithms (e.g., neural networks) to understand if it is possible to improve even more the prediction accuracy, and will extend the regression to other relevant metrics in the network (e.g., the number of handovers, the utilization), to verify the limits of what can be actually predicted in a cellular network.

\bibliographystyle{IEEEtran}
\bibliography{bibl.bib}

\begin{IEEEbiography}
    [{\includegraphics[width=0.9in,clip,keepaspectratio]{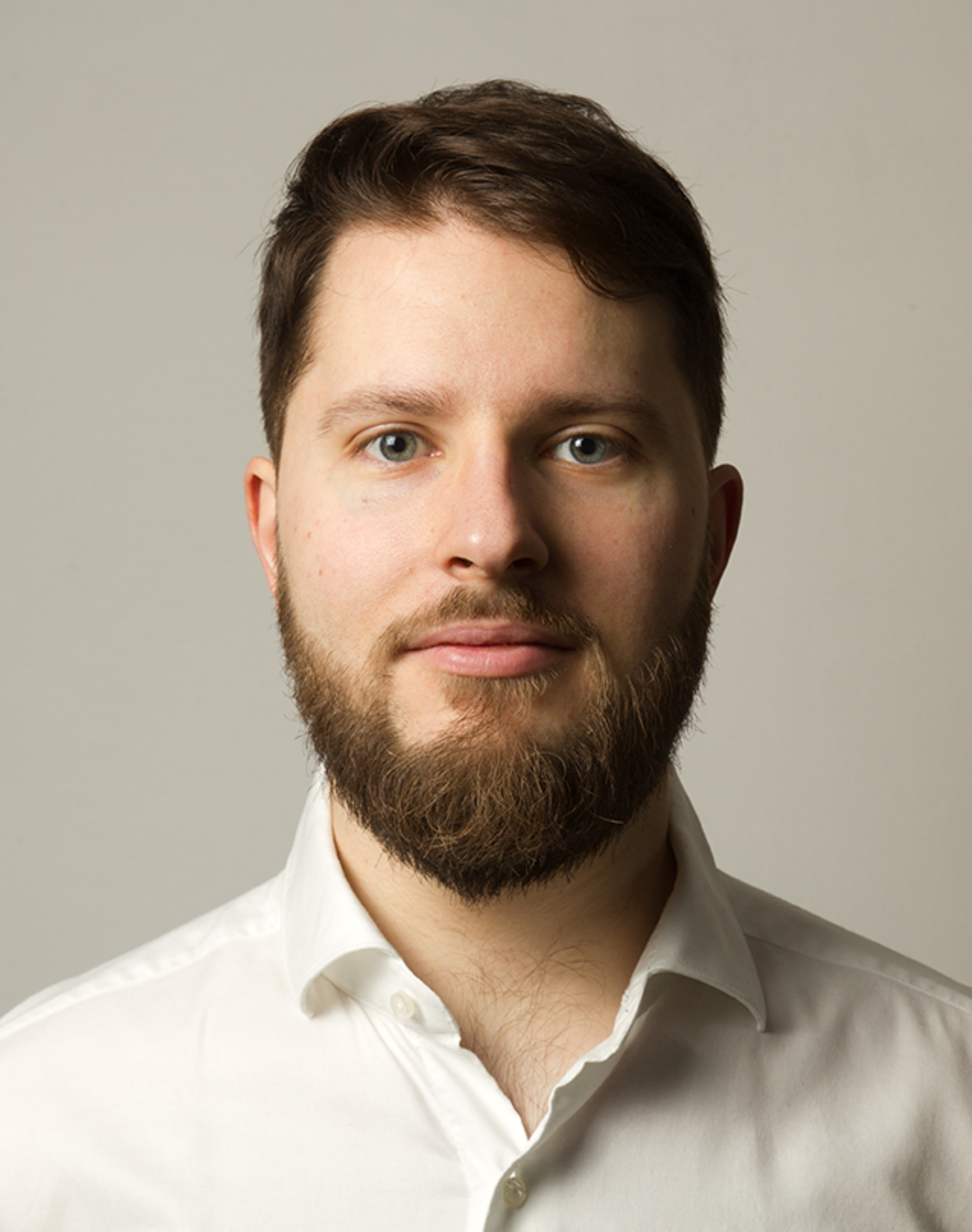}}]{Michele Polese}
[M'20] is an Associate Research Scientist at Northeastern University, Boston, since March 2020. He received his Ph.D. at the Department of Information Engineering of the University of Padova in 2020. He also was an adjunct professor and postdoctoral researcher in 2019/2020 at the University of Padova. During his Ph.D., he visited New York University (NYU), AT\&T Labs in Bedminster, NJ, and Northeastern University, Boston, MA. He collaborated with several academic and industrial research partners, including Intel, InterDigital, NYU, AT\&T Labs, University of Aalborg, King's College and NIST.
He was awarded with an Honorable Mention by the Human Inspired Technology Research Center (HIT) (2018), the Best Journal Paper Award of the IEEE ComSoc CSIM Technical Committee on 2019, and the Best Paper Award at WNS3 2019. His research interests are in the analysis and development of protocols and architectures for future generations of cellular networks (5G and beyond), and in the performance evaluation of complex networks.
\end{IEEEbiography}
\vspace{0cm}
\begin{IEEEbiography}
    [{\includegraphics[width=0.9in,clip,keepaspectratio]{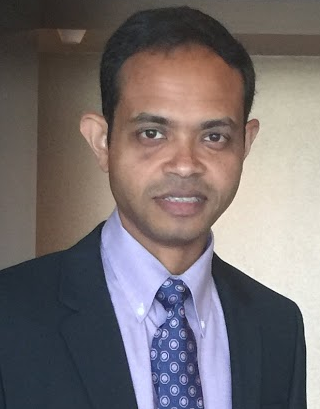}}]{Rittwik Jana}
is a Director of Inventive Science at AT\&T Labs Research. His research interests span the design of a disaggregated RAN Intelligent Controller (RIC), service composition of VNFs using TOSCA, model driven control loop and automation in ONAP, networked video streaming and cellular networks and systems. Rittwik earned a Ph.D. in Telecommunications Engineering from the Australian National University, Australia in 2000. He obtained the AT\&T Science and Technology medal in 2016 and the Jack Neubauer Memorial vehicular technology society award in 2017.
    \end{IEEEbiography}
\vspace{-1cm}
\begin{IEEEbiographynophoto}{Velin Kounev}
    received his doctorate from the University of Pittsburgh, Pittsburgh, PA, USA in 2015, and the M.S. degree in telecommunications from the same, in 2007.  From 2007 to 2011, he was a Software Engineer and a Communication System Architect for driverless real-time train control systems. He is currently working as Principle Inventive Scientist at AT\&T Labs Research.
\end{IEEEbiographynophoto}
\vspace{-0.3cm}
    \begin{IEEEbiography}[{\includegraphics[width=0.9in,clip,keepaspectratio]{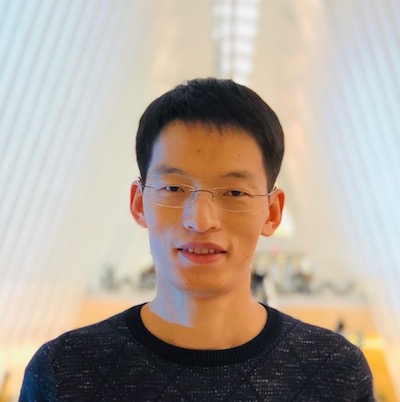}}]{Ke Zhang}
      joined AT\&T Labs as a Senior Member of Technical Staff in Nov 2016. Since then, he has been focusing R\&D on spatial-temporal data mining and machine learning/deep learning for spatially distributed telecommunication network optimization and planning. Before that he received his Ph.D. degree in Information Science from University of Pittsburgh, with research on location-based social media data mining and applied machine learning, to understand and model the social, spatial, temporal and network dynamics of user behaviors as well as its applications to local economy. He also has a MS degree with background of signal processing in wireless sensor networks, and a BS degree in Telecommunication Engineering.
    \end{IEEEbiography}
\vspace{-0.3cm}
\begin{IEEEbiographynophoto}{Supratim Deb}
   had been a researcher with AT\&T Labs and Bell Labs and is
currently employed by Facebook. He obtained his Ph.D. from 
the University of Illinois at Urbana-Champaign in the area of communication networks (2003). 
Following his Ph.D., he had a post-doctoral stint at MIT. His research interests are in the broad 
areas of data-driven system design and networking.
\end{IEEEbiographynophoto}
\vspace{-0.3cm}
    \begin{IEEEbiography}
    [{\includegraphics[width=0.99in,clip,keepaspectratio]{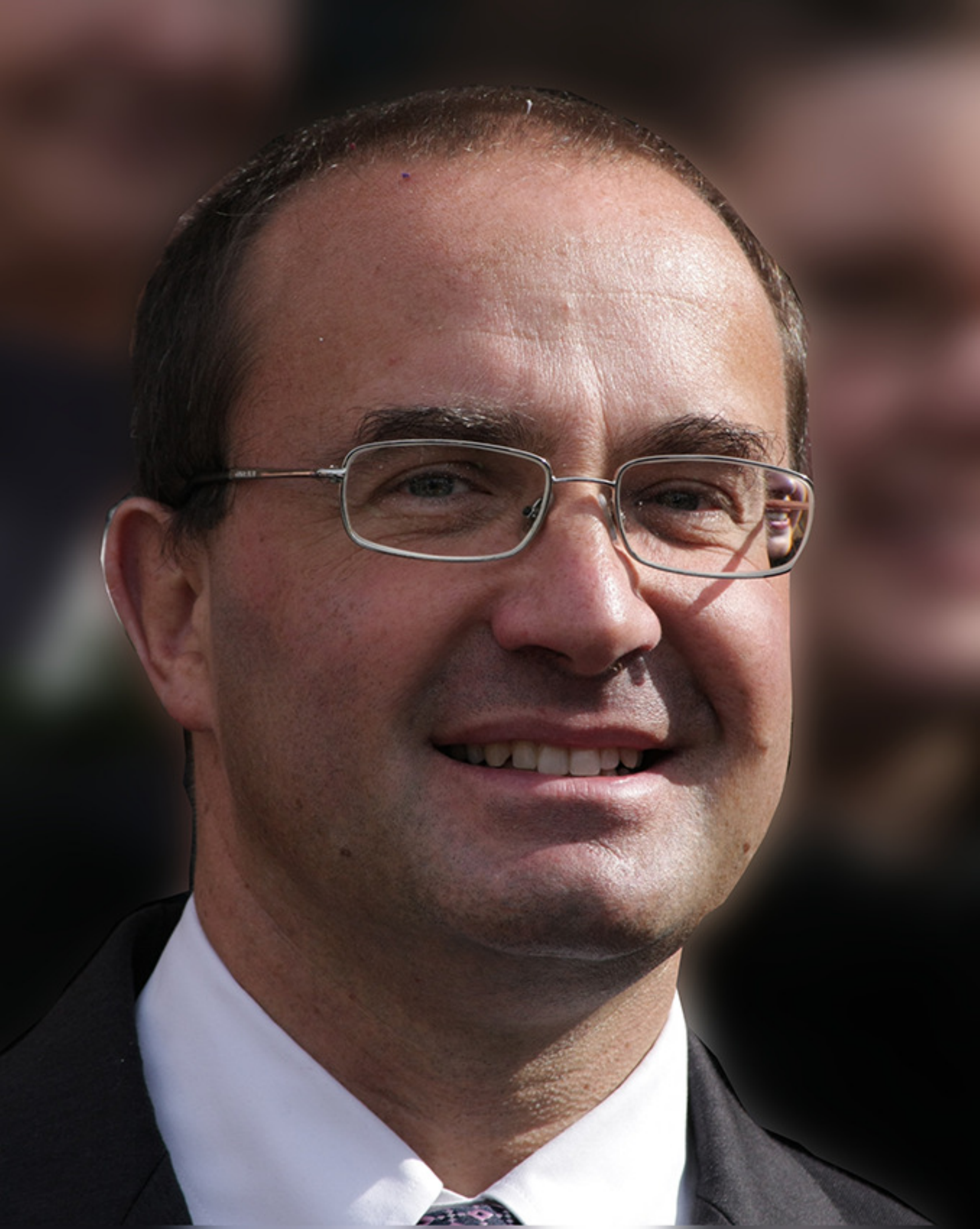}}]{Michele Zorzi}
        [F'07] received his Laurea and PhD degrees in electrical engineering from the University of Padova in 1990 and 1994, respectively. During academic year 1992-1993 he was on leave at the University of California at San Diego (UCSD). In 1993 he joined the faculty of the Dipartimento di Elettronica e Informazione, Politecnico di Milano, Italy. After spending three years with the Center for Wireless Communications at UCSD, in 1998 he joined the School of Engineering of the University of Ferrara, Italy, where he became a professor in 2000. Since November 2003 he has been on the faculty of the Information Engineering Department at the University of Padova. His present research interests include performance evaluation in mobile communications systems, WSN and Internet of Things, cognitive communications and networking, 5G mmWave cellular systems, vehicular networks, and underwater communications and networks. 
        He is the recipient of several awards 
from the IEEE Communications Society, including the Best Tutorial Paper 
Award (2008), the Education Award (2016), and the Stephen O. Rice Best 
Paper Award (2018).
        He was Editor-In-Chief of IEEE Wireless Communications from 2003 to 2005, IEEE Transactions on Communications from 2008 to 2011 and IEEE Transactions on Cognitive Communications and Networking from 2014 to 2018. He served ComSoc as a Member-at-Large of the Board of Governors from 2009 to 2011, as Director of Education and Training from 2014 to 2015, and as Director of Journals from 2020 to 2021.
    \end{IEEEbiography}

\end{document}